\documentclass[12pt]{article}
\usepackage{latexsym}
\usepackage{amsmath,amsfonts}
\usepackage{times}
\allowdisplaybreaks[4]

\newcommand{\rf}[1]{(\ref{#1})}

\renewcommand{\thefootnote}{\fnsymbol{footnote}}
\newcommand{\newsection}{\setcounter{equation}{0}\section}

\def\appendix#1{\addtocounter{section}{1}\setcounter{equation}{0}
\renewcommand{\thesection}{\Alph{section}}
\section*{Appendix \thesection\protect\indent \parbox[t]{11.15cm}{#1}}
\addcontentsline{toc}{section}{Appendix \thesection\ \ \ #1}}

\def\be{\begin{equation}}
\def\ee{\end{equation}}
\def\beq{\begin{eqnarray}}
\def\eeq{\end{eqnarray}}

\def\parline{\,\partial\kern -0.55em /\,\,}

\def\half{{\frac{1}{2}}}

\def\DD{{\cal D}}

\def\FF{{\cal F}}
\def\GG{{\cal G}}
\def\II{{\cal I}}
\def\FF{{\cal F}}
\def\KK{{\cal K}}
\def\LL{{\cal L}}

\def\Scl{{\cal S}}
\def\TT{{\cal T}}

\def\VV{{\cal V}}

\def\Fbf{{\bf F}}

\def\Ibf{{\bf I}}

\def\Tbf{{\bf T}}

\def\noinbf#1{\noindent {\bf #1}}

\def\Dch{\check{D}}
\def\Jch{\check{J}}
\def\Kch{\check{K}}
\def\Pch{\check{P}}
\def\Tch{\check{T}}
\def\phich{\check{\phi}}

\def\smtwo{{\scriptscriptstyle (2)}}

\def\smtwo{{\scriptscriptstyle (2)}}

\def\Dsf{{\sf D}}
\def\Jsf{{\sf J}}
\def\Ksf{{\sf K}}
\def\Psf{{\sf P}}

\def\mun{{\underline{m}}}
\def\nun{{\underline{n}}}

\def\lun{{\underline{l}}}

\def\subsetext{{\,\lower-0.2ex\hbox{${\scriptstyle+}$}}
{\kern-1.3ex\hbox{$\supset$}\,}}

\def\phib{\bar{\phi}}

\def\Ib{\bar{I}}
\def\Fb{\bar{F}}

\def\(A)dS{{\rm (A)dS}}

\def\st{{\rm st}}

\def\lin{{\rm lin}}

\def\Tr{{\rm Tr}}

\def\new{{\rm new}}

\def\genrm{{\rm gen}}

\def\Hdr{{\rm Hdr}}

\def\Tsm{{\scriptscriptstyle T}}
\def\CYMsm{{\scriptscriptstyle \rm CYM}}

\def\addsm{{\scriptscriptstyle \rm add}}

\def\asf{{\sf a}}
\def\bsf{{\sf b}}
\def\csf{{\sf c}}

\def\subsm{{\scriptscriptstyle \rm sub}}

\def\Hdr{{\scriptscriptstyle \rm Hdr}}
\def\Odrsm{{\scriptscriptstyle \rm Odr}}

\def\ppp{p}

\begin{document}


\begin{flushright}
FIAN-TD-2024-09  \hspace{2cm} \ \\
arXiv: 2406.19182 [hep-th],\  V3\\
\end{flushright}

\vspace{1cm}

\begin{center}

{\Large \bf Conformal Yang-Mills field in (A)dS space }

\vspace{2.5cm}

R.R. Metsaev%
\footnote{ E-mail: metsaev@lpi.ru
}

\vspace{1cm}

{\it Department of Theoretical Physics, P.N. Lebedev Physical
Institute, \\ Leninsky prospect 53,  Moscow 119991, Russia }

\vspace{3.5cm}

{\bf Abstract}

\end{center}

Ordinary-derivative (second-derivative) Lagrangian formulation of classical conformal Yang-Mills field in the (A)dS  space of six, eight, and ten dimensions is developed. For such conformal field, we develop two gauge invariant Lagrangian formulations which we refer to as generic formulation and decoupled formulation.
In both formulations, the usual Yang-Mills field is accompanied by additional vector and scalar fields where the scalar fields are realized as Stueckelberg fields. In the generic formulation, the usual Yang-Mills field is realized as  a primary field, while the additional vector fields are realized as auxiliary fields. In the decoupled formulation, the usual Yang-Mills field is realized as massless field, while the additional vector fields together with the Stueckelberg  are realized as massive fields. Some massless/massive fields appear with the wrong sign of kinetic terms, hence demonstrating explicitly that the considered models are not unitary. The use of embedding space method allows us to treat the isometry symmetries of (A)dS space manifestly and obtain conformal transformations of fields in a relatively straightforward way. By accompanying  each vector field by the respective gauge parameter, we introduce an extended gauge algebra. Levy-Maltsev decomposition of such algebra is noted. Use of the extended gauge algebra setup allows us to present concise form for the Lagrangian and gauge transformations of the conformal Yang-Mills field. Higher-derivative representation of the Lagrangian is also obtained.

\vspace{0.5cm}

\phantom{a} \hspace{1.7cm} Embedding space method.


{\bf
\large \noindent Contents\small

\noindent 1 \quad Introduction \hfill 2

\medskip
\noindent 2 \quad Conformal algebra in (A)dS and extended gauge algebra. Notation and conventions \hfill 4

\medskip
\noindent 3 \quad Conformal Yang-Mills field in $(A)dS_6$ space. Generic formulation \hfill 13

\medskip
\noindent 4 \quad Conformal Yang-Mills field in $(A)dS_6$ space. Decoupled formulation \hfill 17

\medskip
\noindent 5 \quad Conformal Yang-Mills field in $(A)dS_8$ space. Generic formulation \hfill 21

\medskip
\noindent 6 \quad Conformal Yang-Mills field in $(A)dS_8$ space. Decoupled formulation \hfill 24

\medskip
\noindent 7 \quad Conformal Yang-Mills field in $(A)dS_{10}$ space. Generic formulation \hfill 29

\medskip
\noindent 8 \quad Conformal Yang-Mills field in $(A)dS_{10}$ space. Decoupled formulation \hfill 34

\medskip
\noindent 9 \quad Conclusions \hfill 40

\medskip
\noindent Appendix A \quad Relations for embedding and intrinsic spaces \hfill 42

\medskip
\noindent Appendix B \quad Extended gauge algebra and field redefinitions \hfill 43

\medskip
\noindent Appendix C \quad Map of fields of generic formulation to fields of decoupled formulation \hfill 44

\medskip
\noindent Appendix D \quad Flat space limit of $K^A$- transformations \hfill 46
}

\renewcommand{\thefootnote}{\arabic{footnote}}
\setcounter{footnote}{0}

\newsection{\large Introduction }

Conformal fields have actively been studied, among other things, in the context of conformal supergravity theories (for review, see Ref.\cite{Fradkin:1985am}). In view of increasing interest in field dynamics in the $(A)dS$ space, conformal fields propagating in the $(A)dS$ space have been investigated by many researchers. As an example we mention the investigations of conformal graviton in $(A)dS_4$ in Refs.\cite{Fradkin:1982xc}-\cite{Deser:2012qg}. Arbitrary spin conformal field in $(A)dS_{d+1}$ was considered in Ref.\cite{Tseytlin:2013jya}, where it was conjectured that higher-derivative kinetic operator of such conformal field is factorized into product of second-derivative kinetic operators of massless, partial-massless, and massive fields (see also Ref.\cite{Joung:2012qy}). This conjecture was supported by the explicit  construction of the second-derivative gauge invariant Lagrangian for arbitrary spin conformal $(A)dS$  field in Ref.\cite{Metsaev:2014iwa} (see also Refs.\cite{Nutma:2014pua,Grigoriev:2018mkp,Hutchings:2024qqf}). Also we mention the study of Weyl action for two-column mixed-symmetry field in $ (A)dS$ in Ref.\cite{Joung:2016naf} and the computation of two-point corellator functions for some special conformal $(A)dS$ fields in  Ref.\cite{Farnsworth:2024iwc}.%
\footnote{In the frame of flat space,  Lagrangian for all mixed-symmetry conformal fields was found in Ref.\cite{Vasiliev:2009ck} (related studies may be found in Refs.\cite{Shaynkman:2004vu,Dobrev:2012mea}).
}
Attempts have been made also in the study of conformal fields in curved backgrounds (see, e.g. Refs.\cite{Grigoriev:2016bzl}-\cite{Kuzenko:2020opc}). List of references on this topic may be found in Ref.\cite{Ponds:2021xvs}.

Kinetic operators of the most conformal fields involve higher-derivatives. Using auxiliary fields, it is possible to obtain a second-derivative formulation which we refer to as ordinary-derivative formulation. The ordinary-derivative gauge invariant and Lorentz covariant formulation of free arbitrary spin conformal fields in $R^{d,1}$ with odd $d\geq 3$ was developed in Refs.\cite{Metsaev:2007fq,Metsaev:2007rw}.%
\footnote{ Generalization of the result in Refs.\cite{Metsaev:2007fq,Metsaev:2007rw} to long and partial-short conformal fields was done in Ref.\cite{Metsaev:2016oic}.
}
Attractive feature of the ordinary-derivative formulation is that the two-derivative contributions to kinetic operators for conformal scalar, vector, graviton, and higher-spin  fields are realized as the respective Klein-Gordon, Maxwell, Einstein-Hilbert, and Fronsdal kinetic operators.

In Ref.\cite{Metsaev:2023qif}, using ordinary-derivative approach, we obtained Lagrangian and $F^3$-vertices  for conformal Yang-Mills (YM) field in $R^{d,1}$ with odd $d\geq 5$. Although the kinetic operators in Ref.\cite{Metsaev:2023qif} are realized as the conventional Klein-Gordon and Maxwell kinetic operators, unfortunately, some kinetic terms turn out to be non-diagonal with respect to fields. On the other hand, it was noticed many years ago in Ref.\cite{Fradkin:1982xc}, that the four-derivative kinetic operator of the conformal graviton in $(A)dS_4$ is factorized into a product of two second-derivative kinetic operators. This factorization property of the four-derivative kinetic operator of conformal  graviton can be realized by using the ordinary-derivative Lagrangian in Ref.\cite{Deser:2012qg}. It turns out that the full kinetic term of the ordinary-derivative Lagrangian of the conformal graviton in $(A)dS_4$ is realized as a sum of the kinetic term of spin-2 massless field and the kinetic term of spin-2 partial-massless field. As noted above, the ordinary-derivative  Lagrangian of free arbitrary spin conformal field in $(A)dS_{d+1}$ was obtained in  Ref.\cite{Metsaev:2014iwa}. The upshot of the discussion in Ref.\cite{Metsaev:2014iwa} is that all second-derivative kinetic terms entering the ordinary-derivative Lagrangian of arbitrary spin conformal field in $(A)dS$ turn out to be diagonal with respect to fields. Motivated by this fact we decided to develop ordinary-derivative Lagrangian formulation of the conformal YM field in $(A)dS$. This is what we do in this paper.
Other motivations beyond the scope of the present paper are discussed in Conclusions.

We now provide a brief review of some results in Ref.\cite{Metsaev:2023qif} and formulate our aim in this paper more explicitly. In higher-derivative approach, the most general Lagrangian of conformal YM field in $R^{d,1}$ can schematically be written as
\be \label{manus-09062024}
\LL_\genrm^\Hdr = \sum_\alpha \sum_{k=2}^{(d+1)/2} g_{\alpha,k}\Ibf_{\alpha, k}^\Hdr\,, \qquad \qquad \Ibf_{\alpha,k}^\Hdr : = \sum_{n=k}^{(d+1)/2} a_{\alpha,k,n} D^{d+1-2n} \Fbf^n\,,
\ee
where $\Ibf_{\alpha,k}^\Hdr$ is a conformal invariant constructed out of field strength $\Fbf$ and covariant derivative $D$. The $a_{\alpha,k,n}$ are some coefficients, while the $g_{\alpha,k}$ are coupling constants. The conformal invariant $\Ibf_{\alpha,k}^\Hdr$ consists of terms of the $k$th order and higher than the $k$th order in the YM field. For $R^{d,1}$, $d\geq5$, there are several conformal invariants $\Ibf_{\alpha,k}^\Hdr$ and the corresponding coupling constants $g_{\alpha,k}$. If $d\geq 5$, then, for each value of $k=2,3$, there is only one conformal invariant which we denote simply as $\Ibf_k^\Hdr$, $k=2,3$.%
\footnote{ For $R^{5,1}$ this statement is well-known, while for $R^{d,1}$, $d>5$, this statement is proved in Ref.\cite{Metsaev:2016rpa}.}
For $R^{5,1}$, the explicit expressions for the conformal invariants $\Ibf_2^\Hdr$ and $\Ibf_3^\Hdr$ are well known, while, for $R^{7,1}$ and $R^{9,1}$, the explicit expressions for these invariants were found in Ref.\cite{Metsaev:2023qif}. As far as we know, explicit expressions for the conformal invariants $\Ibf_2^\Hdr$ and $\Ibf_3^\Hdr$, when $d\geq 11$, and conformal invariants $\Ibf_{\alpha, k}^\Hdr$, $k=4,5,\ldots,(d+1)/2$, when $d\geq 7$, have not been discussed in literature.
In Ref.\cite{Metsaev:2023qif}, by exploiting our ordinary-derivative approach we have found two conformal invariants in $R^{d,1}$ with odd $d\geq 5$. Using the notation $I_2^\Odrsm$, $I_3^\Odrsm$ for those two conformal invariants, we note that $I_2^\Odrsm$, $I_3^\Odrsm$ are cousins of $\Ibf_2^\Hdr$, $\Ibf_3^\Hdr$ above discussed. Using the notation $F$ for a set of field strengths entering our ordinary- derivative approach, we also note that the $I_2^\Odrsm \sim F^2$ and hence involves two derivatives, while $I_3^\Odrsm \sim F^3$ and hence involves three derivatives. In Ref.\cite{Metsaev:2023qif}, by analogy with the usual Poincar\'e invariant theory of YM field, we have suggested refer $I_2^\Odrsm$ to as Lagrangian of the conformal YM field in $R^{d,1}$. Our aim in this paper is to find $I_2^\Odrsm$ for $(A)dS_{d+1}$, when $d=5,7,9$. Such $I_2^\Odrsm$ we refer to as Lagrangian of the classical conformal YM field in $(A)dS$ and denote it as $\LL_\CYMsm$, while, for simplicity, the corresponding coupling constant is fixed to be $g_\CYMsm=1$.

We develop two formulations of the conformal YM field in $(A)dS$, which we refer to as generic and decoupled formulations. The generic formulation has the following two key features: 1) In the flat space limit, Lagrangian and gauge transformations for conformal YM in $(A)dS$ have smooth limit to the ones for conformal YM field in flat space found in Ref.\cite{Metsaev:2023qif} - this simplifies the whole analysis; 2) Field content is given by primary field, auxiliary and Stueckelberg fields - this simplifies finding of the conformal algebra transformations. Undesirable feature of the generic formulation is that some kinetic terms are non-diagonal with respect to fields.
The decoupled formulation has the following two key features: 1) Field content is realized as one massless field and several massive fields; 2) To quadratic order in fields, Lagrangian is realized as a sum of the standard diagonal kinetic terms for massless and massive fields. Undesirable feature of the decoupled formulation is that a procedure of the derivation of Lagrangian, gauge transformations and conformal algebra transformations turns out to be complicated. We find Lagrangian, gauge transformations, and conformal algebra transformations in the decoupled formulation by using the generic formulation and a suitable map of fields of the generic formulation to fields of the decoupled formulation.

This paper is organized as follows.

In Sec.\ref{not-conv}, the notation and conventions used in this paper are collected. We outline a realization of the conformal algebra in the framework of the embedding space method and describe set of fields we use to develop the generic and decoupled formulations of conformal YM field. Setup of an extended gauge algebra which provides us the possibility to represent our results in the concise and elegant form is also presented.

Conformal theories  are very popular in six dimensions.%
\footnote{Conformal Weyl invariants in $6d$ were studied in Refs.\cite{Bonora:1985cq}. Conformal gravity in $6d$ was considered in Ref.\cite{Metsaev:2010kp}, while conformal supergravity models in $6d$ were investigated in Refs.\cite{Linch:2012zh}. Various interesting discussions of $6d$ conformal gravity may be found in Refs.\cite{Lu:2013hx}.}
In Secs.\ref{six-gen},\ref{six-fac}, we start therefore with the presentation of the generic and decoupled formulations of conformal YM field in $(A)dS_6$.  We present our ordinary-derivative Lagrangian, gauge transformations, and conformal algebra transformations of fields and field strengths. Explicit form of the extended gauge algebra is presented.

Secs.\ref{eight-gen}-\ref{ten-fac} are devoted to the conformal YM field in $(A)dS_8$ and $(A)dS_{10}$. Eight and ten dimensions are also popular in various studies of conformal fields.%
\footnote{Conformal Weyl invariants in $8d$ and $10d$ were studied in Refs.\cite{Boulanger:2004zf} and \cite{Joung:2021bhf} (see also Ref.\cite{Boulanger:2007ab}). Attempt to build conformal supergravity in $10d$ was made in Ref.\cite{Bergshoeff:1982az}.
}
For this reason, we decided to present our results for $(A)dS_8$ and $(A)dS_{10}$ separately. Generic and decoupled formulations of conformal YM field in $(A)dS_8$ are presented in the respective Sec.\ref{eight-gen} and Sec.\ref{eight-fac}, while the generic and decoupled formulations of conformal YM field in $(A)dS_{10}$ are presented in the respective Sec.\ref{ten-gen} and Sec.\ref{ten-fac}. For the case of $(A)dS_{10}$, we note that the extended gauge algebra turns out to deformed as compared to its cousin in the flat space. We show that this deformation is related to the parametrization of space of fields in our approach. By a suitable field redefinitions in $(A)dS_{10}$ we can match the extended gauge algebras in $(A)dS_{10}$ and $R^{9,1}$.

In Appendix A, we present some useful relations for various quantities defined in the embedding and intrinsic spaces. In Appendix B, we discuss the extended gauge algebra and field redefinitions. Updated presentation of the extended gauge algebra for conformal YM field in flat space obtained in Ref.\cite{Metsaev:2023qif} is given. In Appendix C, a map of fields of generic formulation to fields of decoupled formulation is described. In Appendix D, we discuss how our conformal algebra transformations in $(A)dS$ are related to the ones in the flat space.

\newsection{\large  Conformal algebra in (A)dS and extended gauge algebra. Notation and conventions } \label{not-conv}

\noinbf{Embedding space and isometry symmetries of $(A)dS_{d+1}$ space}. The $(A)dS_{d+1}$ space can be described as hyperboloid embedded in a $(d+2)$-dimensional pseudo-Euclidean space,
\be \label{manus-08042024-01}
\eta_{_{AB}} y^A y^B = \rho^{-1}\,, \qquad \rho := \epsilon R^{-2}\,, \qquad R>0\,,
\ee
where the flat metric tensor $\eta_{_{AB}}$ of the pseudo-Euclidean space and the symbol $\epsilon$ are given by,
{\small
\beq
\label{manus-08042024-02} && \eta_{_{AB}}=(-,-,+,\ldots,+)\,,\qquad A,B = 0',0,1,\ldots, d\,, \qquad  \epsilon =-1\,,  \hbox{ for AdS}; \qquad
\nonumber\\
&& \eta_{_{AB}}=(+,-,+,\ldots,+)\,,\qquad  A,B = 1',0,1,\ldots, d\,, \qquad \epsilon =1\,,  \ \ \ \hbox{ for dS}\,. \qquad
\eeq
}
\!Isometry symmetries of $(A)dS_{d+1}$ are described by the following algebras:
\be \label{manus-08042024-10}
so(d,2) \qquad \hbox{for } \ AdS_{d+1};\hspace{2cm} so(d+1,1) \quad \hbox{for } \ dS_{d+1}\,.
\ee
In what follows, to simplify our expressions we will drop the metric tensor $\eta_{_{AB}}$ in scalar products, i.e., we use  the convention $X^A Y^A : = \eta_{_{AB}} X^A Y^B$. Generators of $(A)dS$ algebra \rf{manus-08042024-10} denoted as $J^{AB}$  satisfy the following commutation relations:
\be \label{manus-08042024-15}
[J^{AB}, J^{CE}] = \eta^{BC} J^{AE} + 3 \hbox{ terms}.
\ee

Scalar field $\phi$ and vector field $\phi^A$ defined on the hyperboloid \rf{manus-08042024-01} transform under action of the generators of $(A)dS$ algebra as follows
\beq
\label{manus-08042024-16} && \delta_{J^{AB}} \phi \  = \ L^{AB}\phi\,,
\nonumber\\
&& \delta_{J^{AB}} \phi^C = L^{AB}\phi^C + \eta^{CA} \phi^B - \eta^{CB} \phi^A\,,\qquad L^{AB} : = y^A \ppp^B - y^B\ppp^A\,,\qquad
\eeq
where $L^{AB}$ stands for a orbital operator, while $\ppp^A$ stands for a tangential derivative on the hyperboloid \rf{manus-08042024-01}. Various  relations for the tangential derivative $\ppp^A$ may be found in Appendix A.

Field strengths defined on the hyperboloid \rf{manus-08042024-01} will be denoted as $F^A$, $F^{AB}$.  Under action of the generators of $(A)dS$ algebra \rf{manus-08042024-10} the field strengths transform as
\beq
\label{manus-08042024-20} && \delta_{J^{AB}} F^C = L^{AB} F^C + \eta^{CA} F^B - \eta^{CB} F^A\,,
\nonumber\\
&& \delta_{J^{AB}} F^{CE} = L^{AB} F^{CE} + \eta^{EA} F^{CB} - \eta^{EB} F^{CA} + \eta^{CB} F^{EA} - \eta^{CA} F^{EB} \,,
\eeq
where $L^{AB}$ is defined in \rf{manus-08042024-16}. Throughout this paper, all vector fields and all field strengths are assumed to satisfy the $y$-transversality condition given by
\be \label{manus-08042024-25}
y^A \phi^A=0\,, \qquad y^A F^A=0\,, \qquad y^A F^{AB}=0\,.
\ee

\noinbf{Conformal algebra in embedding space approach}. Conformal symmetries of $(A)dS_{d+1}$ are described by the $so(d+1,2)$ algebra whose generators denoted as $J^{\alpha\beta}$ obey the commutators
{\small
\be \label{manus-08042024-30}
[J^{\alpha\beta},J^{\gamma\sigma}]= \eta^{\beta\gamma}J^{\alpha\sigma}+ 3
\hbox{ terms}\,,
\quad \eta^{\alpha\beta} = (-,+,-,+,\ldots,+)\,, \qquad \alpha,\beta,\gamma,\sigma=
0',1',0,1,\ldots,d\,.
\ee
}
\!In the basis of $(A)dS$ algebra \rf{manus-08042024-10}, the conformal algebra generators $J^{\alpha\beta}$ are decomposed as
\be \label{manus-08042024-40}
J^{\alpha\beta} = J^{1'A},\,  J^{AB}\,,  \quad \hbox{for } AdS; \hspace{2cm}  J^{\alpha\beta} = J^{0'A},\,  J^{AB}\,, \quad \hbox{for } dS,
\ee
where the indices $A,B$ take the values as in \rf{manus-08042024-02}. Introducing the notation
\be \label{manus-08042024-50}
K^A := R J^{1'A}\,, \quad  \hbox{for } \ AdS;  \hspace{2cm} K^A := R J^{0'A}\,, \quad  \hbox{for } \  dS;
\ee
we note that the commutators \rf{manus-08042024-40} lead to commutators for the generators $K^A$ and $J^{AB}$ given by
\beq
\label{manus-08042024-60} && [K^A,K^B]= \rho^{-1} J^{AB}\,,
\nonumber\\
&& [K^A,J^{BC}]= \eta^{AB} K^C - \eta^{AC} K^B\,,\qquad [J^{AB}, J^{CE}] = \eta^{BC} J^{AE} + 3 \hbox{ terms}. \qquad
\eeq
To summarize, the algebra of conformal symmetries of $(A)dS_{d+1}$ is described by the  $so(d+1,2)$ algebra. The generators of the conformal algebra $so(d+1,2)$ are decomposed into generators $J^{AB}$, which are the generators of $(A)dS$ algebra \rf{manus-08042024-10} and the generators $K^A$ which describe conformal boost and dilatation transformations in the $(A)dS_{d+1}$ space. Commutators of the conformal algebra we use in this paper are given in  \rf{manus-08042024-60}.%
\footnote{Conformal YM field studied in this paper is associated with non-unitary representation of the conformal algebra. As shown in Ref.\cite{Metsaev:1995jp}, basis of the generators $J^{AB}$, $K^A$ turns out to be convenient also for the study of all unitary irreps of $AdS_{d+1}$ algebra $so(d,2)$ which can be realized as unitary irreps of the conformal algebra $so(d+1,2)$.
}

\noinbf{Field content in generic formulation}. Field content entering our ordinary-derivative generic formulation of conformal YM field in $(A)dS_{d+1}$ consists of the following set of vector and scalar fields of $(A)dS_{d+1}$ algebra \rf{manus-08042024-10}:
\be \label{manus-08042024-70}
\phi_{2n+1}^A\,, \quad n=0,1,\ldots,\frac{d-3}{2}; \hspace{1.5cm} \phi_{2n}\,, \quad n=1,2,\ldots,\frac{d-3}{2}\,.\qquad
\ee
For fields \rf{manus-08042024-70}, we introduce the corresponding field strengths,
{\small
\be  \label{manus-08042024-71}
F_{2n+2}^{BC}\,, \quad n=0,1,\ldots,\frac{d-3}{2}\,; \hspace{1.2cm} F_{2n+1}^B\,, \quad n=1,2,\ldots,\frac{d-3}{2}\,.
\ee
}
\!Explicit relations for the field strengths in terms of fields \rf{manus-08042024-70} are given below. The vector fields and the field strengths are $y$-transversal \rf{manus-08042024-25}.

We use a notation  $t^\asf$ for gauge algebra generators
\be \label{manus-08042024-85}
[t^\asf,t^\bsf] = f^{\asf\bsf\csf} t^\csf\,, \qquad \Tr(t^\asf t^\bsf) = - \KK^{-1}\delta^{\asf\bsf}\,, \qquad \KK > 0\,,
\ee
where, for the flexibility, we introduce some constant $\KK$. We note then that fields \rf{manus-08042024-70} and field strengths \rf{manus-08042024-71} are decomposed in $t^\asf$ as
\beq
\label{manus-08042024-86}
\phi_{2n+1}^A = \phi_{2n+1}^{A\asf} t^\asf\,,   \hspace{0.5cm} \phi_{2n} = \phi_{2n}^\asf t^\asf\,, \qquad F_{2n+2}^{AB} = F_{2n+2}^{AB\asf} t^\asf\,,   \hspace{0.5cm} F_{2n+1}^A = F_{2n+1}^{A\asf} t^\asf\,.
\eeq
Particular choice of the gauge algebra is not important for us.%
\footnote{For higher-spin theories, the discussion of various internal symmetry algebras can be found, e.g., in Refs.\cite{Konstein:1989ij}-\cite{Skvortsov:2020wtf}.
}

In gauge invariant approach, each vector field \rf{manus-08042024-70} is accompanied by gauge parameter. This implies that we use the following set of gauge parameters:
\be   \label{manus-08042024-87}
\xi_{2n}\,, \quad n=0,1,\ldots,\frac{d-3}{2};  \qquad \xi_{2n} = \xi_{2n}^\asf t^\asf\,,
\ee
where in \rf{manus-08042024-87} we show decomposition of $\xi_{2n}$ in the basis of gauge algebra generators \rf{manus-08042024-85}.

The vector field $\phi_1^A$ turns out to be a primary field and this field can be identified with the field entering usual higher-derivative formulation of the conformal YM field. The remaining vector fields $\phi_{2n+1}^a$, $n=1,\ldots, \frac{d-3}{2}$, turn out to be auxiliary fields in our approach. These fields are  required to get our ordinary-derivative Lagrangian formulation. All scalar fields $\phi_{2n}$ appearing in \rf{manus-08042024-70} turn out to be  Stueckelberg fields.  The Stueckelberg fields are required to realize gauge symmetries in our approach.  Throughout this paper, a covariant derivative acting on the auxiliary fields, Stueckelberg fields, and gauge parameters is defined as
{\small
\beq
\label{manus-14092024-01} &&  \DD^A\phi_{2n+1}^B := \ppp^A \phi_{2n+1}^B + \rho y^B \phi_{2n+1}^A+ [\phi_1^A,\phi_{2n+1}^B]\,,
\nonumber\\[-6pt]
&& \DD^A\phi_{2n} := \ppp^A \phi_{2n} + [\phi_1^A,\phi_{2n}]\,,  \hspace{3cm} n=1,\ldots, \frac{d-3}{2}\,;\qquad
\\[-6pt]
\label{manus-14092024-02} &&   \DD^A\xi_{2n} := \ppp^A \xi_{2n} + [\phi_1^A,\xi_{2n}]\,,\hspace{3.2cm} n=0,1,\ldots, \frac{d-3}{2}\,.
\eeq
}
Relations \rf{manus-14092024-01}, \rf{manus-14092024-02} imply that the gauge  field $\phi_1^A$ is realized as the standard YM connection of the covariant derivative $\DD^A$. Also, to simplify our notation we use the following shortcuts for the  products of the field strengths:
\be \label{manus-08042024-87sh}
F_{2m+2} F_{2n+2} : = F_{2m+2}^{AB\asf }F_{2n+2}^{AB\asf }\,, \qquad F_{2m+1} F_{2n+1} : = F_{2m+1}^{A\asf } F_{2n+1}^{A\asf }\,.
\ee

\noinbf{Field content in decoupled formulation}. Field content entering the decoupled formulation is in one-to-one correspondence to the field content entering the generic formulation. This is to say that ordinary-derivative decoupled  formulation of conformal YM field in $(A)dS_{d+1}$ consists of the following set of vector and scalar fields of   $(A)dS_{d+1}$ algebra \rf{manus-08042024-10}:
\be \label{manus-08042024-70add}
\varphi_{2n+1}^A\,, \quad n=0,1,\ldots,\frac{d-3}{2}; \hspace{1.5cm} \varphi_{2n}\,, \hspace{1cm} n=1,2,\ldots,\frac{d-3}{2}\,.\qquad
\ee
Field strengths corresponding to fields \rf{manus-08042024-70add} will be denoted as
{\small
\be  \label{manus-08042024-71add}
\FF_{2n+2}^{BC}\,, \quad n=0,1,\ldots,\frac{d-3}{2}\,; \hspace{1.2cm} \FF_{2n+1}^B\,, \quad n=1,2,\ldots,\frac{d-3}{2}\,.
\ee
}
\!The vector fields and the field strengths \rf{manus-08042024-70add}, \rf{manus-08042024-71add} are $y$-transversal \rf{manus-08042024-25}. Fields \rf{manus-08042024-70add} and field strengths \rf{manus-08042024-71add} are decomposed in $t^\asf$ as
\beq
\label{manus-08042024-86add}
\varphi_{2n+1}^A = \varphi_{2n+1}^{A\asf} t^\asf\,,   \hspace{0.5cm} \varphi_{2n} = \varphi_{2n}^\asf t^\asf\,, \qquad \FF_{2n+2}^{AB} = \FF_{2n+2}^{AB\asf} t^\asf\,,   \hspace{0.5cm} \FF_{2n+1}^A = \FF_{2n+1}^{A\asf} t^\asf\,.
\eeq

Gauge parameters accompanied with vector fields \rf{manus-08042024-70add} will be denoted as:
\be   \label{manus-08042024-87add}
\eta_{2n}\,, \quad n=0,1,\ldots,\frac{d-3}{2};  \qquad \eta_{2n} = \eta_{2n}^\asf t^\asf\,,
\ee
where in \rf{manus-08042024-87add} we show decomposition of $\eta_{2n}$ in the basis of gauge algebra generators \rf{manus-08042024-85}.

The vector field $\varphi_1^A$ turns out to be a massless field, while the remaining vector fields $\varphi_{2n+1}^A$, $n=1,\ldots, \frac{d-3}{2}$, turn out to be massive fields. All scalar fields $\varphi_{2n}$ appearing in \rf{manus-08042024-70add} are realized as Stueckelberg fields. Throughout this paper, a covariant derivative acting on the massive fields, Stueckelberg fields, and gauge parameters is defined as
{\small
\beq
\label{manus-14092024-01add} && \DD^A\varphi_{2n+1}^B := \ppp^A \varphi_{2n+1}^B + \rho y^B\varphi_{2n+1}^A + [\varphi_1^A,\varphi_{2n+1}^B]\,,
\nonumber\\[-6pt]
&& \DD^A\varphi_{2n} := \ppp^A \varphi_{2n} + [\varphi_1^A,\varphi_{2n}]\,, \hspace{3cm} n=1,\ldots, \frac{d-3}{2}\,;\qquad
\\[-6pt]
\label{manus-14092024-02add} && \DD^A\eta_{2n} := \ppp^A \eta_{2n} + [\varphi_1^A,\eta_{2n}]\,,\hspace{3.2cm} n=0,1,\ldots,\frac{d-3}{2}\,.
\eeq
}
Relations \rf{manus-14092024-01add}, \rf{manus-14092024-02add} imply that the gauge  field $\varphi_1^A$ is realized as the standard YM connection of the covariant derivative $\DD^A$.
Cousins of the shortcuts in \rf{manus-08042024-87sh} are defined as
\be \label{manus-08042024-88add}
\FF_{2m+2} \FF_{2n+2} : = \FF_{2m+2}^{AB\asf } \FF_{2n+2}^{AB\asf }\,, \qquad \FF_{2m+1} \FF_{2n+1} : = \FF_{2m+1}^{A\asf } \FF_{2n+1}^{A\asf }\,.
\ee

\noinbf{Realization of $K^A$-transformations on fields and field strengths}. For fields \rf{manus-08042024-70} entering the generic formulation, we introduce an operator $K_\Delta^A$ by the following relations:
{\small
\beq
\label{manus-08042024-75} &&  K_\Delta^A \phi_{2n+1}^B : = (- \rho^{-1} \ppp^A + \Delta y^A)\phi_{2n+1}^B - y^B \phi_{2n+1}^A\,,
\nonumber\\
&& K_\Delta^A \phi_{2n} : = (- \rho^{-1} \ppp^A + \Delta y^A) \phi_{2n}\,,
\eeq
}
\!\!while the definition of the operator $K_\Delta^A$ on the corresponding field strengths is given by
{\small
\beq
\label{manus-08042024-76} && K_\Delta^A F_{2n+2}^{BC} : = (- \rho^{-1} \ppp^A +  \Delta  y^A)F_{2n+2}^{BC} - y^B F_{2n+2}^{AC} - y^C F_{2n+2}^{BA} \,,
\nonumber\\
&& K_\Delta^A F_{2n+1}^B : = (-  \rho^{-1} \ppp^A + \Delta y^A) F_{2n+1}^B - y^B F_{2n+1}^A\,.
\eeq
}
\!\!We verify that the action of the operator $K_\Delta^A$ respects the $y$-transversality constraint \rf{manus-08042024-25}.
For fields \rf{manus-08042024-70add} and the corresponding field strengths \rf{manus-08042024-71add} entering the decoupled formulation, an action of the operator $K_\Delta^A$ is defined by the same relations as in \rf{manus-08042024-75} and \rf{manus-08042024-76}.

In the framework of the generic formulation, the $K^A$-transformations of fields \rf{manus-08042024-70} take the following triangle form:
{\small
\beq
\label{manus-08042024-80} && \delta_{K^A} \phi_{2n+1}^B = \sum_{m=0,\ldots,n} U_{nm}^{ABC} \phi_{2m+1}^C + \sum_{m=1,\ldots,n} U_{nm}^{AB} \phi_{2m}\,,
\nonumber\\
&& \delta_{K^A} \phi_{2n} = \sum_{m=0,\ldots,n} V_{nm}^A\phi_{2m} + \sum_{m=1,\ldots,n-1} V_{nm}^{AB}\phi_{2m+1}^B\,,
\eeq
}
\!where $U$- and $V$-coefficients depend on the derivative $\ppp^A$, the coordinate $y^A$, and the flat metric $\eta^{AB}$. The $K^A$- transformations of the field strengths \rf{manus-08042024-71} have similar triangle form.

In the decoupled formulation, the $K^A$-transformations of fields \rf{manus-08042024-70add} take the form
{\small
\beq
\label{manus-08042024-81} && \hspace{-0.9cm}\delta_{K^A} \varphi_{2n+1}^B = U_{n+}^{ABC} \varphi_{2n+3}^C + U_{n-}^{ABC} \varphi_{2n-1}^C + U_{n+}^{AB} \varphi_{2n+2} + U_{n-}^{AB} \varphi_{2n-2}\,,\hspace{0.5cm} n=0,1,\ldots, N-1\,,\qquad
\nonumber\\
&& \hspace{-0.9cm} \delta_{K^A} \varphi_{2N+1}^B = U_N^{ABC} \varphi_{2N+1}^C + U_{N-}^{ABC} \varphi_{2N-1}^C + U_N^{AB} \varphi_{2N} + U_{N-}^{AB} \varphi_{2N-2}\,,
\nonumber\\
&& \hspace{-0.9cm} \delta_{K^A} \varphi_{2n} =  V_{n+}^A \varphi_{2n+2} + V_{n-}^A \varphi_{2n-2} +  V_{n+}^{AB} \varphi_{2n+3}^B +  V_{n-}^{AB} \varphi_{2n-1}^B\,,
\hspace{1.5cm} n=1,\ldots, N-1\,,
\nonumber\\
&& \hspace{-0.8cm} \delta_{K^A} \varphi_{2N} =  V_N^A \varphi_{2N} + V_{N-}^A \varphi_{2N-2} +  V_N^{AB} \varphi_{2N+1}^B +  V_{N-}^{AB} \varphi_{2N-1}^B\,, \hspace{1.5cm} N:= \frac{d-3}{2}\,,
\eeq
}
\!where $U$- and $V$-coefficients depend on the derivative $\ppp^A$, the coordinate $y^A$, and the flat metric $\eta^{AB}$. The $K^A$- transformations of the field strengths \rf{manus-08042024-71add} have similar form.

Expressions \rf{manus-08042024-80} and \rf{manus-08042024-81} show a general structure of $K^A$-transformations of fields entering the respective generic and decoupled formulations.
Transformations given in \rf{manus-08042024-80} take more complicated form as compared to the ones in \rf{manus-08042024-81}. Transformations \rf{manus-08042024-81} turn out to be more convenient for the check of commutators \rf{manus-08042024-60}.
Explicit expressions of $K^A$-transformations for fields in $(A)dS_{d+1}$ and their field strengths are given below when $d=5,7,9$.

For the systematic study of conformal YM field in $(A)dS_{d+1}$, we use the fields and field strengths defined in \rf{manus-08042024-70}-\rf{manus-08042024-86} and \rf{manus-08042024-70add}-\rf{manus-08042024-86add}. Those fields and field strengths are expanded in the generators of gauge algebra $t^a$ entering the standard formulation of YM field theories. In our approach, we can introduce an extended gauge algebra. Use of the extended gauge algebra allows us to represent many relations obtained in this paper in concise and elegant form. We now outline the extended gauge algebra setup.%
\footnote{For conformal higher-spin gravity, the discussion of symmetry algebra  may be found in Ref.\cite{Basile:2018eac}.
}

\noinbf{Extended gauge algebra setup in generic formulation}.  First, we accompany the set of gauge parameters $\xi_{2n}^\asf$ \rf{manus-08042024-87} by the corresponding set of extended gauge algebra generators denoted as
\be \label{manus-08042024-90}
T_{2n}^\asf, \qquad n=0,1,\ldots,\frac{d-3}{2}\,, \qquad \qquad T_0^\asf:= t^a\,,
\ee
where we show that the generator of the extended gauge algebra $T_0^\asf$ is identified with the generator of gauge algebra $t^a$.
Following the standard setup, the vector fields, their field strengths, and gauge parameters are collected into the following ingredients of the extended gauge algebra:
{\small
\be \label{manus-08042024-91}
\phi^A : = \!\!\!\sum_{n=0,1,\ldots,\frac{d-3}{2}} \phi_{2n+1}^{A\asf} T_{2n}^\asf \,, \qquad
F^{AB} : = \!\!\! \sum_{n=0,1,\ldots,\frac{d-3}{2}} F_{2n+2}^{AB\,\asf} T_{2n}^\asf\,,\qquad  \xi := \!\!\! \sum_{n=0,1,\ldots,\frac{d-3}{2}} \xi_{2n}^\asf T_{2n}^\asf  \,.
\ee
}

Second, using $\phi^A$, $F^{AB}$, and $\xi$, we note that  the expressions for field strengths and the gauge transformations for the fields and field strengths obtained in this paper can be represented as
\beq
\label{manus-08042024-92} && F^{AB} = (\ppp^A - \rho y^A)\phi^B - (\ppp^B - \rho y^B)\phi^A + [\phi^A,\phi^B]\,,
\nonumber\\
&& \delta_\xi \phi^A = \ppp^A \xi + [\phi^A,\xi]\,, \qquad  \delta_\xi F^{AB} = [F^{AB},\xi]\,.
\eeq

Third, as we show below, the generators $T_{2n}^\asf$, $n=1,\ldots,\frac{d-3}{2}$, constitute maximal solvable ideal (radical) of the extended gauge algebra. We find it convenient to represent generators \rf{manus-08042024-90} as
\be \label{manus-08042024-94}
T_0^\asf\,, \quad S_{2n}^\asf\,, \quad n=1,\ldots,\frac{d-3}{2}\,, \qquad  S_{2n}^\asf : = \sum_{m=1,\ldots,\frac{d-3}{2}} V_{2n,2m} T_{2m}^\asf\,,
\ee
where a transformation from the basis of $T_{2n}^\asf$-generators to the basis of $S_{2n}^\asf$-generators, $n=1,\ldots,\frac{d-3}{2}$, is assumed to be invertible. This implies that the set of the generators $S_{2n}^\asf$ also constitutes a maximal solvable ideal (radical) of the algebra \rf{manus-08042024-90}. Using a shortcut $T$ for the generators of the extended gauge algebar \rf{manus-08042024-90}, a shortcut $S$ for the radical spanned by the generators $S_{2n}^\asf$, $n=1,\ldots,\frac{d-3}{2}$, and a shortcut $T_0$ for the gauge algebra spanned by $T_0^\asf$, we note then the Levy-Maltsev decomposition of the algebra $T$,
\be
T = S \subsetext T_0\,, \qquad T := \{\,T_{2n}^\asf\,\}_{n=0,1,\ldots,\frac{d-3}{2}} \,, \qquad S := \{\,S_{2n}^\asf\,\}_{n=1,\ldots,\frac{d-3}{2}}\,, \qquad T_0 := T_0^\asf\,.
\ee
In order to present gauge transformation of the Stueckelberg fields and the corresponding field strengths we now introduce
{\small
\beq
\label{manus-08042024-95} && \phi := \sum_{n=1,\ldots,\frac{d-3}{2}}  \phi_{2n}^\asf  S_{2n}^\asf \,,\hspace{1.8cm} F^A := \sum_{n=1,\ldots,\frac{d-3}{2}} F_{2n+1}^{A\asf}  S_{2n}^\asf\,,
\nonumber\\
&& \xi_\subsm :=  \sum_{n=1,\ldots,\frac{d-3}{2}} \xi_{2n}^\asf S_{2n}^\asf\,, \hspace{1.5cm} \phi_\subsm^A := \sum_{n=1,\ldots,\frac{d-3}{2}} \phi_{2n+1}^{A\asf} S_{2n}^\asf\,.
\eeq
}

Fourth, using $\phi^A$ and $\xi$ \rf{manus-08042024-94} and the quantities defined in \rf{manus-08042024-95}, we find that the expression for $F^A$, gauge transformations of $\phi$, and $F^A$ obtained in this paper can be represented as
\beq
\label{manus-08042024-96} && F^A = \phi_\subsm^A + \ppp^A \phi + [\phi^A,\phi]\,, \qquad
\nonumber\\
&& \delta_\xi \phi = - \xi_\subsm + [\phi,\xi]\,,\qquad   \delta_\xi F^A = [F^A,\xi]\,.
\eeq

The following representation for the gauge transformation of $\phi_\subsm^A$  \rf{manus-08042024-95} turns out to be convenient for the derivation of $\delta_\xi F^A$ in \rf{manus-08042024-96},
\be \label{manus-08042024-120-xxx}
\delta_\eta \phi_\subsm^A = \ppp^A \xi_\subsm + [\phi_\subsm^A,\xi]+ [\phi^A,\xi_\subsm]\,.
\ee
Finally, Lagrangian obtained in this paper by using the generic formulation can be presented as
\be \label{manus-08042024-97}
\frac{1}{\KK} \LL_\CYMsm = \frac{1}{4} \langle F^{AB}, F^{AB}\rangle + \half \langle F^A , F^A\rangle_\subsm\,,
\ee
where the notation $\langle ...\rangle$ and $\langle ...\rangle_\subsm$ is used for two invariant bilinear forms defined for the respective generators of the extended gauge algebra \rf{manus-08042024-90} and the generators of the radical \rf{manus-08042024-94},
\beq
\label{manus-08042024-98} && G_{2m\,2n}^{\asf\bsf} := \langle T_{2m}^\asf,T_{2n}^\bsf\rangle \,, \hspace{1.7cm} G_{2m\,2n}^{\subsm\,\asf\bsf} : = \langle S_{2m}^\asf,S_{2n}^\bsf\rangle_\subsm \,,
\nonumber\\
&& G_{2m\,2n}^{\asf\bsf} = - \frac{1}{\KK}\,\, G_{2m\,2n} \delta^{\asf\bsf}\,, \hspace{1cm} G_{2m\,2n}^{\subsm\,\asf\bsf} = - \frac{1}{\KK}\,\, G_{2m\,2n}^\subsm \delta^{\asf\bsf}\,.
\eeq
In \rf{manus-08042024-98}, we show that in a suitable basis the bilinear forms are diagonal with respect to the indices $\asf$, $\bsf$. Using \rf{manus-08042024-98}, Lagrangian \rf{manus-08042024-97} can be represented as
\be \label{manus-08042024-100}
\LL_\CYMsm =  - \frac{1}{4}\sum_{m,n=0,1,\ldots,\frac{d-3}{2}} G_{2m\, 2n} F_{2m+2} F_{2n+2} - \half \sum_{m,n= 1,\ldots,\frac{d-3}{2}} G_{2m\,2n}^\subsm F_{2m+1} F_{2n+1}\,.
\ee
Our results in this paper suggest the following solution for the bilinear forms \rf{manus-08042024-98},
{\small
\beq
\label{manus-08042024-100-a1} && G_{2m\, 2n} = G_{n+m}(-\rho)^{\frac{d-3}{2}-n-m}\theta(\frac{d-3}{2}-n-m)\,,
\nonumber\\
&& G_{2m\, 2n}^\subsm =  G_{n+m-1}(-\rho)^{\frac{d-1}{2}-n-m}\theta(\frac{d-1}{2}-n-m)\,, \qquad
\eeq
}
\!\!where $\theta(n)=1$ for $n\geq 0$ and $\theta(n)=0$ for $n<0$. Coefficients $G_n$ depend on $d$ and $n$ and do not depend on $\rho$. Relations \rf{manus-08042024-100}, \rf{manus-08042024-100-a1} lead to
\beq
\label{manus-08042024-100-a3} && \LL_\CYMsm =  - \sum_{k=0,1,\ldots,\frac{d-3}{2}} (-\rho)^{\frac{d-3}{2}-k} G_k I_{2k+4}\,,
\nonumber\\
&& \hspace{1cm} I_{2k+4} : =  \frac{1}{4}\sum_{m,n=0,1,\ldots,\frac{d-3}{2}\atop m+n=k} F_{2m+2} F_{2n+2} + \half \sum_{m,n= 1,\ldots,\frac{d-3}{2}\atop m+n=k+1} F_{2m+1} F_{2n+1}\,.
\eeq
We are motivated to use the quantities $I_{2k+4}$ because it is  combination of field strengths appearing in $I_{2k+4}$ \rf{manus-08042024-100-a3} that is realized in the Lagrangian of conformal YM field in the flat space. Namely, in the flat space limit, the minus $I_{2k+4}$, when $k=\frac{d-3}{2}$, coincides with ordinary derivative Lagrangian of the conformal YM field in $R^{d,1}$ found in Ref.\cite{Metsaev:2023qif}.

For $(A)dS_{d+1}$ with $d=5,7,9$, relations \rf{manus-08042024-100}-\rf{manus-08042024-100-a3} are proved in this paper, while, for $d\geq 11$, the relations \rf{manus-08042024-100}-\rf{manus-08042024-100-a3}  should be considered as our conjecture.

\noinbf{Extended gauge algebra setup in decoupled formulation}. First, we accompany the set of gauge parameters $\eta_{2n}^\asf$ \rf{manus-08042024-87add} by the corresponding set of the extended gauge algebra generators given by
\be \label{manus-08042024-110}
\TT_{2n}^\asf, \qquad n=0,1,\ldots,\frac{d-3}{2}\,, \qquad \qquad \TT_0^\asf:= t^a\,,
\ee
where we show that the generator $T_0^\asf$ is identified with the generator of gauge algebra $t^a$.
Most of the remaining relations of the extended gauge setup for the decoupled formulation are similar to the ones for the generic formulation.
Namely, the vector fields, the field strengths, and the gauge parameters are collected into the following ingredients of the extended gauge algebra:
{\small
\be \label{manus-08042024-91ab}
\varphi^A : = \!\!\! \sum_{n=0,1,\ldots,\frac{d-3}{2}} \,\, \varphi_{2n+1}^{A\asf} \TT_{2n}^\asf \,, \qquad  \FF^{AB} : = \!\!\! \sum_{n=0,1,\ldots,\frac{d-3}{2}}\,\, \FF_{2n+2}^{AB\,\asf} \TT_{2n}^\asf\,,
\qquad \eta :=  \!\!\! \sum_{n=0,1,\ldots,\frac{d-3}{2}}\,\, \eta_{2n}^\asf \TT_{2n}^\asf  \,.
\ee
}

Second, using the $\varphi^A$, $\FF^{AB}$, and $\eta$, we note that  the expressions for field strengths and the gauge transformations for the fields and field strengths obtained in this paper can be represented as
\beq
\label{manus-08042024-120} && \FF^{AB} = (\ppp^A - \rho y^A)\varphi^B - (\ppp^B - \rho y^B)\varphi^A + [\varphi^A,\varphi^B]\,,
\nonumber\\
&& \delta_\eta \varphi^A = \ppp^A \eta + [\varphi^A,\eta]\,, \qquad  \delta_\eta \FF^{AB} = [\FF^{AB},\eta]\,.
\eeq

Third, as we demonstrate below, in place of the basis of generators \rf{manus-08042024-110}, we can introduce the basis of generators given by
\be \label{manus-08042024-94ab}
\TT_0^\asf\,, \quad \Scl_{2n}^\asf\,, \quad n=1,\ldots,\frac{d-3}{2}\,, \qquad  \Scl_{2n}^\asf : = \sum_{m=0,1,\ldots,\frac{d-3}{2}} \VV_{2n,2m} \TT_{2m}^\asf\,,
\ee
where the generators $\Scl_{2n}^\asf$ constitute a maximal solvable ideal (radical) of the algebra \rf{manus-08042024-110}. Using the shortcut $\TT$ for algebra spanned by the generators \rf{manus-08042024-110}, the shortcut $\Scl$ for the radical spanned by the generators $\Scl_{2n}^\asf$, $n=1,\ldots,\frac{d-3}{2}$, and the shortcut $\TT_0$ for the algebra spanned by $\TT_0^\asf$, we note then the Levy-Maltsev decomposition of the algebra $\TT$,
\be
\TT = \Scl \subsetext \TT_0\,, \qquad \TT := \{\,\TT_{2n}^\asf\,\}_{n=0,1,\ldots,\frac{d-3}{2}}\,, \qquad \Scl := \{\,\Scl_{2n}^\asf\,\}_{n=1,\ldots,\frac{d-3}{2}}\,, \qquad \TT_0 := \TT_0^\asf\,.
\ee
In order to present gauge transformations of the Stueckelberg fields and the corresponding field strengths we now introduce the following quantities:
{\small
\beq \label{manus-08042024-95ab}
&& \varphi := \sum_{n=1,\ldots,\frac{d-3}{2}}  \varphi_{2n}^\asf  \Scl_{2n}^\asf \,,\hspace{1.8cm} \FF^A :=\!\!\! \sum_{n=1,\ldots,\frac{d-3}{2}} \FF_{2n+1}^{A\asf}  S_{2n}^\asf\,,
\nonumber\\
&& \varphi_\subsm^A := \sum_{n=1,\ldots,\frac{d-3}{2}} \varphi_{2n+1}^{A\asf} \Scl_{2n}^\asf\,, \hspace{1.2cm} \eta_\subsm :=  \sum_{n=1,\ldots,\frac{d-3}{2}} \eta_{2n}^\asf \Scl_{2n}^\asf\,.
\eeq
}

Fourth, using $\varphi^A$ and $\eta$ \rf{manus-08042024-91ab} and the quantities defined in \rf{manus-08042024-95ab}, we find that the expression for $\FF^A$, gauge transformations of $\varphi$ and $\FF^A$ obtained in this paper can be represented as
\beq
\label{manus-08042024-126} && \FF^A = \varphi_\subsm^A + \ppp^A \varphi + [\varphi^A,\varphi]\,, \qquad
\nonumber\\
&& \delta_\eta \varphi = - \eta_\subsm + [\varphi,\eta]\,, \qquad
\delta_\eta \FF^A = [\FF^A,\eta]\,.
\eeq
The following representation for the gauge transformation of $\varphi_\subsm^A$  \rf{manus-08042024-95ab} turns out to be convenient for the derivation of $\delta_\xi \FF^A$ in \rf{manus-08042024-126},
\be \label{manus-08042024-105add}
\delta_\eta \varphi_\subsm^A = \ppp^A \eta_\subsm + [\varphi_\subsm^A,\eta]+ [\varphi^A,\eta_\subsm]\,.
\ee

Finally, Lagrangian obtained by using the decoupled formulation can be presented as
\be  \label{manus-08042024-117}
\frac{1}{\KK} \LL_\CYMsm = \frac{1}{4} \langle \FF^{AB}, \FF^{AB}\rangle + \half \langle \FF^A, \FF^A\rangle_\subsm\,,
\ee
where the notation $\langle ...\rangle$ and $\langle ...\rangle_\subsm$ is used to for two invariant bilinear forms defined for the respective generators of the extended gauge algebra \rf{manus-08042024-110} and the generators of radical \rf{manus-08042024-94ab},
\beq
\label{manus-08042024-98ab} && \GG_{2m\,2n}^{\asf\bsf} := \langle \TT_{2m}^\asf, \TT_{2n}^\bsf\rangle \,, \hspace{1.7cm} \GG_{2m\,2n}^{\subsm\,\asf\bsf} : = \langle \Scl_{2m}^\asf, \Scl_{2n}^\bsf\rangle_\subsm \,,
\nonumber\\
&& \GG_{2m\,2n}^{\asf\bsf} = - \frac{1}{\KK}\,\, \GG_{2m\,2n} \delta^{\asf\bsf}\,, \hspace{1cm} \GG_{2m\,2n}^{\subsm\,\asf\bsf} = - \frac{1}{\KK}\,\, \GG_{2m\,2n}^\subsm \delta^{\asf\bsf}\,.
\eeq
In \rf{manus-08042024-98ab}, we show that in a suitable basis the bilinear forms are diagonal with respect to the indices $\asf$, $\bsf$. The constant $\KK$ in \rf{manus-08042024-98ab} coincides with the one in \rf{manus-08042024-98}. Using \rf{manus-08042024-98ab}, Lagrangian \rf{manus-08042024-117} can be represented as
\be \label{manus-08042024-100add}
\LL_\CYMsm =  - \frac{1}{4}\sum_{m,n=0,1,\ldots,\frac{d-3}{2}} \GG_{2m\, 2n} \FF_{2m+2} \FF_{2n+2} - \half \sum_{m,n= 1,\ldots,\frac{d-3}{2}} \GG_{2m\,2n}^\subsm \FF_{2m+1} \FF_{2n+1}\,.
\ee
Our results in this paper suggest the following solution for the bilinear forms \rf{manus-08042024-98ab},
\beq
\label{manus-08042024-106} && \GG_{2m\, 2n}= (-)^n \delta_{mn} k_\rho\,, \hspace{1.6cm}  \GG_{2m\, 2n}^\subsm= (-)^n m_{\varphi_{2n+1}}^2 \delta_{mn} k_\rho\,,
\nonumber\\
&& m_{\varphi_{2n+1}}^2 = \rho n(d-2-n)\,, \qquad  k_\rho := (-\rho)^{ \frac{d-3}{2} }(d-3)!\,,
\eeq
where $\delta_{nn}=1$ for $n\geq 0$ and $\delta_{mn}=0$ for $m\ne n$. Relations \rf{manus-08042024-100add}, \rf{manus-08042024-106} lead to
\beq
\label{manus-08042024-106-b1} && \hspace{-1cm} k_\rho^{-1} \LL_\CYMsm =  \sum_{n=0,1,\ldots,\frac{d-3}{2}} (-)^{n+1}\II_{4n+4}\,,
\nonumber\\
&& \II_4:= \frac{1}{4} \FF_2 \FF_2\,, \hspace{1cm} \II_{4n+4}: = \frac{1}{4} \FF_{2n+2}\FF_{2n+2} + \frac{m_{\varphi_{2n+1}}^2}{2} \FF_{2n+1}\FF_{2n+1}\,,\quad n>0\,.\qquad
\eeq
We are motivated to use the quantities $\II_{4n+4}$ because, to the second order in the fields, the $\II_4$ describes free massless $(A)dS$ field, while the $\II_{4n+4}$, $n>0$, describes free massive $(A)dS$ field having the mass square as in \rf{manus-08042024-106}. This can be seen by using the explicit expressions for linearized field strengths given by
\be \label{manus-08042024-106-b2}
\FF_{2n+2}^{\lin\, AB} = (\ppp^A - \rho y^A)\varphi_{2n+1}^B  - (\ppp^B - \rho y^B)\varphi_{2n+1}^A\,, \qquad  \FF_{2n+1}^{\lin\, A} =  \varphi_{2n+1}^A  + \ppp^A\varphi_{2n}\,.
\ee
Using the linearized field strengths, we find that, to quadratic order in the fields, the Lagrangian \rf{manus-08042024-106-b1} takes the form
{\small
\beq
\label{manus-08042024-106-b3} && \hspace{-1.2cm} k_\rho^{-1} \LL_\CYMsm^\smtwo =  \sum_{n=0,1,\ldots,\frac{d-3}{2}} (-)^{n+1}\II_{4n+4}^\smtwo \,,
\nonumber\\
&& - \II_4^\smtwo := \half \varphi_1^A \big(\ppp^2 - (d-1)\rho \big) \varphi_1^A + \half L_2 L_2\,, \qquad  L_2 : = \ppp^A \varphi_1^A\,,
\nonumber\\
&& -\II_{4n+4}^\smtwo := \half \varphi_{2n+1}^A \big( \ppp^2- (d-1) \rho - m_{\varphi_{2n+1}}^2 \big) \varphi_{2n+1}^A
\nonumber\\
&& \hspace{1.4cm} +\,\,  \frac{m_{\varphi_{2n+1}}^2}{2} \varphi_{2n} \big(\ppp^2  - m_{\varphi_{2n+1}}^2 \big) \varphi_{2n} + \half L_{2n+2} L_{2n+2} \,,
\nonumber\\
&&  \hspace{1.8cm}  L_{2n+2} : = \ppp^A \varphi_{2n+1}^A + m_{\varphi_{2n+1}}^2 \varphi_{2n}\,, \qquad \hbox{for}\ \ n>0\,.
\eeq
}
\!The $\II_4^\smtwo$ and $\II_{4n+4}^\smtwo$, $n>0$, are invariant under the linearized gauge transformations,
\be \label{manus-08042024-106-b4}
\delta_\eta^{\,\lin} \varphi_1^A = \ppp^A \eta_0 \,, \qquad \delta_\eta^{\,\lin} \varphi_{2n+1}^A = \ppp^A \eta_{2n}\,, \qquad \delta_\eta^{\,\lin} \varphi_{2n} = - \eta_{2n}\,,\qquad n > 0\,.
\ee
Relations \rf{manus-08042024-106-b3}, \rf{manus-08042024-106-b4} imply that the $\II_4^\smtwo$ describes massless field in $(A)dS_{d+1}$, while $\II_{4n+4}^\smtwo$, $n>0$, describes massive field in $(A)dS_{d+1}$ with mass square given in \rf{manus-08042024-106}.

For $(A)dS_{d+1}$ with $d=5,7,9$, relations \rf{manus-08042024-100add}, \rf{manus-08042024-106} are proved in this paper, while, for $d\geq 11$, the relations \rf{manus-08042024-100add}, \rf{manus-08042024-106}  should be considered as our conjecture.

\noinbf{Canonical (mass) dimensions and action}. Canonical (mass) dimensions of Lagrangian $\LL_\CYMsm$, the generators of gauge algebra $t^\asf$ and the parameter $\rho$ are given by
\be \label{manus-08042024-106-b5}
[\LL_\CYMsm]= d+1\,, \quad [t^\asf]=0\,, \quad [\rho]=2\,.
\ee
Canonical dimensions of coordinates, momenta, and generators of conformal algebra are as follows
\be \label{manus-08042024-106-b6}
[y^A]=-1,\qquad [\ppp^A]=1\,, \quad [K^A]=-1\,, \quad [J^{AB}]=0\,.
\ee
For the generic formulation, canonical dimensions of field, field strengths, gauge parameters, the generators of the extended gauge algebra and the radical are given by
{\small
\beq
\label{manus-08042024-106-b7} && \hspace{-1cm} [\phi_{2n+1}^A]= 2n+1\,,\quad [\phi_{2n}]=2n\,, \quad [F_{2n+2}^{AB}]= 2n+2\,, \quad [F_{2n+1}^A]=2n+1\,, \quad [\xi_{2n}]= 2n\,,\qquad
\nonumber\\
&& \hspace{-1cm} [T_{2n}^\asf]= -2n\,,\hspace{1.2cm} [S_{2n}^\asf]=-2n\,.
\eeq
}
\!\!In the flat space limit, the canonical dimensions of field, field strengths, and gauge parameters in \rf{manus-08042024-106-b7} are realized as the conformal dimensions. Therefore, in what follows, we refer the canonical dimensions in \rf{manus-08042024-106-b7} to as conformal dimensions. For the decoupled formulation, we use the following canonical dimensions:
{\small
\beq
&& [\varphi_{2n+1}^A]= 1\,,\qquad [\varphi_{2n}]=0\,, \quad [\FF_{2n+2}^{AB}]= 2\,, \quad [\FF_{2n+1}^A]=1\,, \quad [\eta_{2n}]= 0\,,
\nonumber\\
&& [\TT_{2n}^\asf]= 0\,, \hspace{1.2cm} [\Scl_{2n}^\asf] = 0\,.
\eeq
}
Using the notation in \rf{12062024-10}, we note that, in the embedding approach, the action is given by
\be
S_\CYMsm = \int d\sigma\,\, \LL_\CYMsm\,.
\ee

\newsection{ \large Conformal Yang-Mills field in $(A)dS_6$ space. Generic formulation } \label{six-gen}

{\bf Field content}. In the framework of gauge invariant generic
formulation, the conformal YM field in $(A)dS_6$ is described by two  vector fields and one scalar field:
\beq
\label{manus-09042024-01} && \phi_1^A \qquad \phi_3^A
\nonumber\\
&&
\\[-35pt]
&& \hspace{0.7cm} \phi_2
\nonumber
\eeq
These fields are decomposed in the gauge algebra generators $t^\asf$ as in \rf{manus-08042024-86}. Transformation rules of the vector fields $\phi_1^A$, $\phi_3^A$ and the scalar field  $\phi_2$ under $(A)dS_6$ algebra \rf{manus-08042024-10} take the form presented in \rf{manus-08042024-16}.  The vector fields are $y$-transversal \rf{manus-08042024-25}. Conformal dimensions of the fields in \rf{manus-09042024-01} are given by
\be \label{manus-16082023-02}
\Delta(\phi_1^A) = 1\,,\qquad \Delta(\phi_3^A) = 3\,,\qquad \Delta(\phi_2) = 2\,.
\ee
The vector field $\phi_1^A$ has the same conformal dimension as the primary conformal YM field entering the commonly used higher-derivative approach. Therefore the vector field $\phi_1^A$ can be identified with the primary conformal YM field. Then the remaining field $\phi_3^A$ turns out to be an auxiliary field. Below we show that the scalar field $\phi_2$ turns out to be a Stueckelberg field in our approach.

\noinbf{Gauge invariant Lagrangian}. The ordinary-derivative Lagrangian of the conformal YM field in $(A)dS_6$ we find takes the form
\be \label{manus-09042024-06}
\LL_\CYMsm =  - I_6 + 2\rho I_4\,, \qquad  I_6 :=  \half F_2 F_4 + \half F_3 F_3\,, \hspace{1cm}  I_4 : =  \frac{1}{4} F_2 F_2 \,,
\ee
where we use the shortcuts defined in \rf{manus-08042024-87sh}. Expressions for the field strengths are given by
\beq
\label{manus-09042024-10} && F_2^{AB} := (\ppp^A - \rho y^A)\phi_1^B - (\ppp^B - \rho  y^B)\phi_1^A + [\phi_1^A,\phi_1^B]\,,
\nonumber\\
&& F_4^{AB}: = \DD^A \phi_3^B -  \DD^B\phi_3^A \,,
\nonumber\\
&& F_3^A := \phi_3^A + \DD^A\phi_2\,,
\eeq
while the covariant derivative $\DD^A$ is given in \rf{manus-14092024-01} when $d=5$.
Expression for $F_2^{AB}$ in \rf{manus-09042024-10} tells us that $F_2^{AB}$ is a field strength for the conformal YM field $\phi_1^A$. The $y^A$-transversality constraint for gauge fields $\phi_1^A$, $\phi_3^A$ imply that the field strengths \rf{manus-09042024-10} are also $y$-transversal.

\noinbf{Gauge transformations}. In order to describe gauge symmetries of Lagrangian given in \rf{manus-09042024-06} we introduce two gauge parameters,
\be \label{manus-09042024-20}
\xi_0\,, \qquad \xi_2\,,
\ee
which are decomposed in  $t^\asf$ as in \rf{manus-08042024-87}. Gauge transformations of the fields we find take the form
\beq
\label{manus-09042024-21} && \delta_\xi\phi_1^A = \DD^A \xi_0\,,
\nonumber\\
&& \delta_\xi\phi_3^A = \DD^A \xi_2 + [\phi_3^A,\xi_0]\,,
\nonumber\\
&& \delta_\xi\phi_2 = - \xi_2 + [\phi_2,\xi_0]\,,
\eeq
where the covariant derivative $\DD^A$ is given in \rf{manus-14092024-02} when $d=5$.
Gauge transformations of field strengths \rf{manus-09042024-10} we find can be presented as
\beq
\label{manus-09042024-30} && \delta_\xi F_2^{AB} = [F_2^{AB},\xi_0]\,,
\nonumber\\
&& \delta_\xi F_4^{AB} = [F_4^{AB},\xi_0] + [F_2^{AB},\xi_2]\,,
\nonumber\\
&& \delta_\xi F_3^A =  [F_3^A,\xi_0]\,.
\eeq
From \rf{manus-09042024-21}, we see that the vector fields are realized as gauge fields, while the scalar field $\phi_2$  transforms as Stueckelberg field.

Field strengths \rf{manus-09042024-10} and gauge transformations \rf{manus-09042024-21}, \rf{manus-09042024-30} are simply obtained by covariantization of their flat space cousins found in Sec.3 in Ref.\cite{Metsaev:2023qif}. We verified that: 1) Plugging gauge transformations of fields \rf{manus-09042024-21} into field strengths \rf{manus-09042024-10} leads to gauge transformations of field strengths \rf{manus-09042024-30}; 2) Lagrangian \rf{manus-09042024-06} is invariant under gauge transformations \rf{manus-09042024-30}.
Note however that the gauge transformations do not fix the Lagrangian uniquely. The Lagrangian is determined uniquely by $K^A$-symmetries.%
\footnote{ For derivation of the results presented in this paper, we could used conformal flatness of $(A)dS$ space and known results for conformal YM field in flat space of arbitrary dimension obtained in Ref.\cite{Metsaev:2023qif}. Use of conformal flatness of $(A)dS$ for the derivation of metric-like Lagrangian for free field in $(A)dS_{d+1}$ may be found in Ref.\cite{Metsaev:2014iwa}. We note however that embedding space method we use in the present paper turns out to be much elegant, powerful, and pragmatical for the study of interacting conformal YM field in $(A)dS$.}

\noinbf{$K^A$-transformations of fields and field strengths}. We find the following $K^A$-transformations of fields \rf{manus-09042024-01}:
\beq
\label{manus-09042024-50} && \delta_{K^A} \phi_1^B = K_1^A \phi_1^B\,,
\nonumber\\
&& \delta_{K^A} \phi_3^B = K_3^A\phi_3^B - 2\theta^{AB} \phi_2 + 2\rho K_1^A\phi_1^B \,,
\nonumber\\
&& \delta_{K^A} \phi_2 = K_2^A\phi_2 + 2 \phi_1^A\,,
\eeq
while the $K^A$-transformations of the corresponding field strengths \rf{manus-09042024-10} take the form
\beq
\label{manus-09042024-51} && \delta_{K^A} F_2^{BC} = K_2^A F_2^{BC}\,,\qquad
\nonumber\\
&& \delta_{K^A} F_4^{BC} = K_4^A F_4^{BC} + 2\theta^{AB} F_3^C - 2 \theta^{AC} F_3^B + 2 \rho K_2^A F_2^{BC}\,,\qquad
\nonumber\\
&& \delta_{K^A} F_3^B  = K_3^A F_3^B - 2F_2^{AB}\,,
\eeq
where operator $K_\Delta^A$ appearing in \rf{manus-09042024-50}, \rf{manus-09042024-51} is given in \rf{manus-08042024-75}, \rf{manus-08042024-76}, while the symbol $\theta^{AB}$ is given in \rf{12062024-01}. From \rf{manus-09042024-50}, we see that the field $\phi_1^A$   transforms as expected for a primary field.

We verified that: 1) $K^A$-transformations \rf{manus-09042024-50}, \rf{manus-09042024-51} obey commutators \rf{manus-08042024-60}; 2) Plugging $K^A$- transformations of fields \rf{manus-09042024-50} into field strengths \rf{manus-09042024-10} leads to $K^A$- transformations of field strengths \rf{manus-09042024-51}; 3) Once $K^A$- transformations \rf{manus-09042024-51} are fixed, Lagrangian \rf{manus-09042024-06} is determined uniquely. We found $K^A$-transformations of fields and field strengths \rf{manus-09042024-50}, \rf{manus-09042024-51} by adopting to $(A)dS$ the procedure we used for the derivation of conformal transformations of fields and field strengths in flat space in Refs.\cite{Metsaev:2007fq,Metsaev:2023qif}. For the interrelation between $K^A$-transformations \rf{manus-09042024-50}, \rf{manus-09042024-51} and the conformal transformations in flat space used in Refs.\cite{Metsaev:2007fq,Metsaev:2023qif}, see Appendix D.

The following remarks are in order.

\noinbf{i}) Lagrangian  \rf{manus-09042024-06} does not consist of higher than second-order terms in the derivatives. Two-derivative contributions to Lagrangian \rf{manus-09042024-06} given by (up to normalization factors)
\be \label{manus-09042024-60}
\phi_3^A (\eta^{AB} \ppp^2  - \ppp^A \ppp^B)\phi_1^B\,, \hspace{1cm} \phi_1^A (\eta^{AB} \ppp^2  - \ppp^A \ppp^B)\phi_1^B\,, \hspace{1cm} \phi_2 \ppp^2 \phi_2\,,
\ee
are governed by the Klein-Gordon and Maxwell kinetic operators. The last two kinetic terms in \rf{manus-09042024-60} are diagonal with respect to the fields, while the first kinetic term in \rf{manus-09042024-60} is not diagonal with respect to the fields.  Appearance of the non-diagonal kinetic terms is characteristic feature of the generic formulation of conformal YM field. Lagrangian formulation of conformal YM field having the kinetic and mass terms that are diagonal with respect to all fields is realized in the framework of a formulation we refer to as decoupled formulation (see Sec.\ref{six-fac}).

\noinbf{ii})   As the quantities $I_6$,$I_4$ given in \rf{manus-09042024-06}, \rf{manus-09042024-10} are regular in the flat space limit, $\rho\rightarrow 0$, Lagrangian \rf{manus-09042024-06} is also regular in the flat space limit. We use the shortcuts $I_6^{\rho=0}$, $I_4^{\rho=0}$ for the flat space limit of the $I_6$, $I_4$ and note that, in $R^{5,1}$, only $I_6^{\rho=0}$ is invariant under conformal transformations, while the $I_4^{\rho=0}$ is invariant under conformal transformations only upon reduction to $R^{3,1}$. The reduction of minus $I_4^{\rho=0}$  to $R^{3,1}$ coincides with the standard Lagrangian of conformal YM field in $R^{3,1}$, while the minus $I_6^{\rho=0}$  coincides with the ordinary-derivative Lagrangian of conformal YM field in $R^{5,1}$ found in Ref.\cite{Metsaev:2023qif}.
Therefore, roughly speaking, the Lagrangian of conformal YM field in $(A)dS_6$ is obtained from the sum of Lagrangian of conformal YM field in $R^{5,1}$ and Lagrangian of conformal YM field in $R^{3,1}$ uplifted to $(A)dS_6$.

\noinbf{Extended gauge algebra}. Gauge symmetries are governed by two gauge parameters $\xi_0^\asf$ and $\xi_2^\asf$ entering $\xi_0$ and $\xi_2$ \rf{manus-09042024-20} respectively. Therefore, the extended gauge algebra consists of two generators denoted as
\be  \label{manus-09042024-70}
T_0^\asf\,,\quad T_2^\asf\,.
\ee
In order to use the extended gauge algebra setup described in Sec.\ref{not-conv} we need commutators for generators \rf{manus-09042024-70}. To this end, using gauge transformations \rf{manus-09042024-21} or \rf{manus-09042024-30} and considering commutators of two gauge  transformations, we find that generators \rf{manus-09042024-70} should obey the commutators
\be \label{manus-09042024-73}
[T_0^\asf,T_0^\bsf] = f^{\asf\bsf\csf} T_0^\csf\,,\qquad [T_0^\asf,T_2^\bsf] = f^{\asf\bsf\csf} T_2^\csf\,, \qquad [T_2^\asf,T_2^\bsf] = 0\,.
\ee
From \rf{manus-09042024-73} we see that the generator $T_2^\asf$ constitutes a radical of the extended gauge algebra.

Now we can use the general setup described in Sec.\ref{not-conv}. Setting $d=5$ in  \rf{manus-08042024-91}, we introduce
\be  \label{manus-09042024-72}
\phi^A : = \phi_1^{A\asf} T_0^\asf + \phi_3^{A\asf} T_2^\asf  \,,   \qquad F^{AB} : = F_2^{AB\asf} T_0^\asf + F_4^{AB\asf} T_2^\asf  \,,\qquad \xi := \xi_0^\asf T_0^\asf +  \xi_2^\asf T_2^\asf  \,.
\ee
Using $\phi^A$, $F^{AB}$, and $\xi$ above defined, we verify that  the expressions for $F_2^{AB}$ and $F_4^{AB}$ given in \rf{manus-09042024-10}, and the expressions for the gauge transformations  $\phi_1^A$, $\phi_3^A$, and $F_2^{AB}$, $F_4^{AB}$ given in \rf{manus-09042024-21} and \rf{manus-09042024-30} amount to expressions  \rf{manus-08042024-92}.

Following the general pattern in Sec.\ref{not-conv}, we then represent the set of the generators \rf{manus-09042024-70} as
\be \label{manus-09042024-72ad}
T_0^\asf\,, \quad S_2^\asf\,,  \qquad  S_2^\asf : = T_2^\asf\,,
\ee
where, in \rf{manus-09042024-72ad}, we identify the generator $S_2^\asf$ which has not been fixed in \rf{manus-08042024-94}. In the basis of $T_0^\asf$, $S_2^\asf$, the commutators \rf{manus-09042024-73} are represented as
\be \label{manus-09042024-73abc}
[T_0^\asf,T_0^\bsf] = f^{\asf\bsf\csf} T_0^\csf\,,\qquad [T_0^\asf,S_2^\bsf] = f^{\asf\bsf\csf} S_2^\csf\,, \qquad [S_2^\asf,S_2^\bsf] = 0\,.
\ee
From \rf{manus-09042024-73abc}, we see that the $S_2^\asf$ constitutes the radical of the extended gauge algebra. In other words, we have the Levy-Maltsev decomposition
\be
T_0^\asf\,, \ S_2^\asf \,\, = \,\,  S_2^\asf\,\, \subsetext\,\, T_0^\asf\,.
\ee

In order to represent the gauge transformation of the Stueckelberg field and the corresponding field strength \rf{manus-09042024-21}, \rf{manus-09042024-30} in the framework of the extended gauge algebra we use the general relations in \rf{manus-08042024-95} when $d=5$,
\be
\label{manus-09042024-75} \phi := \phi_2^\asf  S_2^\asf \,, \qquad F^A: = F_3^{A\asf}  S_2^\asf  \,, \qquad \phi_\subsm^A : = \phi_3^{A\asf} S_2^\asf  \,, \qquad\xi_\subsm : =  \xi_2^\asf S_2^\asf \,.
\ee
Using \rf{manus-09042024-73}, \rf{manus-09042024-75}, we note that the $F_3^A$ given in \rf{manus-09042024-10} and the gauge transformations of $\phi_2$, $F_3^A$ given in \rf{manus-09042024-21}, \rf{manus-09042024-30} can be represented as in \rf{manus-08042024-96}.

Finally, we note that the Lagrangian given in \rf{manus-09042024-06} can be represented as in \rf{manus-08042024-97}, \rf{manus-08042024-100}, where all non-zero values of the invariant bilinear forms are given by
\be \label{manus-09042024-76}
G_{02} = 1\,, \quad G_{00} = - 2\rho\,,\qquad G_{22}^\subsm = 1\,.
\ee
For $d=5$, the conjectured expressions in \rf{manus-08042024-100-a1} are in agreement with \rf{manus-09042024-76}.

\noinbf{Higher-derivative Lagrangian of conformal YM field in $(A)dS_6$}. We now demonstrate how a higher-derivative Lagrangian is obtained in our approach. To this end we gauge away the Stueckelberg field $\phi_2$ and represent Lagrangian \rf{manus-09042024-06} as (up to total derivative)
{\small
\be \label{manus-09042024-78}
\LL_\CYMsm  = -\KK \Tr\big( \phi_3^A \DD^B F_2^{BA} - \half \phi_3^A \phi_3^A + \half \rho F_2^{AB} F_2^{AB})\,, \qquad \DD^B F_2^{BA}:=\ppp^B F_2^{BA} + [\phi_1^B,F_2^{BA}]\,,
\ee
}
where we use  convention for $\Tr$-operation given in  \rf{manus-08042024-85}.
Equations of motion for the auxiliary field $\phi_3^A$ give solution for $\phi_3^A$ which we denote as $\phib_3^A$,
\be \label{manus-09042024-80}
\phib_3^A = \DD^B F_2^{BA}\,.
\ee
Plugging $\phib_3^A$ \rf{manus-09042024-80} into Lagrangian \rf{manus-09042024-78} and using \rf{manus-08042024-85}, we get a higher-derivative Lagrangian
\be \label{manus-09042024-81}
\LL_\CYMsm^\Hdr = -\KK\, \Tr L_\CYMsm^\Hdr\,, \qquad L_\CYMsm^\Hdr: =   \half \DD^A F_2^{AC}\DD^B F_2^{BC} + \half \rho F_2^{AB} F_2^{AB}\,.
\ee
For the comparison with results in $(A)dS_8$ and $(A)dS_{10}$ we find it convenient to represent $L_\CYMsm^\Hdr$ \rf{manus-09042024-81} as
\be \label{manus-09042024-82}
L_\CYMsm^\Hdr: =  I_6^\Hdr + 2\rho I_4\,,\qquad  I_6^\Hdr: = \half \phib_3^A \phib_3^A\,, \qquad I_4: = \frac{1}{4} F_2^{AB} F_2^{AB}\,.
\ee
From \rf{manus-09042024-82} we see that, the higher-derivative Lagrangian of conformal YM field in $(A)dS_6$ is obtained from the sum of Lagrangian of conformal YM field in $R^{5,1}$ and Lagrangian of conformal YM field in $R^{3,1}$ uplifted to $(A)dS_6$.

\newsection{\large Conformal Yang-Mills field in $(A)dS_6$ space. Decoupled formulation } \label{six-fac}

{\bf Field content}. In the framework of gauge invariant decoupled
formulation, the conformal YM field in $(A)dS_6$ is described by two  vector fields and one scalar field:
\beq
\label{manus-12042024-01} && \varphi_1^A \qquad \varphi_3^A
\nonumber\\
&&
\nonumber\\[-35pt]
&& \hspace{0.7cm} \varphi_2
\eeq
The vector fields are $y$-transversal \rf{manus-08042024-25}. Fields \rf{manus-12042024-01} are decomposed in $t^a$ as in \rf{manus-08042024-86add}.
Realization of $(A)dS_6$ algebra \rf{manus-08042024-10} on the vector fields and the scalar field  takes the same form as in \rf{manus-08042024-16}.  Below, we show that the vector field $\varphi_1^A$ describes massless field, while the vector field $\varphi_3^A$ and the scalar field $\varphi_2$ describe massive vector field. In other words, the vector field $\varphi_3^A$ and the scalar field $\varphi_2$ provide the gauge invariant description of massive field in $(A)dS_6$, where the scalar field $\varphi_2$ plays the role of a Stueckelberg field.
The mass squares are given by,
\be \label{manus-12042024-02}
m^2_{\varphi_1} = 0\,,\qquad m^2_{\varphi_3} = 2\rho\,.
\ee

\noinbf{Gauge invariant Lagrangian}. In the decoupled formulation, the ordinary-derivative Lagrangian of the conformal YM field in $(A)dS_6$ we find takes the form
\be
\label{manus-12042024-06}
k_\rho^{-1} \LL_\CYMsm =  -  \frac{1}{4} \FF_2 \FF_2  + \frac{1}{4}  \FF_4 \FF_4 +  \rho  \FF_3 \FF_3\,,\qquad k_\rho := -2\rho\,,
\ee
where we use the shortcuts for the scalar products  defined in \rf{manus-08042024-88add}, while expressions for field strengths are given below. It is instructive to represent  Lagrangian \rf{manus-12042024-06} as
\be \label{manus-12042024-06add}
k_\rho^{-1} \LL_\CYMsm =  - \II_4  + \II_8 \,,  \qquad \II_4 := \frac{1}{4} \FF_2 \FF_2\,, \hspace{1cm} \II_8 : = \frac{1}{4} \FF_4 \FF_4 + \frac{m_{\varphi_3}^2}{2} \FF_3\FF_3\,,
\ee
where the $m_{\varphi_3}^2$ is given in \rf{manus-12042024-02}. The field strengths entering Lagrangian \rf{manus-12042024-06} are given by
\beq
\label{manus-12042024-10} && \FF_2^{AB} = (\ppp^A - \rho y^A)\varphi_1^B -  (\ppp^B -  \rho y^B)\varphi_1^A  +    [\varphi_1^A,\varphi_1^B]  -  [\varphi_3^A,\varphi_3^B]\,,\qquad
\nonumber\\
&& \FF_4^{AB} = \DD^A \varphi_3^B -  \DD^B\varphi_3^A   - 2 [\varphi_3^A,\varphi_3^B] \,,\qquad
\nonumber\\
&& \FF_3^A =  \varphi_3^B + \DD^A \varphi_2  - [\varphi_3^A,\varphi_2] \,,
\eeq
where the covariant derivative $\DD^A$ is given in \rf{manus-14092024-01add} when $d=5$.
The unusual expression for the field strength $\FF_2^{AB}$ in \rf{manus-12042024-10} can be represented as
{\small
\be
\FF_2^{AB} = \FF_{2\,\st}^{AB} -  [\varphi_3^A,\varphi_3^B]\,, \qquad  \FF_{2\,\st}^{AB} := (\ppp^A - \rho y^A)\varphi_1^B -  (\ppp^B -  \rho y^B)\varphi_1^A  + [\varphi_1^A,\varphi_1^B]\,,
\ee
}
\!\!where $\FF_{2\, \st}^{AB}$ is a standard field strength for the YM field $\varphi_1^A$. The appearance of the $\FF_2^{AB}$ in place of the $\FF_{2\,\st}^{AB}$ is the interesting feature of the decoupled formulation. It is the use of the  $\FF_2^{AB}$ that allows us to write the expression for the Lagrangian $\LL_\CYMsm$ in the simple form given in \rf{manus-12042024-06}. In view of the $y^A$-transversality of the vector fields, field strengths \rf{manus-12042024-10} are also $y$-transversal.

\noinbf{Gauge transformations}. In order to describe gauge symmetries of Lagrangian given in \rf{manus-12042024-06} we introduce two gauge parameters:
\be \label{manus-12042024-20}
\eta_0\,, \qquad \eta_2\,,
\ee
which are decomposed in $t^a$ as in \rf{manus-08042024-87add}. Gauge transformations of the fields we find take the form
\beq
\label{manus-12042024-21} && \delta_\eta \varphi_1^A = \ppp^A \eta_0 +   [\varphi_1^A,\eta_0] -  [\varphi_3^A,\eta_2]\,,
\nonumber\\
&& \delta_\eta \varphi_3^A = \ppp^A \eta_2 +  [\varphi_1^A,\eta_2] +   [\varphi_3^A,\eta_0] -   2 [\varphi_3^A,\eta_2]\,,
\nonumber\\
&& \delta_\eta \varphi_2 = - \eta_2 +  [\varphi_2,\eta_0] -   [\varphi_2,\eta_2] \,,
\eeq
where the covariant derivative $\DD^A$ is given in \rf{manus-14092024-02add} when $d=5$.
Plugging gauge transformations of fields \rf{manus-12042024-21} into field strengths \rf{manus-12042024-10}, we find gauge transformations of the field strengths,
\beq
\label{manus-12042024-30}  &&   \delta_\eta \FF_2^{AB} = \half [\FF_2^{AB},\eta_0] -   \half [\FF_4^{AB},\eta_2]\,,
\nonumber\\
&& \delta_\eta \FF_4^{AB} =   [\FF_4^{AB},\eta_0] +    [\FF_2^{AB},\eta_2] -    2 [\FF_4^{AB},\eta_2]\,,
\nonumber\\
&& \delta_\eta \FF_3^A \ \ = \ \   [\FF_3^A,\eta_0] -  [\FF_3^A,\eta_2] \,.
\eeq
It easy to see that Lagrangian \rf{manus-12042024-02} is invariant under gauge transformations \rf{manus-12042024-30}.

\noinbf{$K^A$-transformations of fields and field strengths}. We find the following $K^A$-transformations of fields \rf{manus-12042024-01}:
\beq
\label{manus-12042024-50} && \delta_{K^A} \varphi_1^B = K_3^A \varphi_3^B - 2\theta^{AB}\varphi_2\,,
\nonumber\\
&& \delta_{K^A} \varphi_3^B = 2 K_2^A \varphi_3^B - K_1^A \varphi_1^B - 2\theta^{AB}\varphi_2 \,,
\nonumber\\
&& \delta_{K^A} \varphi_2 = K_2^A \varphi_2  + \frac{1}{\rho} \varphi_3^A - \frac{1}{\rho}  \varphi_1^A\,,
\eeq
while the $K^A$-transformations of the field strengths \rf{manus-12042024-10} take the form
\beq
\label{manus-12042024-51} && \hspace{-1.2cm} \delta_{K^A} \FF_2^{BC} = K_4^A \FF_4^{BC} + 2\theta^{AB} \FF_3^C - 2 \theta^{AC} \FF_3^B \,,\qquad
\nonumber\\
&& \hspace{-1.2cm} \delta_{K^A} \FF_4^{BC} = 2 K_3^A \FF_4^{BC} - K_2^A \FF_2^{BC} + 2\theta^{AB} \FF_3^C - 2 \theta^{AC} \FF_3^B \,,\qquad
\nonumber\\
&& \hspace{-1.2cm} \delta_{K^A} \FF_3^B  = K_3^A \FF_3^B   - \frac{1}{\rho} \FF_4^{AB} + \frac{1}{\rho} \FF_2^{AB}\,,
\eeq
where $\theta^{AB}$ is defined in \rf{12062024-01}. Action of operator $K_\Delta^A$ on the fields and fields strengths in \rf{manus-12042024-50}, \rf{manus-12042024-51} is defined in the same way as in \rf{manus-08042024-75}, \rf{manus-08042024-76}.
The following remarks are in order.

\noinbf{i}) To quadratic order in the fields, Lagrangian \rf{manus-12042024-06} takes the form
{\small
\beq
\label{manus-12042024-54} && \hspace{-1.5cm} k_\rho^{-1} \LL_\CYMsm^\smtwo =  - \II_4^\smtwo  +  \II_8^\smtwo \,,
\nonumber\\
&& - \II_4^\smtwo := \half \varphi_1^A \big(\ppp^2 - 4\rho \big) \varphi_1^A + \half L_2 L_2\,, \qquad  L_2 : = \ppp^A \varphi_1^A\,,
\nonumber\\
&& -\II_8^\smtwo := \half \varphi_3^A \big( \ppp^2- 4\rho - m_{\varphi_3}^2 \big) \varphi_3^A  +   \frac{m_{\varphi_3}^2}{2} \varphi_2 \big(\ppp^2  - m_{\varphi_3}^2 \big) \varphi_2 + \half L_4 L_4 \,,
\nonumber\\
&&  \hspace{1.8cm}  L_4 : = \ppp^A \varphi_3^A + m_{\varphi_3}^2 \varphi_2\,, \qquad m_{\varphi_3}^2 = 2\rho\,, \qquad k_\rho = - 2\rho\,.
\eeq
}
\!The  $\II_4^\smtwo$ depends only the field $\varphi_1^A$, while the $\II_8^\smtwo$ depends only on the fields $\varphi_3^A$, $\varphi_2$. Thus, to quadratic order in the fields, the two set of fields given by $\varphi_1^A$ and $\varphi_3^A$, $\varphi_2$ are decoupled.  At the level of interaction vertices these two set of fields are coupled.
The metric-like cousin of  Lagrangian \rf{manus-12042024-54} was obtained in Ref.\cite{Metsaev:2014iwa}.

\noinbf{ii}) From \rf{manus-12042024-54}, we see that the kinetic operators of the scalar and vector fields take the form of the respective standard Klein-Gordon and Maxwell kinetic operators. Moreover, using the gauge conditions $L_2=0$, $L_4=0$, we see that the two-derivative and derivative-independent terms are diagonal with respect to all fields. This implies that, for the quantization of the conformal YM field in $(A)dS_6$, we can try to use the standard quantization methods in $(A)dS$.

\noinbf{iii}) The minus $\II_4^\smtwo$ coincides with a Lagrangian for free massless vector field $\varphi_1^A$ in $(A)dS_6$, while the minus $\II_8^\smtwo$ coincides a Lagrangian of free massive vector field in $(A)dS_6$ described by the fields $\varphi_3^A$, $\varphi_2$. The signs of $\II_4^\smtwo$ and $\II_4^\smtwo$ in the expression for $\LL_\CYMsm^\smtwo$ \rf{manus-12042024-54} depend on $k_\rho$. For $AdS_6$, we note $k_\rho > 0 $, and hence the massless field is realized with the correct sign of kinetic term, while the massive field is realized with the wrong sign of the kinetic term and a negative value of the mass square $m_{\varphi_3}^2$.  For $dS_6$, we note $k_\rho < 0 $, hence the massless field is realized with the wrong sign of the kinetic term, while the massive field is realized with the correct sign of the kinetic term and a positive value of the mass square $m_{\varphi_3}^2$.

\noinbf{iv}) Lagrangian $\LL_\CYMsm^\smtwo$ \rf{manus-12042024-54} is invariant under the linearized gauge transformations
\be \label{manus-12042024-55}
\delta_\eta^{\,\lin} \varphi_1^A = \ppp^A \eta_0 \,, \qquad \delta_\eta^{\,\lin} \varphi_3^A = \ppp^A \eta_2\,, \qquad \delta_\eta^{\,\lin} \varphi_2 = - \eta_2\,.
\ee
Gauge transformations \rf{manus-12042024-55} tell us the that the vector fields $\varphi_1^A$, $\varphi_3^A$ are realized as gauge fields, while the scalar field $\varphi_2$ transforms as Stueckelberg field. In other words, massive field appearing in our approach is realized in the same way as in the framework of gauge invariant formulation of massive fields. Various gauge invariant formulations of arbitrary spin massive fields and list of references may be found, e.g., in Refs.\cite{Zinoviev:2001dt}-\cite{Delplanque:2024enh}. Comparison of various metric-like formulations of arbitrary spin massive fields may be found in Ref.\cite{Fegebank:2024yft}.

\noinbf{v}) The Lagrangian, the gauge transformations, and the $K^A$-transformations presented in this section are obtained from the ones in Sec.\ref{six-gen} by using the linear transformation from the fields entering the generic formulation to the fields entering the decoupled formulation. Up to normalization factors, the linear transformation is fixed uniquely by requiring the generic Lagrangian \rf{manus-09042024-06} to be diagonal with respect to the field strengths $\FF_2^{AB}$, $\FF_4^{AB}$, $\FF_3^A$. Explicit form of the linear transformation is presented in Appendix C.

\noinbf{vi}) $K^A$-transformations \rf{manus-12042024-50} are linear with respect to  fields \rf{manus-12042024-01}. From \rf{manus-12042024-50}, we see that on a space of fields \rf{manus-12042024-01} there is no invariant subspace under $K^A$- transformations. In other words, at the level of the $K^A$-transformations \rf{manus-12042024-50}, the fields $\varphi_1^A$ and $\varphi_3^A$, $\varphi_2$ are coupled. Only the combination of  the  $\II_4$ and $\II_8$ appearing in \rf{manus-12042024-06add} is invariant under $K^A$-transformations \rf{manus-12042024-51}.

\noinbf{vii}) To consider the flat space limit, $\rho\rightarrow 0$, we introduce new fields by the relations
\be \label{manus-12042024-60} \varphi_1^A =  |\rho|^{-1/2} \varphi_{1\,\new}^A\,, \qquad \varphi_3^A = |\rho|^{-1/2} \varphi_{3\,\new}^A\,, \qquad \varphi_2 = |\rho|^{-1/2} \varphi_{2\,\new}\,.
\ee
We see then that, to the second order in the new fields, Lagrangian \rf{manus-12042024-54} is regular in the flat space limit.
Note however that, in terms of the new fields \rf{manus-12042024-60}, the interaction vertices of Lagrangian \rf{manus-12042024-06} are singular in the flat space limit.

\noinbf{Extended gauge algebra}. Gauge symmetries are governed by two
gauge parameters $\eta_0^\asf$ and $\eta_2^\asf$ entering $\eta_0$ and $\eta_2$ \rf{manus-12042024-20} respectively. Therefore, the extended gauge
algebra consists of two generators,
\be \label{manus-12042024-71}
\TT_0^\asf\,, \qquad \TT_2^\asf  \,.
\ee
Considering commutators of gauge transformations \rf{manus-12042024-21} or \rf{manus-12042024-30}, we find the commutators
\be \label{manus-12042024-72}
[\TT_0^\asf,\TT_0^\bsf] = f^{\asf\bsf\csf} \TT_0^\csf\,, \hspace{1cm} [\TT_0^\asf,\TT_2^\bsf] = f^{\asf\bsf\csf} \TT_2^\csf\,,\qquad [\TT_2^\asf,\TT_2^\bsf] = - f^{\asf\bsf\csf} \TT_0^\csf -  2 f^{\asf\bsf\csf} \TT_2^\csf\,.
\ee

Using commutators \rf{manus-12042024-72}, we can apply the general setup of extended gauge algebra described in Sec.\ref{not-conv}. Namely, setting $d=5$ in \rf{manus-08042024-91ab}, we introduce
\be
\label{manus-12042024-73} \varphi^A : = \varphi_1^{A\asf} \TT_0^\asf + \varphi_3^{A\asf} \TT_2^\asf  \,,   \qquad \FF^{AB} : = \FF_2^{AB\asf} \TT_0^\asf + \FF_4^{AB\asf} \TT_2^\asf  \,,\qquad \eta : = \eta_0^{A\asf} \TT_0^\asf + \eta_2^{A\asf} \TT_2^\asf \,.
\ee
Using $\varphi^A$, $\FF^{AB}$, and $\eta$, we verify that the expressions for $\FF_2^{AB}$ and $\FF_4^{AB}$ given in \rf{manus-12042024-06} and the gauge transformations for the $\varphi_1^A$, $\varphi_3^A$ and $\FF_2^{AB}$, $\FF_4^{AB}$ given in \rf{manus-12042024-21} and \rf{manus-12042024-30} can be represented in a more compact way as in \rf{manus-08042024-120}.

Following the general pattern in Sec.\ref{not-conv}, we then represent the set of the generators \rf{manus-12042024-71} as
\be \label{manus-12042024-73ad}
\TT_0^\asf\,, \quad \Scl_2^\asf\,,  \qquad  \Scl_2^\asf : = \TT_0^\asf + \TT_2^\asf\,,
\ee
where, in \rf{manus-12042024-73ad}, we identify the generator $\Scl_2^\asf$ which has not been fixed in \rf{manus-08042024-94ab}. In the basis of $\TT_0^\asf$, $\Scl_2^\asf$, the commutators \rf{manus-12042024-72} are represented as
\be \label{manus-12042024-76ad}
[\TT_0^\asf,\TT_0^\bsf] =  f^{\asf\bsf\csf} \TT_2^\csf\,, \qquad [S_2^\asf,\TT_0^\bsf] =  f^{\asf\bsf\csf} \Scl_2^\csf\,, \qquad [\Scl_2^\asf,\Scl_2^\bsf] = 0\,.
\ee
From \rf{manus-12042024-76ad}, we see that the $\Scl_2^\asf$ constitutes the radical of the extended gauge algebra. In other words, we have the Levy-Maltsev decomposition
\be
\TT_0^\asf\,, \ \Scl_2^\asf \,\, =\,\, \Scl_2^\asf\, \subsetext\, \TT_0^\asf\,.
\ee

In order to represent the gauge transformation of the Stueckelberg field and the corresponding field strength \rf{manus-12042024-21}, \rf{manus-12042024-30} in the framework of the extended gauge algebra we use the general relations in \rf{manus-08042024-91ab} when $d=5$,
\be \label{manus-12042024-75}
\varphi := \varphi_2^\asf  \Scl_2^\asf \,, \qquad \FF^A: = \FF_3^{A\asf}  \Scl_2^\asf  \,,\qquad \varphi_\subsm^A : = \varphi_3^{A\asf} \Scl_2^\asf  \,, \qquad \eta_\subsm : =  \eta_2^\asf \Scl_2^\asf\,. \qquad
\ee

Using the gauge algebra ingredients defined in \rf{manus-12042024-71}, \rf{manus-12042024-73}, and \rf{manus-12042024-75}, the expression for $\FF_3^A$ in \rf{manus-12042024-10} and the gauge transformations of $\varphi_2$ and $\FF_3^A$ given in \rf{manus-12042024-21}, \rf{manus-12042024-30} can be represented in a more short way as in \rf{manus-08042024-126}.

Finally, we note that the Lagrangian given in \rf{manus-12042024-06} can be represented as in \rf{manus-08042024-117}, \rf{manus-08042024-100add}, where all non-zero values of the invariant bilinear forms are given by
\be \label{manus-12042024-76}
\GG_{00} = k_\rho\,,\quad  \quad \GG_{22} = - k_\rho \,, \qquad
\GG_{22}^\subsm = - 2\rho k_\rho\,,  \qquad k_\rho := -2\rho\,.
\ee
For $d=5$, the conjectured expressions in \rf{manus-08042024-106} are in agreement with \rf{manus-12042024-76}.

\newsection{\large Conformal Yang-Mills field in $(A)dS_8$ space. Generic formulation} \label{eight-gen}

{\bf Field content}. In the framework of gauge invariant generic formulation, the conformal YM field in $(A)dS_8$ is described by three  vector fields and two scalar fields:
\beq
\label{manus-19042024-01}  && \phi_1^A \qquad \phi_3^A \qquad\quad \phi_5^A
\nonumber\\
&&
\\[-35pt]
&& \hspace{0.8cm} \phi_2 \qquad \phi_4
\nonumber
\eeq
Transformation rules of the vector fields and the scalar fields under $(A)dS_8$ algebra \rf{manus-08042024-10} take the form presented in \rf{manus-08042024-16}.  Fields in \rf{manus-19042024-01} are decomposed in $t^a$ as in \rf{manus-08042024-86}. The vector fields are $y$-transversal \rf{manus-08042024-25}. Conformal dimensions of the fields in \rf{manus-19042024-01} are given by
\be \label{manus-19042024-02}
\Delta(\phi_{2n+1}^a) = 2n+1\,, \hspace{0.5cm} n=0,1,2\,; \hspace{2cm} \Delta(\phi_{2n}) = 2n\,, \hspace{0.7cm} n=1,2\,.\qquad
\ee
The vector field $\phi_1^A$ has the same conformal dimension as the primary conformal YM field entering the commonly used higher-derivative approach. Therefore the vector field $\phi_1^A$ can be identified with the primary conformal YM field. Below we show that  remaining two vector fields $\phi_3^A$ and $\phi_5^A$ turn out to be an auxiliary fields, while two scalar fields $\phi_2$, $\phi_4$ turn out to be Stueckelberg fields.

\noinbf{Gauge invariant Lagrangian}. The ordinary-derivative Lagrangian of the conformal YM field in $(A)dS_8$ we find takes the form
\beq
\label{manus-19042024-06}  &&\hspace{-1cm}  \LL_\CYMsm = - I_8 + 10 \rho I_6 - 24 \rho^2 I_4\,,
\nonumber\\
&& I_8 :=  \half F_2 F_6 + \frac{1}{4} F_4 F_4 + F_3 F_5\,,
\nonumber\\
&& I_6 :=  \half  F_2 F_4 + \half F_3 F_3\,, \hspace{1cm} I_4 :=  \frac{1}{4} F_2 F_2\,,
\eeq
where we use the shortcuts defined in \rf{manus-08042024-87sh}. Expressions for the field strengths are given by
\beq
\label{manus-19042024-10} && F_2^{AB} := (\ppp^A - \rho y^A)\phi_1^B - (\ppp^B - \rho y^B)\phi_1^A + [\phi_1^A,\phi_1^B]\,,
\nonumber\\
&& F_4^{AB}: = \DD^A \phi_3^B -  \DD^B \phi_3^A \,,
\nonumber\\
&& F_6^{AB}: = \DD^A \phi_5^B -  \DD^B\phi_5^A +[\phi_3^A,\phi_3^B] \,,
\nonumber\\
&& F_3^A := \phi_3^A + \DD^A\phi_2\,,
\nonumber\\
&& F_5^A := \phi_5^A + \DD^A\phi_4 + \half [\phi_3^A,\phi_2]\,,
\eeq
where the covariant derivative $\DD^A$ is given in \rf{manus-14092024-01} when $d=7$. Note that the $y^A$-transversality of the vector fields imply that field strengths \rf{manus-19042024-10} are also $y$-transversal. Lagrangian \rf{manus-19042024-06} is built entirely in terms of the quantities $I_{2n+4}$, $n=0,1,2$, which depend on the field strengths in the same way as their flat space cousins studied in Ref.\cite{Metsaev:2023qif}.

\noinbf{Gauge transformations}. In gauge invariant approach, each vector field \rf{manus-19042024-01} is accompanied by gauge parameter. This implies that we use three gauge parameters,
\be \label{manus-19042024-20}
\xi_0\,, \qquad \xi_2\,, \qquad \xi_4\,,
\ee
which are decomposed in  $t^\asf$ as in \rf{manus-08042024-87}. Gauge transformations of the fields we find take the form
\beq
\label{manus-19042024-21} &&  \delta_\xi\phi_1^A = \DD^A \xi_0\,,
\nonumber\\
&& \delta_\xi\phi_3^A = \DD^A \xi_2 + [\phi_3^A,\xi_0]\,,
\nonumber\\
&& \delta_\xi \phi_5^A  = \DD^A \xi_4 + [\phi_5^A,\xi_0] + [\phi_3^A,\xi_2] \,,
\nonumber\\
&& \delta_\xi\phi_2 = - \xi_2 + [\phi_2,\xi_0]\,,
\nonumber\\
&& \delta_\xi \phi_4 = - \xi_4 + [\phi_4,\xi_0] + \half   [\phi_2,\xi_2] \,,
\eeq
where the covariant derivative $\DD^A$ is given in \rf{manus-14092024-02} when $d=7$. Gauge transformations of the corresponding field strengths \rf{manus-19042024-10} we find can be presented as
\beq
\label{manus-19042024-30} &&  \delta_\xi F_2^{AB} = [F_2^{AB},\xi_0]\,,
\nonumber\\
&& \delta_\xi F_4^{AB} = [F_4^{AB},\xi_0] + [F_2^{AB},\xi_2]\,,
\nonumber\\
&& \delta_\xi F_6^{AB} =  [F_6^{AB},\xi_0] + [F_4^{AB},\xi_2] + [F_2^{AB},\xi_4] \,,
\nonumber\\
&& \delta_\xi F_3^A =  [F_3^A,\xi_0]\,,
\nonumber\\
&& \delta_\xi F_5^A = [F_5^A,\xi_0] + \half [F_3^A,\xi_2] \,.
\eeq

From \rf{manus-19042024-21}, we see that the vector fields are realized as gauge fields, while the scalar fields are realized as Stueckelberg fields.

\noinbf{$K^A$-transformations of fields and field strengths}. We find the following $K^A$-transformations of fields \rf{manus-19042024-01}:
{\small
\beq
\label{manus-19042024-50} && \delta_{K^A} \phi_1^B = K_1^A \phi_1^B\,,
\nonumber\\
&& \delta_{K^A} \phi_3^B = K_3^A \phi_3^B - 2\theta^{AB} \phi_2 + 4\rho K_1^A\phi_1^B \,,
\nonumber\\
&& \delta_{K^A} \phi_5^B = K_5^A \phi_5^B - 4\theta^{AB} \phi_4 + 4 \rho K_3^A\phi_3^B - 12 \rho y^A \phi_3^B + 4\rho \theta^{AB}\phi_2 + 16 \rho^2 K_1^A  \phi_1^B \,,\qquad
\nonumber\\
&& \delta_{K^A} \phi_2 = K_2^A \phi_2 + 4 \phi_1^A\,,
\nonumber\\
&& \delta_{K^A} \phi_4 = K_4^A \phi_4 + 2 \phi_3^A + 2\rho K_2^A \phi_2 - 8\rho y^A \phi_2 + 16\rho \phi_1^A\,,
\eeq
}
\!\!while the $K^A$-transformations of field strengths \rf{manus-19042024-10} take the form
{\small
\beq
\label{manus-19042024-51} &&  \delta_{K^A} F_2^{BC} = K_2^A F_2^{BC}\,,\qquad
\nonumber\\
&& \delta_{K^A} F_4^{BC} = K_4^A F_4^{BC}  + 2\theta^{AB} F_3^C - 2 \theta^{AC} F_3^B + 4 \rho K_2^A F_2^{BC}\,,\qquad
\nonumber\\
&& \delta_{K^A} F_6^{BC} = K_6^A F_6^{BC} + 4\theta^{AB} F_5^C - 4 \theta^{AC} F_5^B + 4 \rho K_4^A F_4^{BC}
\nonumber\\
&& \hspace{1.5cm} - \,\, 12\rho y^A F_4^{BC} - 4\rho \theta^{AB} F_3^C + 4\rho \theta^{AC} F_3^B + 16\rho^2 K_2^A F_2^{BC}\,,\qquad
\nonumber\\
&& \delta_{K^A} F_3^B  = K_3^A F_3^B  -  4F_2^{AB}\,,
\nonumber\\
&& \delta_{K^A} F_5^B  = K_5^A F_5^B  - 2F_4^{AB} + 2\rho K_3^A F_3^B -  8\rho y^A F_3^B - 16\rho F_2^{AB}\,,
\eeq
}
\!\!where $K_\Delta^A$ in \rf{manus-19042024-50}, \rf{manus-19042024-51} is defined in \rf{manus-08042024-75}, \rf{manus-08042024-76}, while the symbol $\theta^{AB}$ is defined in \rf{12062024-01}.

The following two remarks are in order.

\noinbf{i}) Lagrangian  \rf{manus-19042024-06} does not consist of higher than second-order terms in the derivatives. Two-derivative contributions to Lagrangian \rf{manus-19042024-06} take the forms (up to normalization factors):
\beq
\label{manus-19042024-60} &&  \phi_{2m+1}^A (\eta^{AB} \ppp^2  - \ppp^A \ppp^B)\phi_{2n+1}^B\,, \hspace{0.3cm} m,n=0,1,2\,, \qquad n+m\leq 2;
\nonumber\\
&& \phi_{2m} \ppp^2 \phi_{2n}\,,   \hspace{3.8cm} m,n=1,2\,, \hspace{1.2cm} n+m\leq 3\,.
\eeq
From \rf{manus-19042024-60}, we see that, for the scalar and vector fields, the two-derivative terms in Lagrangian \rf{manus-19042024-06} are governed by the respective Klein-Gordon and Maxwell kinetic operators. Note that the Lagrangian involves both  diagonal and non-diagonal kinetic terms. Lagrangian of conformal YM field in $(A)dS_8$ involving only diagonal kinetic and mass-like terms is presented in Sec.\ref{eight-fac}.

\noinbf{ii}) The quantities $I_{2n+4}$, $n=0,1,2$ and hence the Lagrangian \rf{manus-19042024-06} are  regular in the flat space limit. Using the shortcut $I_{2n+4}^{\rho=0}$ for the flat space limit of $I_{2n+4}$, we note that, in $R^{7,1}$, only $I_8^{\rho=0}$ is invariant under conformal transformations. The $I_6^{\rho=0}$ and $I_4^{\rho=0}$ are invariant under conformal transformations only upon reduction to $R^{5,1}$ and $R^{3,1}$ respectively.
The minus $I_8^{\rho=0}$ coincides with ordinary-derivative Lagrangian of conformal YM field in $R^{7,1}$ found in Ref.\cite{Metsaev:2023qif}, while the reduction of the minus $I_6^{\rho=0}$ to $R^{5,1}$ coincides with the ordinary-derivative Lagrangian of conformal YM field in $R^{5,1}$ found in Ref.\cite{Metsaev:2023qif}.
Reduction of the minus $I_4^{\rho=0}$ to $R^{3,1}$ coincides with the standard Lagrangian of conformal YM field in $R^{3,1}$.

\noinbf{Extended gauge algebra}.  Gauge symmetries are governed by three gauge parameters $\xi_{2n}^\asf$ entering $\xi_{2n}$, $n=0,1,2$ respectively. Therefore, the extended gauge algebra consists of three generators denoted as
\be  \label{manus-19042024-70}
T_0^\asf\,, \qquad T_2^\asf\,, \qquad T_4^\asf\,.
\ee
In order to apply the general setup of the extended gauge algebra described in Sec.\ref{not-conv}  we should present commutation relations for the generators in \rf{manus-19042024-70}. To this end, using gauge transformations given in \rf{manus-19042024-21} or \rf{manus-19042024-30} and considering commutators of two gauge  transformations, we find that generators \rf{manus-19042024-70} should obey the commutators
{\small
\beq
\label{manus-19042024-72} && [T_0^\asf,T_0^\bsf] = f^{\asf\bsf\csf} T_0^\csf\,,\qquad [T_0^\asf,T_2^\bsf] = f^{\asf\bsf\csf} T_2^\csf\,,\qquad   [T_0^\asf,T_4^\bsf] = f^{\asf\bsf\csf} T_4^\csf\,,
\nonumber\\
&& [T_2^\asf,T_2^\bsf] = f^{\asf\bsf\csf} T_4^\csf\,,\qquad [T_{2m}^\asf,T_{2n}^\bsf] = 0\,, \quad \hbox{ for } \ m+n > 2\,.
\eeq
}
\!From \rf{manus-19042024-72}, we see that $T_2^\asf$, $T_4^\asf$ constitute a radical of the extended gauge algebra \rf{manus-19042024-70}.

Now we can use the general setup described in Sec.\ref{not-conv}. Setting $d=7$ in  \rf{manus-08042024-91}, we introduce
\beq
\label{manus-19042024-71} && \phi^A : = \phi_1^{A\asf} T_0^\asf + \phi_3^{A\asf} T_2^\asf + \phi_5^{A\asf} T_4^\asf  \,,   \qquad F^{AB} : = F_2^{AB\asf} T_0^\asf + F_4^{AB\asf} T_2^\asf  + F_6^{AB\asf} T_4^\asf  \,,
\nonumber\\
&& \hspace{3cm} \xi := \xi_0^\asf T_0^\asf +  \xi_2^\asf T_2^\asf +  \xi_4^\asf T_4^\asf  \,.
\eeq
Using $\phi^A$, $F^{AB}$, and $\xi$ above defined, we note that  the expressions for field strengths given in \rf{manus-19042024-10}, and the gauge transformations for the fields and field strengths given in \rf{manus-19042024-21} and \rf{manus-19042024-30}  amount to the expressions given in \rf{manus-08042024-92}.

Following the general pattern in Sec.\ref{not-conv}, we then represent the set of generators \rf{manus-19042024-70} as
\be \label{manus-19042024-72ad}
T_0^\asf\,, \quad S_2^\asf\,, \quad S_4^\asf\,;  \qquad  S_2^\asf : = T_2^\asf\,, \qquad  S_4^\asf : = 2 T_2^\asf\,,
\ee
where, in \rf{manus-19042024-72ad}, we identify the generators $S_2^\asf$, $S_4^\asf$ which have not been fixed in \rf{manus-08042024-94}. In the basis \rf{manus-19042024-72ad}, the commutators \rf{manus-19042024-72} are represented as
{\small
\beq
\label{manus-19042024-73abc} && [T_0^\asf,T_0^\bsf] = f^{\asf\bsf\csf} T_0^\csf\,,\qquad [T_0^\asf,S_2^\bsf] = f^{\asf\bsf\csf} S_2^\csf\,,\qquad   [T_0^\asf,S_4^\bsf] = f^{\asf\bsf\csf} S_4^\csf\,,
\nonumber\\
&& [S_2^\asf,S_2^\bsf] = \half f^{\asf\bsf\csf} S_4^\csf\,, \qquad [S_{2m}^\asf,S_{2n}^\bsf] = 0\,, \qquad \hbox{ for } \ m+n > 2\,.
\eeq
}
\!From \rf{manus-19042024-73abc}, we see that the $S_2^\asf$, $S_4^\asf$ constitute a radical of the extended gauge algebra. In other words, we have the Levy-Maltsev decomposition,
\be
T_0^\asf\,, \ S_2^\asf\,, \  S_4^\asf \,\, =\,\,  S_2^\asf\,, S_4^\asf \,\, \subsetext\,\, T_0^\asf\,.
\ee

To represent the gauge transformations of the Stueckelberg fields and field strengths \rf{manus-19042024-21}, \rf{manus-19042024-30} in the framework of the extended gauge algebra we use the general relations in \rf{manus-08042024-95} when $d=7$,
{\small
\beq
\label{manus-19042024-75} && \phi := \phi_2^\asf  S_2^\asf + \phi_4^\asf S_4^\asf\,,\hspace{1.8cm} F^A := F_3^{A\asf}  S_2^\asf + F_5^{A\asf} S_4^\asf\,,
\nonumber\\
&& \phi_\subsm^A := \phi_3^{A\asf} S_2^\asf + \phi_5^{A\asf} S_4^\asf\,, \hspace{1cm} \xi_\subsm :=  \xi_2^\asf S_2^\asf +  \xi_4^\asf S_4^\asf\,.
\eeq
}
\!Using \rf{manus-19042024-71}, \rf{manus-19042024-75}, we verify that the expressions for $F_3^A$, $F_5^A$ given in \rf{manus-19042024-10} and the expressions for gauge transformations of $\phi_2$, $\phi_4$, $F_3^A$, $F_5^A$ given in \rf{manus-19042024-21}, \rf{manus-19042024-30} amount to the expressions given in \rf{manus-08042024-96}.
To this end the following commutator turns out to be helpful
{\small
\be
[S_2^\asf,T_2^\bsf] = \half f^{\asf\bsf\csf} S_4^\csf\,.
\ee
}
\ \ Finally we note that the Lagrangian given in \rf{manus-19042024-06} can be represented as in \rf{manus-08042024-97}, \rf{manus-08042024-100}, where all non-zero values of the invariant bilinear forms are given by
{\small
\beq
\label{manus-19042024-76} &&  G_{04} = 1\,,\quad  \quad G_{22} = 1\,,\quad
G_{02} = - 10\rho \,, \quad  \quad G_{00} = 24\rho^2\,,
\nonumber\\
&& G_{24}^\subsm = 1\,, \qquad G_{22}^\subsm = - 10\rho\,.
\eeq
}
\!For $d=7$, the conjectured expressions in \rf{manus-08042024-100-a1} are in agreement with \rf{manus-19042024-76}.

\noinbf{Higher-derivative Lagrangian of conformal YM field in $(A)dS_8$}. Higher-derivative Lagrangian in eight dimensions is obtained by using the same procedure as the one we used in six dimensions. Namely, we gauge away the Stueckelberg fields $\phi_2$, $\phi_4$ and plug a solution to equations of motion for the auxiliary fields $\phi_3^A$ and $\phi_5^A$ in ordinary-derivative Lagrangian  \rf{manus-19042024-06}. Doing so and using conventions in \rf{manus-08042024-85}, we find the following higher-derivative Lagrangian:
\be \label{manus-16092924-01}
\LL_\CYMsm^\Hdr = -\KK\, \Tr L_\CYMsm^\Hdr\,, \qquad L_\CYMsm^\Hdr: =  - I_8^\Hdr - 10 \rho I_6^\Hdr - 24 \rho^2 I_4\,,
\ee
where we use the notation
{\small
\beq
\label{manus-16092924-02} && I_8^\Hdr := \frac{1}{4} \Fb_4^{AB} \Fb_4^{AB} - \phib_3^A F_2^{AB} \phib_3^B\,, \qquad
I_6^\Hdr =:  \half \phib_3^A \phib_3^A\,,\qquad  I_4 = \frac{1}{4} F_2^{AB} F_2^{AB}\,,
\nonumber\\
&& \Fb_4^{AB} : = \DD^A  \phib_3^B - \DD^B  \phib_3^A\,, \qquad \phib_3^A :=  \DD^B F_2^{BA} \,,
\eeq
}
\!and $\phib_3^A$ stands for the solution to the auxiliary field $\phi_3^A$.
The $I_8^\Hdr$ and minus $I_6^\Hdr$ \rf{manus-16092924-02} are equal to the respective quantities $I_8$ and $I_6$ \rf{manus-19042024-06} evaluated on the solutions of equations for the auxiliary fields. Field strength $F_2^{AB}$ \rf{manus-19042024-10} is entirely expressed in terms of the primary field. Therefore higher-derivative Lagrangian \rf{manus-16092924-01} is also entirely expressed in terms of the primary field $\phi_1^A$.

In the flat limit, $I_6^\Hdr$ and $I_4$ are similar to the known Lagrangians of conformal YM field in $R^{5,1}$ and $R^{3,1}$ respectively, while $I_8^\Hdr$ is similar to Lagrangian of conformal YM in $R^{7,1}$ in Ref.\cite{Metsaev:2023qif}. In other words, the higher-derivative Lagrangian of conformal YM field in $(A)dS_8$ is obtained from the sum of Lagrangians of conformal YM field in $R^{7,1}$, $R^{5,1}$, and $R^{3,1}$ uplifted to $(A)dS_8$.

\newsection{\large Conformal Yang-Mills field in $(A)dS_8$ space. Decoupled formulation} \label{eight-fac}

\vspace{-0.1cm}

{\bf Field content}. In the framework of gauge invariant decoupled
formulation, the conformal YM field in $(A)dS_8$ space is described by three  vector fields and two scalar fields:
\vspace{-0.1cm}
\beq
\label{manus-01052024-01} && \varphi_1^A \qquad \varphi_3^A\qquad \varphi_5^A
\nonumber\\
&&
\nonumber\\[-35pt]
&& \hspace{0.7cm} \varphi_2 \qquad \varphi_4
\eeq
The vector fields are $y$-transversal \rf{manus-08042024-25}. All fields in \rf{manus-01052024-01} are decomposed in $t^a$ as in \rf{manus-08042024-86add}.
Realization of $(A)dS_8$ algebra \rf{manus-08042024-10} on the vector fields and the scalar fields takes the same form as in \rf{manus-08042024-16}.  Below, we show that the vector field $\varphi_1^A$ describes massless field with a mass parameter $m_{\varphi_1}^2 = 0$, while the vector fields $\varphi_3^A$, $\varphi_5^A$ and scalar fields $\varphi_2$, $\varphi_4$ describe massive vector fields with mass squares $m_{\varphi_3}^2$, $m_{\varphi_5}^2$,
\be \label{manus-01052024-02}
m_{\varphi_1}^2 =0\,, \qquad m_{\varphi_3}^2 =  4\rho\,, \qquad m_{\varphi_5}^2 = 6\rho\,,
\ee
where $\rho$ is given in \rf{manus-08042024-01}. Namely the vector field $\varphi_3^A$ and the scalar field $\varphi_2$ provide a gauge invariant description of massive field with the mass square $m_{\varphi_3}^2$, while the vector field $\varphi_5^A$ and the scalar field $\varphi_4$ provide the gauge invariant description of massive field with the mass square $m_{\varphi_5}^2$. The scalar fields $\varphi_2$ and $\varphi_4$ play a role of the Stueckelberg fields.

\noinbf{Gauge invariant Lagrangian}. In the decoupled formulation, the ordinary-derivative Lagrangian of the conformal YM field in $(A)dS_8$ we find takes the form
\be \label{manus-01052024-03}
k_\rho^{-1} \LL_\CYMsm =  - \frac{1}{4} \FF_2 \FF_2  + \frac{1}{4} \FF_4 \FF_4 - \frac{1}{4} \FF_6 \FF_6  + 2\rho \FF_3 \FF_3 - 3 \rho \FF_5 \FF_5\,,\qquad k_\rho : = 24 \rho^2\,,
\ee
where we use the shortcuts for the scalar products  defined in \rf{manus-08042024-88add}, while expressions for field strengths are given below. It is instructive to represent the Lagrangian \rf{manus-01052024-03} as
\beq
\label{manus-01052024-03add} && \hspace{-0.7cm} k_\rho^{-1} \LL_\CYMsm =  - \II_4  + \II_8 - \II_{12}\,,  \qquad
\nonumber\\
&& \II_4 := \frac{1}{4} \FF_2 \FF_2\,, \hspace{1cm} \II_8 : = \frac{1}{4} \FF_4 \FF_4 + \frac{m_{\varphi_3}^2}{2} \FF_3\FF_3\,,\qquad \II_{12} : = \frac{1}{4} \FF_6 \FF_6 + \frac{m_{\varphi_5}^2}{2} \FF_5\FF_5\,,\qquad
\eeq
where the mass square are given in \rf{manus-01052024-02} and we note that the $\II_8$, $\II_{12}$ resemble expressions for Lagrangians of massive fields. The field strengths entering Lagrangian \rf{manus-01052024-03} are given by
\beq
\label{manus-01052024-04} && \FF_2^{AB} := (\ppp^A - \rho y^A)\varphi_1^B - (\ppp^B - \rho y^B)\varphi_1^A +  [\varphi_1^A,\varphi_1^B] +  \FF_{2,\,\addsm}^{AB}\,,
\nonumber\\
&& \FF_4^{AB}: = \DD^A \varphi_3^B -  \DD^B \varphi_3^A +  \FF_{4,\,\addsm}^{AB}\,,
\nonumber\\
&& \FF_6^{AB}: = \DD^A  \varphi_5^B -  \DD^B  \varphi_5^A +  \FF_{6,\,\addsm}^{AB} \,,
\nonumber\\
&& \FF_3^A := \varphi_3^A + \DD^A\varphi_2  + \FF_{3,\,\addsm}^A \,,
\nonumber\\
&& \FF_5^A := \varphi_5^A + \DD^A\varphi_4 +  \FF_{5,\,\addsm}^A \,,
\eeq
where the covariant derivative $\DD^A$ is given in \rf{manus-14092024-01add} when $d=7$,
while the expressions for $\FF_{2n+2,\addsm}^{AB}$ and $\FF_{2n+1,\addsm}^A$ appearing in \rf{manus-01052024-04} are given by
\beq
\label{manus-01052024-06} &&   \FF_{2,\, \addsm}^{AB}  =  -   [\varphi_3^A,\varphi_3^B] +   [\varphi_5^A,\varphi_5^B]\,,
\nonumber\\
&&    \FF_{4,\,\addsm}^{AB}  = - c_8 [\varphi_3^A,\varphi_5^B] - c_8  [\varphi_5^A,\varphi_3^B]  + c_{27} [\varphi_5^A,\varphi_5^B]\,,
\nonumber\\
&&  \FF_{6,\,\addsm}^{AB}  = c_8 [\varphi_3^A,\varphi_3^B] - c_{27}   [\varphi_3^A,\varphi_5^B] -  c_{27}   [\varphi_5^A,\varphi_3^B]  + c_{50}  [\varphi_5^A,\varphi_5^B]\,, \qquad
\nonumber\\
&&   \FF_{3,\,\addsm}^A =   - c_{\frac{9}{2}} [\varphi_3^A,\varphi_4] - c_\half [\varphi_5^A,\varphi_2]   + c_{\frac{27}{4}} [\varphi_5^A,\varphi_4]\,,
\nonumber\\
&&   \FF_{5,\,\addsm}^A =   c_2 [\varphi_3^A,\varphi_2] - c_{12} [\varphi_3^A,\varphi_4] -  c_3 [\varphi_5^A,\varphi_2]  + c_{ \frac{25}{2} } [\varphi_5^A,\varphi_4]\,,
\nonumber\\
&& \hspace{1cm} c_q : = \sqrt{q}\,.
\eeq

Lagrangian \rf{manus-01052024-03} is diagonal with respect to all field strengths. This attractive representation for the Lagrangian \rf{manus-01052024-03} is achieved by finding a suitable linear transformation from the fields of the generic formulation \rf{manus-19042024-01} to the fields of decoupled formulation \rf{manus-01052024-01} (see Appendix C).

\noinbf{Gauge transformations}. In order to describe gauge symmetries of the Lagrangian given in \rf{manus-01052024-03} we introduce three gauge parameters,
\be \label{manus-01052024-07}
\eta_0\,, \qquad \eta_2\,,\qquad \eta_4\,,
\ee
which are decomposed in $t^a$ \rf{manus-08042024-87add}. We find the following gauge transformations of fields \rf{manus-01052024-01}:
\beq
\label{manus-01052024-08} && \delta_\eta\varphi_1^A = \DD^A \eta_0 + \delta_{\eta,\addsm} \varphi_1^A \,,
\nonumber\\
&& \delta_\eta\varphi_{2n+1}^A = \DD^A \eta_{2n} +  [\varphi_{2n+1}^A,\eta_0] + \delta_{\eta,\addsm} \varphi_{2n+1}^A \,,
\nonumber\\
&& \delta_\eta\varphi_{2n} \,\, =  \,\, - \,\,\eta_{2n}\,\, +\, [\varphi_{2n},\eta_0] \,\, + \,\, \delta_{\eta,\addsm} \varphi_{2n} \,, \qquad n=1,2\,,
\eeq
where the covariant derivative $\DD^A$ is given in \rf{manus-14092024-02add} when $d=7$,
while the $\delta_{\eta,\addsm}$-contributions entering \rf{manus-01052024-08} are given by
\beq
\label{manus-01052024-09} &&  \hspace{1cm}  \delta_{\eta,\addsm} \varphi_1^A =  -  [\varphi_3^A,\eta_2] +  [\varphi_5^A,\eta_4]\,,
\nonumber\\
&&  \hspace{1cm}  \delta_{\eta,\addsm} \varphi_3^A =  - c_8 [\varphi_5^A,\eta_2] - c_8 [\varphi_3^A,\eta_4]   + c_{27} [\varphi_5^A,\eta_4]  \,,
\nonumber\\
&&  \hspace{1cm}   \delta_{\eta,\addsm} \varphi_5^A =  c_8 [\varphi_3^A,\eta_2] - c_{27} [\varphi_5^A,\eta_2] - c_{27} [\varphi_3^A,\eta_4]   + c_{50} [\varphi_5^A,\eta_4]\,, \qquad
\nonumber\\
&& \hspace{1cm}  \delta_{\eta,\addsm} \varphi_2 =   - c_{\frac{9}{2}} [\varphi_4,\eta_2] - c_\half [\varphi_2,\eta_4]  + c_{\frac{27}{4}} [\varphi_4,\eta_4] \,,
\nonumber\\
&&  \hspace{1cm}  \delta_{\eta,\addsm} \varphi_4 =   c_2 [\varphi_2,\eta_2] - c_{12} [\varphi_4,\eta_2] - c_3 [\varphi_2,\eta_4]  + c_{\frac{25}{2}} [\varphi_4,\eta_4] \,.
\eeq
The $c_q$ is defined in \rf{manus-01052024-06}. Gauge transformations of the corresponding field strengths \rf{manus-01052024-04} we find can be presented as
\beq
\label{manus-01052024-10} && \delta_\eta \FF_{2n+2}^{AB} = [\FF_{2n+2}^{AB},\eta_0] + \delta_{\eta,\addsm} \FF_{2n+2}^{AB}\,, \hspace{1cm} n=0,1,2\,;
\nonumber\\
&& \delta_\eta \FF_{2n+1}^A = [\FF_{2n+1}^A,\eta_0] + \delta_{\eta,\addsm} \FF_{2n+1}^A\,, \hspace{1cm}  n=1,2\,,
\eeq
where the $\delta_{\eta,\addsm}$-contributions entering \rf{manus-01052024-10} are given by
{\small
\beq
&& \hspace{-0.8cm} \delta_{\eta,\addsm} \FF_2^{AB} = -  [\FF_4^{AB},\eta_2] +  [\FF_6^{AB},\eta_4]\,,
\nonumber\\
&&\hspace{-0.8cm} \delta_{\eta,\addsm} \FF_4^{AB} = [\FF_2^{AB},\eta_2] - c_8 [\FF_6^{AB},\eta_2] - c_8 [\FF_4^{AB},\eta_4]   + c_{27} [\FF_6^{AB},\eta_4]  \,,
\nonumber\\
&&\hspace{-0.8cm} \delta_{\eta,\addsm} \FF_6^{AB} =   [\FF_2^{AB},\eta_4] + c_8 [\FF_4^{AB},\eta_2] - c_{27} [\FF_6^{AB},\eta_2]     - c_{27} [\FF_4^{AB},\eta_4]   + c_{50} [\FF_6^{AB},\eta_4]  \,,\qquad
\nonumber\\
&& \hspace{-0.8cm} \delta_{\eta,\addsm} \FF_3^A \ \ = \ \ - c_{\frac{9}{2}} [\FF_5^A,\eta_2] - c_\half  [\FF_3^A,\eta_4]  + c_{\frac{27}{4}} [\FF_5^A,\eta_4] \,,
\nonumber\\
&& \hspace{-0.8cm} \delta_{\eta,\addsm} \FF_5^A \ \ = \ \   c_2 [\FF_3^A,\eta_2] - c_{12} [\FF_5^A,\eta_2] - c_3 [\FF_3^A,\eta_4]  + c_{\frac{25}{2}} [\FF_5^A,\eta_4] \,.
\eeq
}
\!From \rf{manus-01052024-08} we learn  that the vector fields are realized as gauge fields, while the scalar fields are realized as Stueckelberg fields.

\noinbf{$K^A$-transformations of fields and field strengths}. We find the following $K^A$-transformations of fields \rf{manus-01052024-01} :
{\small
\beq
\label{manus-01052024-11} && \delta_{K^A} \varphi_1^B = c_{\frac{1}{3}} K_5^A \varphi_3^B - 4c_{\frac{1}{3}} \theta^{AB}\varphi_2\,,
\nonumber\\
&& \delta_{K^A} \varphi_3^B = 4c_{\frac{1}{6}}  K_4^A \varphi_5^B  -  c_{\frac{1}{3}} K_1^A \varphi_1^B - 2c_6\, \theta^{AB}\varphi_4 \,,
\nonumber\\
&& \delta_{K^A} \varphi_5^B = 3 K_3^A \varphi_5^B - 4c_{\frac{1}{6}} K_2^A \varphi_3^B - 6 \theta^{AB}\varphi_4 +  4c_{\frac{1}{6}} \theta^{AB} \varphi_2\,,
\nonumber\\
&& \delta_{K^A} \varphi_2 =  \half c_6 K_4^A \varphi_4 + \frac{1}{\rho} c_{\frac{1}{6}} \varphi_5^A - \frac{1}{\rho} c_{\frac{1}{3}}\varphi_1^A\,,
\nonumber\\
&& \delta_{K^A} \varphi_4 = 2 K_3^A \varphi_4 -  2c_{\frac{1}{6}}   K_2^A\varphi_2 + \frac{1}{\rho} \varphi_5^A  -    \frac{2}{\rho}c_{\frac{1}{6}}  \varphi_3^A\,,
\eeq
}
\!while the $K^A$-transformations of the field strengths \rf{manus-01052024-04} take the form
{\small
\beq
\label{manus-01052024-12} && \delta_{K^A} \FF_2^{BC} =  c_{\frac{1}{3}} K_6^A \FF_4^{BC} + 4c_{\frac{1}{3}} \big(\theta^{AB} \FF_3^C - \theta^{AC} \FF_3^B\big)\,,
\nonumber\\
&& \delta_{K^A} \FF_4^{BC} = 4 c_{\frac{1}{6}} K_5^A \FF_6^{BC} - c_{\frac{1}{3}} K_2^A \FF_2^{BC} +  2c_6\big(\theta^{AB} \FF_5^C -  \theta^{AC} \FF_5^B\big)\,,
\nonumber\\
&& \delta_{K^A} \FF_6^{BC} = 3 K_4^A \FF_6^{BC} - 4c_{\frac{1}{6}} K_3^A \FF_4^{BC} + 6 \theta^{AB} \FF_5^C - 6 \theta^{AC} \FF_5^B   - 4c_{\frac{1}{6}} \big( \theta^{AB} \FF_3^C -\theta^{AC} \FF_3^B \big)\,, \qquad
\nonumber\\
&& \delta_{K^A} \FF_3^B =  \half c_6 K_5^A \FF_5^B - \frac{1}{\rho} c_{\frac{1}{6}} \FF_6^{AB} + \frac{1}{\rho}c_{\frac{1}{3}} \FF_2^{AB}\,,
\nonumber\\
&& \delta_{K^A} \FF_5^B = 2 K_4^A \FF_5^B - 2c_{\frac{1}{6}} K_3^A \FF_3^B - \frac{1}{\rho} \FF_6^{AB} +  \frac{2}{\rho} c_{\frac{1}{6}}  \FF_4^{AB}\,,
\eeq
}
\!\!where $c_q$ and $\theta^{AB}$ are defined in \rf{manus-01052024-06} and \rf{12062024-01} respectively. Action of operator $K_\Delta^A$ on the fields and fields strengths  \rf{manus-01052024-11}, \rf{manus-01052024-12} is defined in the same way as in \rf{manus-08042024-75}, \rf{manus-08042024-76}.

The following remarks are in order

\noinbf{i}) To quadratic order in the fields, Lagrangian \rf{manus-01052024-03} takes the form
\be \label{manus-01052024-12-a1}
k_\rho^{-1} \LL_\CYMsm^\smtwo =  - \II_4^\smtwo  +  \II_8^\smtwo -  \II_{12}^\smtwo \,,
\ee
where $\II_4^\smtwo$, $\II_8^\smtwo$, and $\II_{12}^\smtwo$ are obtained by setting $d=7$ in \rf{manus-08042024-106-b3}. From \rf{manus-01052024-12-a1}, we see that the free Lagrangian describes three decoupled sets of fields. The first set is given by the single field $\varphi_1^A$ which describes massless field, the
second set is given by the fields $\varphi_3^A$, $\varphi_2$ which describe massive field with the mass square $m_{\varphi_3}^2$, and the third
set is given by the fields $\varphi_5^A$, $\varphi_4$ which describes massive field with the mass square $m_{\varphi_5}^2$. At the level of the interaction vertices entering Lagrangian \rf{manus-01052024-03}, the three sets of fields are not decoupled.

\noinbf{ii}) For $(A)dS_8$, we note $k_\rho>0$ for both $AdS_8$ and $dS_8$. This implies that, for both $AdS_8$ and $dS_8$, the massless field and the massive field with the mass square $m_{\varphi_5}^2$ enter the Lagrangian with the correct sign of the kinetic terms,
while the massive field with the mass square $m_{\varphi_3}^2$ enter the Lagrangian with the wrong sign of the kinetic term. Note also that, for $AdS_8$, the mass squares $m_{\varphi_3}^2$ and $m_{\varphi_5}^2$ are negative, while, for $dS_8$, they are  positive.

\noinbf{Extended gauge algebra}.  For the case under consideration, the extended gauge
algebra consists of three generators denoted as
\be \label{manus-01052024-15}
\TT_0^\asf\,, \quad \TT_2^\asf\,, \quad \TT_4^\asf\,.
\ee
Considering the commutators of gauge transformations \rf{manus-01052024-08} or \rf{manus-01052024-10}, we get the following commutators for generators \rf{manus-01052024-15}:
\beq
\label{manus-01052024-16} &&  [\TT_0^a,\TT_0^b] = f^{\asf\bsf\csf} \TT_0^\csf  \,, \qquad [\TT_0^a,\TT_2^b] = f^{\asf\bsf\csf} \TT_2^\csf  \,,\qquad [\TT_0^a,\TT_4^b] = f^{\asf\bsf\csf} \TT_4^\csf  \,,\qquad
\nonumber\\
&&  [\TT_2^a,\TT_2^b] = -  f^{\asf\bsf\csf} \TT_0^\csf +  c_8 f^{\asf\bsf\csf} \TT_4^\csf \,,
\nonumber\\
&&  [\TT_2^a,\TT_4^b] = - c_8 f^{\asf\bsf\csf} \TT_2^\csf -  c_{27}  f^{\asf\bsf\csf} \TT_4^\csf \,,
\nonumber\\
&&  [\TT_4^a,\TT_4^b] =  f^{\asf\bsf\csf} \TT_0^\csf +  c_{27} f^{\asf\bsf\csf} \TT_2^\csf + c_{50} f^{\asf\bsf\csf} \TT_4^\csf \,,
\eeq
where $c_q$ is defined in \rf{manus-01052024-06}. Using commutators \rf{manus-01052024-16}, we can apply the general setup of extended gauge algebra described in Sec.\ref{not-conv} to the case under consideration. Namely, setting $d=7$ in \rf{manus-08042024-91ab}, we introduce
\beq
\label{manus-01052024-17} && \varphi^A : = \varphi_1^{A\asf} \TT_0^\asf + \varphi_3^{A\asf} \TT_2^\asf + \varphi_5^{A\asf} \TT_4^\asf  \,,   \qquad \FF^{AB} : = \FF_2^{AB\asf} \TT_0^\asf + \FF_4^{AB\asf} \TT_2^\asf + \FF_6^{AB\asf} \TT_4^\asf  \,,\qquad
\nonumber\\
&& \hspace{3.5cm} \eta : =  \eta_0^\asf \TT_0^\asf + \eta_2^\asf \TT_2^\asf + \eta_4^\asf \TT_4^\asf\,.
\eeq
Now, using $\varphi^A$, $\FF^{AB}$, and $\eta$ above-defined, we note that the expressions for $\FF_{2n+2}^{AB}$, $n=0,1,2$, given in \rf{manus-01052024-04} and the gauge transformations for $\varphi_{2n+1}^A$ and $\FF_{2n+2}^{AB}$, $n=0,1,2$ given in \rf{manus-01052024-08}, \rf{manus-01052024-10} can be represented in a more compact way as in \rf{manus-08042024-120}.

Following the general pattern in Sec.\ref{not-conv}, we then represent the set of the generators \rf{manus-01052024-15} as
\beq
\label{manus-01052024-17ad} && \hspace{-1.2cm}\TT_0^\asf\,, \quad \Scl_2^\asf\,, \quad \Scl_4^\asf\,,
\nonumber\\
&& \Scl_2^\asf : =  c_{\frac{1}{3}} \TT_0^\asf -   \TT_2^\asf - c_{\frac{8}{3}}  \TT_4^\asf\,,  \qquad \Scl_4^\asf :=   c_\half \TT_0^\asf + c_{\frac{27}{2}} \TT_2^\asf + 4\TT_4^\asf\,,
\eeq
where, in \rf{manus-01052024-17ad}, we identify the generators $\Scl_2^\asf$, $\Scl_4^\asf$ which have not been fixed in \rf{manus-08042024-94ab}. In the basis \rf{manus-01052024-17ad}, the commutators \rf{manus-01052024-16} are represented as
\beq
\label{manus-01052024-27} &&  [\TT_0^a,\TT_0^b] = f^{\asf\bsf\csf} \TT_0^\csf  \,, \hspace{1.2cm} [\TT_0^a,\Scl_2^b] = f^{\asf\bsf\csf} \Scl_2^\csf\,, \hspace{2cm}    [\TT_0^a,\Scl_4^b] = f^{\asf\bsf\csf} \Scl_4^\csf\,,
\nonumber\\
&& [\Scl_2^\asf,\Scl_2^\bsf] = f^{\asf\bsf\csf} \Scl_\new^\csf\,, \hspace{1cm} [\Scl_2^\asf,\Scl_4^\bsf] = - c_{\frac{3}{2}} f^{\asf\bsf\csf} \Scl_\new^\csf\,, \hspace{1cm}  [\Scl_4^\asf,\Scl_4^\bsf] =  c_{\frac{9}{4}} f^{\asf\bsf\csf} \Scl_\new^\csf\,, \qquad
\nonumber\\
&& \hspace{4cm} \Scl_\new^\asf : = c_3 \Scl_2^\asf + c_2 \Scl_4^\asf\,.
\eeq
Using commutators \rf{manus-01052024-27}, we find $[\Scl_\new^\asf,\Scl_\new^\bsf]=0$. This implies that the generators $\Scl_2^\asf$, $\Scl_4^\asf$ form a radical of the extended gauge algebra. In other words, we have the Levy-Maltsev decomposition,
\be
\TT_0^\asf\,, \ \Scl_2^\asf\,, \ \Scl_4^\asf\,\, =\,\, \Scl_2^\asf\,, \ \Scl_4^\asf \,\, \subsetext\,\, \TT_0^\asf\,.
\ee

In order to represent gauge transformation of the Stueckelberg fields $\varphi_2$, $\varphi_4$ and the corresponding field strengths $\FF_3^A$,  $\FF_5^A$ in the framework of the extended gauge algebra we use the general relations \rf{manus-08042024-95ab} to introduce
\beq
\label{manus-01052024-25} &&  \varphi = \varphi_2^\asf \Scl_2^\asf + \varphi_4^\asf \Scl_4^\asf\,, \hspace{1.6cm} \FF^A = \FF_3^{A\asf}  \Scl_2^\asf + \FF_5^{A\asf} \Scl_4^\asf\,,
\nonumber\\
&& \varphi_\subsm^A = \varphi_3^{A\asf} \Scl_2^\asf + \varphi_5^{A\asf} \Scl_4^\asf\,, \qquad \eta_\subsm =  \eta_2^\asf \Scl_2^\asf +  \eta_4^\asf \Scl_4^\asf\,.\qquad
\eeq

Using the definitions given in \rf{manus-01052024-17}, \rf{manus-01052024-25}, we find that the expressions for $\FF_3^A$, $\FF_5^A$ in \rf{manus-01052024-04} and the gauge transformations of $\varphi_2$, $\varphi_4$ and $\FF_3^A$, $\FF_5^A$ given in \rf{manus-01052024-08}, \rf{manus-01052024-10} can be represented in a more short way as in \rf{manus-08042024-126}. To this end the following commutators turn out to be helpful:
\beq
&&   [\Scl_2^a,\TT_2^b] = c_2 f^{\asf\bsf\csf} \Scl_4^\csf\,, \hspace{3.6cm}  [\Scl_2^a,\TT_4^b] = - c_\half  f^{\asf\bsf\csf} \Scl_2^\csf -  c_3 f^{\asf\bsf\csf} \Scl_4^\csf\,,
\nonumber\\
&&   [\Scl_4^a,\TT_2^b] = - c_{\frac{9}{2}} f^{\asf\bsf\csf} \Scl_2^\csf -  c_{12} f^{\asf\bsf\csf} \Scl_4^\csf\,, \hspace{1cm}  [\Scl_4^a,\TT_4^b] =  c_{\frac{27}{4}} f^{\asf\bsf\csf} \Scl_2^\csf +  c_{\frac{25}{2}} f^{\asf\bsf\csf} \Scl_4^\csf\,.\qquad
\eeq

Finally, we note that Lagrangian \rf{manus-01052024-03} can be represented as in \rf{manus-08042024-117}, \rf{manus-08042024-100add}, where all non-zero values of the invariant bilinear forms are given by
\be \label{manus-01052024-27-xxx}
\GG_{00} = k_\rho\,,\qquad \GG_{22} = - k_\rho \,,\qquad \GG_{44} = k_\rho \,,\qquad  \GG_{22}^\subsm = - 4\rho k_\rho\,, \qquad \GG_{44}^\subsm = 6\rho k_\rho\,,
\ee
where $k_\rho$ is given in \rf{manus-01052024-03}. For $d=7$, the conjectured expressions in \rf{manus-08042024-106} agree with \rf{manus-01052024-27-xxx}.

\newsection{\large Conformal Yang-Mills field in $(A)dS_{10}$ space. Generic formulation } \label{ten-gen}

{\bf Field content}. In the framework of generic formulation, the conformal YM field in $(A)dS_{10}$ space is described by four vector fields and three scalar fields:
\beq
\label{manus-02052024-01}   && \phi_1^A \qquad  \phi_3^A \qquad  \phi_5^A \qquad \phi_7^a
\nonumber\\
&&
\\[-35pt]
&& \hspace{0.8cm} \phi_2 \qquad   \phi_4\qquad   \phi_6
\nonumber
\eeq
Transformation rules of the vector fields and the scalar fields under $(A)dS_{10}$ algebra \rf{manus-08042024-10} take the form presented in \rf{manus-08042024-16}.  The vector fields are subject to $y$-transversality constraint \rf{manus-08042024-25}. Conformal dimensions of the fields in \rf{manus-02052024-01} are given by
\be \label{manus-02052024-02}
\Delta(\phi_{2n+1}^a) = 2n+1\,, \quad n=0,1,2,3\,; \hspace{1.5cm}  \Delta(\phi_{2n}) = 2n\,, \quad  n=1,2,3\,.
\ee
The vector field $\phi_1^A$ has the same conformal dimension as the primary conformal YM field entering the commonly used higher-derivative approach. Therefore the vector field $\phi_1^A$ can be identified with the primary conformal YM field. The remaining three vector fields $\phi_3^A$, $\phi_5^A$, and $\phi_7^A$ are realized as auxiliary fields, while the three scalar fields $\phi_2$, $\phi_4$, and  $\phi_6$ are realized as Stueckelberg fields.

\noinbf{Gauge invariant Lagrangian}. In the generic formulation, the ordinary-derivative  Lagrangian of the conformal YM field in $(A)dS_{10}$ we find takes the form
\beq
\label{manus-02052024-03}  &&  \hspace{-1cm} \LL_\CYMsm = - I_{10} + 28\rho I_8 -  252 \rho^2 I_6 + 720\rho^3 I_4\,,
\nonumber\\
&& I_{10} := \half F_2  F_8  + \half F_4  F_6  + \half F_5  F_5  + F_3  F_7 \,, \qquad  I_8 :=  \half F_2 F_6 + \frac{1}{4} F_4 F_4 + F_3 F_5\,,\qquad
\nonumber\\
&& I_6 :=  \half  F_2 F_4 + \half F_3 F_3\,, \hspace{4cm} I_4 :=  \frac{1}{4} F_2 F_2\,,
\eeq
where we use the shortcuts for the scalar products defined in \rf{manus-08042024-87sh} and  the field strengths are expressed in terms of fields \rf{manus-02052024-01} as
{\small
\beq
\label{manus-02052024-03add} && F_2^{AB} = (\ppp^A - \rho y^A) \phi_1^B - (\ppp^B - \rho  y^B) \phi_1^A +  [\phi_1^A,\phi_1^B]\,,
\nonumber\\
&& F_4^{AB} = \DD^A \phi_3^B - \DD^B  \phi_3^A\,,
\nonumber\\
&& F_6^{AB}  = \DD^A \phi_5^B - \DD^B  \phi_5^A +  \frac{4}{3}[\phi_3^A,\phi_3^B]\,,
\nonumber\\
&& F_8^{AB}  = \DD^A  \phi_7^B - \DD^B  \phi_7^A +  [\phi_3^A,\phi_5^B] + [\phi_5^A,\phi_3^B] + \frac{28}{3} \rho [\phi_3^A,\phi_3^B]\,, \qquad
\nonumber\\
&& F_3^A = \phi_3^A + \DD^A\phi_2\,,
\nonumber\\
&& F_5^A = \phi_5^A + \DD^A\phi_4  + \frac{2}{3} [\phi_3^A,\phi_2]\,,
\nonumber\\
&& F_7^A = \phi_7^A + \DD^A\phi_6  +  \frac{2}{3}  [\phi_3^A,\phi_4] + \frac{1}{3} [\phi_5^A,\phi_2] + \frac{14}{3} \rho [\phi_3^A,\phi_2]\,.
\eeq
}
The covariant derivative $\DD^A$ acting on the auxiliary and Stueckelberg fields is given in \rf{manus-14092024-01}  when $d=9$. Expression for $F_2^{AB}$ in \rf{manus-02052024-03add} tells us that the $F_2^{AB}$ is a field strength for the conformal YM field $\phi_1^A$. The Lagrangian is built entirely in terms of the quantities $I_{2n+4}$, $n=0,1,2,3$. The  $I_{2n+4}$ are expressed in terms of the field strengths in the same way as in the flat space (see Ref.\cite{Metsaev:2023qif}).

\noinbf{Gauge transformations}. In gauge invariant approach, each vector field \rf{manus-02052024-01} is accompanied by gauge parameter. This implies that we use four gauge parameters,
\be \label{manus-02052024-05}
\xi_0\,, \quad \xi_2\,, \quad \xi_4\,, \quad \xi_6\,,
\ee
which are decomposed in  $t^\asf$ as in \rf{manus-08042024-87}. Gauge transformations of fields we find take the form
{\small
\beq
\label{manus-02052024-06}  && \delta_\xi \phi_1^B = \DD^B\xi_0\,,
\nonumber\\
&& \delta_\xi \phi_3^B = \DD^B\xi_2 + [\phi_3^B,\xi_0]\,,
\nonumber\\
&& \delta_\xi \phi_5^B = \DD^B\xi_4 + [\phi_5^B,\xi_0] + \frac{4}{3} [\phi_3^B,\xi_2]\,,
\nonumber\\
&& \delta_\xi \phi_7^B = \DD^B\xi_6 + [\phi_7^B,\xi_0] +  [\phi_5^B,\xi_2]+ [\phi_3^B,\xi_4]  + \frac{28}{3} \rho [\phi_3^B,\xi_2]\,,
\nonumber\\
&& \delta_\xi \phi_2 = - \xi_2 + [\phi_2,\xi_0]\,,
\nonumber\\
&& \delta_\xi \phi_4 = - \xi_4 + [\phi_4,\xi_0] + \frac{2}{3} [\phi_2,\xi_2]\,,
\nonumber\\
&& \delta_\xi \phi_6 = - \xi_6 + [\phi_6,\xi_0] + \frac{2}{3} [\phi_4,\xi_2] + \frac{1}{3} [\phi_2,\xi_4] + \frac{14}{3} \rho [\phi_2,\xi_2]\,,
\eeq
}
where the covariant derivative $\DD^A$ is given in \rf{manus-14092024-02} when $d=9$.
Gauge transformations of fields \rf{manus-02052024-06} are accompanied by the following gauge transformations of the corresponding field strengths:
{\small
\beq
\label{manus-02052024-08} && \delta_\xi F_2^{AB} = [F_2^{AB},\xi_0]\,,
\nonumber\\
&& \delta_\xi F_4^{AB} = [F_4^{AB},\xi_0] + [F_2^{AB},\xi_2]\,,
\nonumber\\
&& \delta_\xi F_6^{AB} = [F_6^{AB},\xi_0] + \frac{4}{3} [F_4^{AB},\xi_2]+ [F_2^{AB},\xi_4]\,,
\nonumber\\
&& \delta_\xi F_8^{AB} = [F_8^{AB},\xi_0] +  [F_6^{AB},\xi_2] +   [F_4^{AB},\xi_4] + [F_2^{AB},\xi_6]  + \frac{28}{3} \rho [F_4^{AB},\xi_2]\,,
\nonumber\\
&& \delta_\xi F_3^A = [F_3^A,\xi_0]\,,
\nonumber\\
&& \delta_\xi F_5^A = [F_5^A,\xi_0] + \frac{2}{3}[F_3^A,\xi_2]\,,
\nonumber\\
&& \delta_\xi F_7^A = [F_7^A,\xi_0] + \frac{2}{3} [F_5^A,\xi_2]+ \frac{1}{3} [F_3^A,\xi_4] + \frac{14}{3} \rho [F_3^A,\xi_2]\,.
\eeq
}
\!\!Gauge transformations \rf{manus-02052024-06} tell us that the vector fields are realized as gauge fields, while the scalar fields are realized as the Stueckelberg fields.

\noinbf{$K^A$-transformations of fields and field strengths}. We find the following $K^A$-transformations of fields \rf{manus-02052024-01}:
{\small
\beq
\label{manus-02052024-09} && \delta_{K^A} \phi_1^B = K_1^A \phi_1^B\,,
\nonumber\\
&& \delta_{K^A} \phi_3^B = K_3^A \phi_3^B - 2\theta^{AB} \phi_2 + 6 \rho K_1^A\phi_1^B \,,
\nonumber\\
&& \delta_{K^A} \phi_5^B = K_5^A \phi_5^B - 4\theta^{AB} \phi_4 + 8  \rho K_3^A\phi_3^B - 20  \rho  y^A \phi_3^B + 4  \rho \theta^{AB}\phi_2 + 36  \rho^2 K_1^A \phi_1^B \,,\qquad
\nonumber\\
&&  \delta_{K^A} \phi_7^B  =  K_7^A \phi_7^B - 6\theta^{AB} \phi_6 + 6  \rho  K_5^A \phi_5^B - 44 \rho  y^A \phi_5^B + 20 \rho  \theta^{AB}\phi_4
\nonumber\\
&& \hspace{1.3cm} +\,\, 92\rho^2 K_3^A\phi_3^B - 80 \rho^2 y^A \phi_3^B + 104\rho^3 \theta^{AB}\phi_2  + 216 \rho^3 K_1^A \phi_1^B \,,\qquad
\nonumber\\
&& \delta_{K^A} \phi_2 = K_2^A \phi_2 + 6 \phi_1^A\,,
\nonumber\\
&& \delta_{K^A} \phi_4 = K_4^A \phi_4 + 4 \phi_3^A + 4 \rho K_2^A \phi_2 - 12 \rho  y^A \phi_2 + 36  \rho \phi_1^A\,,
\nonumber\\
&& \delta_{K^A} \phi_6 = K_6^A \phi_6 + 2 \phi_5^A + 4 \rho  K_4^A \phi_4 - 36 \rho  y^A \phi_4 + 52 \rho  \phi_3^A
\nonumber\\
&& \hspace{1.2cm} +\,\,  40 \rho^2 K_2^A \phi_2 + 24 \rho^2 y^A \phi_2 + 216 \rho^2 \phi_1^A\,,
\eeq
}
\!while the $K^A$-transformations of the corresponding field strengths \rf{manus-02052024-03add} take the form
{\small
\beq
\label{manus-02052024-10} &&  \delta_{K^A} F_2^{BC} = K_2^A F_2^{BC}\,,\qquad
\nonumber\\
&& \delta_{K^A} F_4^{BC} = K_4^A F_4^{BC}  + 2\theta^{AB} F_3^C - 2 \theta^{AC} F_3^B + 6 \rho K_2^A F_2^{BC}\,,\qquad
\nonumber\\
&& \delta_{K^A} F_6^{BC} = K_6^A F_6^{BC} + 4\theta^{AB} F_5^C - 4 \theta^{AC} F_5^B + 8 \rho K_4^A F_4^{BC} \qquad
\nonumber\\
&& \hspace{1.6cm} - \,\, 20 \rho  y^A F_4^{BC}  - 4 \rho \big(\theta^{AB} F_3^C - \theta^{AC} F_3^B\big) + 36 \rho^2 K_2^A F_2^{BC}\,,\qquad
\nonumber\\
&& \delta_{K^A} F_8^{BC} = K_8^A F_8^{BC} + 6\theta^{AB} F_7^C - 6 \theta^{AC} F_7^B + 6  \rho  K_6^A F_6^{BC} - 44 \rho   y^A F_6^{BC}  \qquad
\nonumber\\
&& \hspace{1.6cm} - \,\, 20 \rho \big(\theta^{AB} F_5^C  - \theta^{AC} F_5^B) + 92 \rho^2 K_4^A F_4^{BC} - 80 \rho ^2 y^A F_4^{BC}
\nonumber\\
&& \hspace{1.6cm} +\,\, 104 \rho^2\big(\theta^{AB} F_3^C - \theta^{AC} F_3^B\big) + 216 \rho ^3 K_2^A F_2^{BC}\,,\qquad
\nonumber\\
&& \delta_{K^A} F_3^B  = K_3^A F_3^B   -6 F_2^{AB}\,,
\nonumber\\
&& \delta_{K^A} F_5^B  = K_5^A F_5^B  - 4 F_4^{AB} + 4  \rho  K_3^A F_3^B - 12 \rho  y^A F_3^B - 36 \rho   F_2^{AB}\,,
\nonumber\\
&& \delta_{K^A} F_7^B  = K_7^A F_7^B  - 2 F_6^{AB}  4 \rho  K_5^A F_5^B -  36 \rho  y^A F_5^B
\nonumber\\
&& \hspace{1.6cm} -\,\, 52 \rho  F_4^{AB} + 40 \rho^2 K_3^A F_3^B + 24 \rho^2 y^A F_3^B - 216 \rho^2  F_2^{AB}\,,
\eeq
}
\!\!where operator $K_\Delta^A$ appearing in \rf{manus-02052024-09} and \rf{manus-02052024-10} is defined in \rf{manus-08042024-75} and \rf{manus-08042024-76} respectively, while the symbol $\theta^{AB}$ is defined in \rf{12062024-01}.

The following remarks are in order.

\noinbf{i}) Two-derivative terms in Lagrangian \rf{manus-02052024-03} take the forms (up to normalization factors):
\beq
\label{manus-02052024-15} &&  \phi_{2m+1}^A (\eta^{AB} \ppp^2  - \ppp^A \ppp^B)\phi_{2n+1}^B\,, \hspace{0.3cm} m,n=0,1,2,3\,, \qquad n+m\leq 3;
\nonumber\\
&& \phi_{2m} \ppp^2 \phi_{2n}\,,   \hspace{3.8cm} m,n=1,2,3\,, \hspace{1.2cm} n+m\leq 4\,.
\eeq
From \rf{manus-02052024-15}, we see that, for the scalar and vector fields, the two-derivative terms are constructed out of the respective Klein-Gordon and Maxwell kinetic operators. With respect to the fields, some kinetic terms are diagonal, while the remaining ones are non-diagonal. Lagrangian of conformal YM field  in $(A)dS_{10}$ involving only diagonal kinetic and mass-like terms is presented in  Sec.\ref{ten-fac}.

\noinbf{ii})  As the quantities $I_{2n+4}$ defined in \rf{manus-02052024-03}, \rf{manus-02052024-03add} are regular in the flat space limit, Lagrangian \rf{manus-02052024-03} is also regular in the flat space limit. To clarify a notion of the $I_{2n+4}$ in the flat space limit, we use a shortcut $I_{2n+4}^{\rho=0}$ for the flat space limit of the $I_{2n+4}$ and note that the minus $I_{10}^{\rho=0}$ coincides with the ordinary-derivative Lagrangian of conformal YM field in $R^{9,1}$ found in Ref.\cite{Metsaev:2023qif}, while a reduction of the minus $I_6^{\rho=0}$ to $R^{5,1}$ coincides with the ordinary-derivative Lagrangian of conformal YM field in $R^{5,1}$  found in Ref.\cite{Metsaev:2023qif}. Reduction of the minus $I_4^{\rho=0}$ to $R^{3,1}$ coincides with the standard Lagrangian of conformal YM field in $R^{3,1}$. Note however that a reduction of $I_8^{\rho=0}$ \rf{manus-02052024-03}, \rf{manus-02052024-03add} to $R^{7,1}$ is not equal to $I_8^{\rho=0}$ defined in $R^{7,1}$ by the relations \rf{manus-19042024-06}, \rf{manus-19042024-10}. This is easily seen by comparing  \rf{manus-19042024-06}, \rf{manus-19042024-10} against \rf{manus-02052024-03}, \rf{manus-02052024-03add}. Namely, though the dependence of $I_8$ in \rf{manus-19042024-06} and $I_8$ in \rf{manus-02052024-03} on the field strengths is identical, the expressions for the field strengths $F_6^{AB}$ in \rf{manus-19042024-10} and \rf{manus-02052024-03add}  are different.

\noinbf{Extended gauge algebra}. As seen from \rf{manus-02052024-05}, gauge symmetries are governed by four gauge parameters $\xi_{2n}= \xi_{2n}^\asf t^\asf$, $n=0,1,2,3$. Therefore, the extended gauge algebra consists of four generators denoted as
\be \label{manus-02052024-24}
T_0^\asf\,, \qquad T_2^\asf\,, \qquad T_4^\asf\,, \qquad T_6^\asf\,.
\ee
The general setup of extended gauge algebra was discussed in Sec.\ref{not-conv}. In order to apply the general setup to the case under consideration we should present commutation relation for the generators in \rf{manus-02052024-24}. Using gauge transformations given in \rf{manus-02052024-06} or \rf{manus-02052024-08} and considering commutators of two gauge transformations, we find
that generators \rf{manus-02052024-24} should obey the commutators%
\footnote{ Commutator $[T_2^\asf,T_2^\bsf]$ in \rf{manus-02052024-26} differ from its flat space cousin by the $\rho f^{\asf\bsf\csf}T_6^\csf$-term. The appearance of parameter $\rho$ \rf{manus-08042024-01} in \rf{manus-02052024-26} is related to a basis of the fields we use. A change of a basis of the fields amounts to a change of a basis for generators of the extended gauge algebra. In Appendix B, we show that by a suitable change of the basis for the generators of the extended gauge algebra we can remove the $\rho f^{\asf\bsf\csf}T_6^\csf$-term from \rf{manus-02052024-26}.}
{\small
\beq
\label{manus-02052024-26}   && \hspace{-1cm} [T_0^\asf,T_0^\bsf] = f^{\asf\bsf\csf} T_0^\csf\,,\quad [T_0^\asf,T_2^\bsf] = f^{\asf\bsf\csf} T_2^\csf\,,\quad   [T_0^\asf,T_4^\bsf] = f^{\asf\bsf\csf} T_4^\csf\,,
\quad   [T_0^\asf,T_6^\bsf] = f^{\asf\bsf\csf} T_6^\csf\,,
\\
&& \hspace{-1cm} [T_2^\asf,T_2^\bsf] = \frac{4}{3} f^{\asf\bsf\csf} T_4^\csf + \frac{28}{3} \rho  f^{\asf\bsf\csf} T_6^\csf\,,\hspace{1cm} [T_2^\asf,T_4^\bsf] = f^{\asf\bsf\csf} T_6^\csf\,,\qquad [T_{2m}^\asf,T_{2n}^\bsf] = 0\,, \ \hbox{for} \ m+n > 3\,.
\nonumber
\eeq
}
\!From \rf{manus-02052024-26}, we see that $T_2^\asf$, $T_4^\asf$, $T_6^\asf$ constitute a radical of the extended gauge algebra \rf{manus-02052024-24}.

Following \rf{manus-08042024-91}, we introduce then
\be \label{manus-02052024-25}
\phi^A : = \sum_{n=0,1,2,3} \phi_{2n+1}^{A\asf} T_{2n}^\asf   \,,   \qquad F^{AB} : = \sum_{n=0,1,2,3} F_{2n+2}^{AB\asf} T_{2n}^\asf  \,, \qquad \xi := \sum_{n=0,1,2,3} \xi_{2n}^\asf T_{2n}^\asf\,. \qquad
\ee

Now, using $\phi^A$, $F^{AB}$, and $\xi$ above defined, we note that  the expressions for field strengths \rf{manus-02052024-03add}, and the gauge transformations for the fields and field strengths given in \rf{manus-02052024-06} and \rf{manus-02052024-08}  amount to expressions given in \rf{manus-08042024-92}.

Following the general pattern in Sec.\ref{not-conv}, we then represent the set of the generators \rf{manus-02052024-24} as
\be \label{manus-02052024-25ad}
T_0^\asf\,, \quad S_2^\asf\,, \quad S_4^\asf\,,  \quad S_6^\asf\,: \qquad S_2: =  T_2 + \frac{14}{3} \rho T_4\,,\hspace{1cm} S_4: = 2 T_4\,,\hspace{1cm} S_6 : = 3 T_6\,,
\ee
where, in \rf{manus-02052024-25ad}, we identify the generators $S_2^\asf$, $S_4^\asf$,  $S_6^\asf$ which have not been fixed in \rf{manus-08042024-94}. In the basis \rf{manus-02052024-25ad}, the commutators \rf{manus-02052024-26} are represented as
{\small
\beq
\label{manus-02052024-27ad} && \hspace{-0.7cm} [T_0^\asf,T_0^\bsf] = f^{\asf\bsf\csf} T_0^\csf \qquad [T_0^\asf,S_2^\bsf] =  f^{\asf\bsf\csf} S_2^\csf \,, \qquad
[T_0^\asf,S_4^\bsf] =  f^{\asf\bsf\csf} S_4^\csf \,,\qquad
[T_0^\asf,S_6^\bsf] =  f^{\asf\bsf\csf} S_6^\csf \,.
\\
&& \hspace{-0.7cm}  [S_2^\asf,S_2^\bsf] = \frac{2}{3} f^{\asf\bsf\csf} S_4^\csf + \frac{56}{3} \rho f^{\asf\bsf\csf} S_6^\csf \,,\hspace{1cm} [S_2^\asf,S_4^\bsf] = \frac{2}{3} f^{\asf\bsf\csf} S_6^\csf \,, \qquad [S_{2m}^\asf,S_{2n}^\bsf] = 0\,, \ \hbox{for} \ m+n > 3\,.
\nonumber
\eeq
}
\!From \rf{manus-02052024-27ad}, we learn that the generators$ S_2^\asf$, $S_4^\asf$, $S_6^\asf$ form radical of the extended gauge algebra \rf{manus-02052024-25ad}.
In other words, we obtain the Levy-Maltsev decomposition given by
\be
T_0^\asf\,, \ S_2^\asf\,, \  S_4^\asf \,, \  S_6^\asf \,\, =\,\,  S_2^\asf\,, S_4^\asf\,, \  S_6^\asf\,\, \subsetext\,\, T_0^\asf\,.
\ee

In order to represent the gauge transformations of the Stueckelberg fields and the corresponding field strengths \rf{manus-02052024-06}, \rf{manus-02052024-08} in the framework of the extended gauge algebra we use the general relations in \rf{manus-08042024-95} when $d=9$,
\beq
\label{manus-02052024-30} && \phi := \phi_2^\asf  S_2^\asf + \phi_4^\asf S_4^\asf + \phi_6^\asf S_6^\asf\,,\hspace{2cm} F^A := F_3^{A\asf}  S_2^\asf + F_5^{A\asf} S_4^\asf+ F_7^{A\asf} S_6^\asf\,,\qquad
\nonumber\\
&& \phi_\subsm^A := \phi_3^{A\asf} S_2^\asf + \phi_5^{A\asf} S_4^\asf+ \phi_7^{A\asf} S_6^\asf\,, \hspace{1cm} \xi_\subsm :=  \xi_2^\asf S_2^\asf +  \xi_4^\asf S_4^\asf+  \xi_6^\asf S_6^\asf\,.
\eeq

Using the quantities defined in  \rf{manus-02052024-25}, \rf{manus-02052024-30}, we note that the expressions for $F_{2n+1}^A$ given in \rf{manus-02052024-03add} and the expressions for gauge transformations of $\phi_{2n}$, $F_{2n+1}^A$ given in \rf{manus-02052024-06}, \rf{manus-02052024-08} amount to the expressions given in \rf{manus-08042024-96}.  To this end the following commutators turn out to be helpful:
{\small
\be
[S_2^\asf,T_2^\bsf] = \frac{2}{3} f^{\asf\bsf\csf} S_4^\csf +
\frac{14}{3} \rho f^{\asf\bsf\csf} S_6^\csf \,,\qquad  [S_2^\asf,T_4^\bsf] = \frac{1}{3} f^{\asf\bsf\csf} S_6^\csf \,,\qquad [S_4^\asf,T_2^\bsf] = \frac{2}{3} f^{\asf\bsf\csf} S_6^\csf \,.
\ee
}

Finally, we note that the Lagrangian given in \rf{manus-02052024-03} can be represented as in \rf{manus-08042024-97}, \rf{manus-08042024-100}, where all non-zero values of the invariant bilinear forms are given by
{\small
\beq
\label{manus-02052024-34} &&  \hspace{-1cm}  G_{06} = 1\,,\quad  \quad G_{24} = 1\,,\quad
G_{04} = - 28 \rho \,, \quad  G_{22} = - 28 \rho \,, \quad G_{02} = 252\rho^2\,, \quad G_{00} = - 720\rho^3\,,\qquad
\nonumber\\
&&  \hspace{-1cm}  G_{26}^\subsm = 1\,, \quad G_{44}^\subsm = 1\,, \quad G_{24}^\subsm =  - 28\rho \,, \quad G_{22}^\subsm =  252 \rho^2\,.
\eeq
}
\!For $d=9$, the conjectured expressions in \rf{manus-08042024-100-a1} are in agreement with \rf{manus-02052024-34}.

\noinbf{Higher-derivative Lagrangian of conformal YM field in $(A)dS_{10}$}. Higher-derivative Lagrangian is obtained by using the same procedure as the one we used in six and eight dimensions. Namely, we gauge away the Stueckelberg fields $\phi_2$, $\phi_4$, $\phi_6$ and plug solution to equations of motion for the auxiliary fields $\phi_3^A$, $\phi_5^A$, and $\phi_7^A$ in ordinary-derivative Lagrangian \rf{manus-02052024-03}. Doing so and using conventions in \rf{manus-08042024-85}, we find the following higher-derivative Lagrangian:
\be \label{manus-16092924-01add}
\LL_\CYMsm^\Hdr = -\KK\, \Tr L_\CYMsm^\Hdr\,, \qquad L_\CYMsm^\Hdr:  =  - \Ib_{10} + 28 \rho \Ib_8 -  252 \rho^2 \Ib_6 + 720 \rho^3 I_4\,,
\ee
where we use the notation
\beq
\label{manus-16092924-02add}
&& \hspace{-1cm} \Ib_{10} : = - \half \phib_5^A\phib_5^A - \frac{4}{3} \phib_3^A \Fb_4^{AB} \phib_3^B  - \frac{28}{3} \rho \phib_3^A F_2^{AB} \phib_3^B \,,
\nonumber\\
&& \hspace{-1cm} \Ib_8 : = \frac{1}{4} \Fb_4^{AB} \Fb_4^{AB} - \frac{4}{3} \phib_3^A F_2^{AB} \phib_3^B\,, \qquad \Ib_6 : = -\half \phib_3^A \phib_3^A\,,\qquad I_4 := \frac{1}{4} F_2^{AB} F_2^{AB}\,,
\nonumber\\
&& \hspace{-1cm} \phib_3^A =  \DD^B F_2^{BA} \,,\qquad \phib_5^A =  \DD^B \Fb_4^{BA} \,, \qquad \Fb_4^{AB} : = \DD^A \phib_3^B - \DD^B  \phib_3^A\,.
\eeq
\!while $\phib_3^A$, $\phib_3^A$ stand for the solution to the auxiliary fields $\phi_3^A$, $\phi_5^A$. The quantities $\Ib_{10}$, $\Ib_8$, and $\Ib_6$ \rf{manus-16092924-02add} are equal to the respective quantities $I_{10}$, $I_8$, and $I_6$ \rf{manus-02052024-03} evaluated on the solutions of equations for the auxiliary fields. Field strength $F_2^{AB}$  \rf{manus-02052024-03add} depends only on the primary field $\phi_1^A$. Therefore the higher-derivative Lagrangian \rf{manus-16092924-01add} is entirely expressed in terms of the field $\phi_1^A$.

In $L_\CYMsm^\Hdr$ \rf{manus-16092924-01add}, the $\rho\phib_3^A F_2^{AB} \phib_3^B$-term enters the both $\Ib_{10}$ and $\rho \Ib_8$. To avoid of such double appearance of the $\rho\phib_3^A F_2^{AB} \phib_3^B$-term we represent $L_\CYMsm^\Hdr$ \rf{manus-16092924-01add} as the following Taylor series expansion in $\rho$:
\be \label{manus-16092924-03add}
L_\CYMsm^\Hdr =   I_{10}^\Hdr + 28 \rho I_8^\Hdr +  252 \rho^2 I_6^\Hdr + 720 \rho^3 I_4\,,
\ee
where new basis is defined as
\beq
\label{manus-16092924-04add}
&& \hspace{-1cm} I_{10}^\Hdr :=  \half \phib_5^A\phib_5^A + \frac{4}{3} \phib_3^A \Fb_4^{AB} \phib_3^B\,,
\nonumber\\
&& \hspace{-1cm} I_8^\Hdr : = \frac{1}{4} \Fb_4^{AB} \Fb_4^{AB} -  \phib_3^A F_2^{AB} \phib_3^B\,, \qquad I_6^\Hdr := \half \phib_3^A \phib_3^A\,,\qquad I_4 := \frac{1}{4} F_2^{AB} F_2^{AB}\,.
\eeq
Note that the $I_8^\Hdr$, $I_6^\Hdr$, and $I_4$ in \rf{manus-16092924-04add} take the same form as their cousins in $(A)dS_6$ and $(A)dS_8$ given in  \rf{manus-09042024-82} and \rf{manus-16092924-02}.

In the flat limit, $I_6^\Hdr$ and $I_4$ take the same form as the known Lagrangians of conformal YM field in the respective six and four dimensions, while $I_{10}^\Hdr$ and $I_8^\Hdr$ take the same form as the Lagrangians of conformal YM in the respective ten and eight dimensions found in Ref.\cite{Metsaev:2023qif}. In other words, the higher-derivative Lagrangian of conformal YM field in $(A)dS_{10}$ is obtained from the sum of Lagrangians of conformal YM field in flat space of ten, eight, six, and four dimensions uplifted to $(A)dS_{10}$.
All in all we are led to the following

\noinbf{Conjecture}: {\bf a}) Higher-derivative Lagrangian of conformal YM field in $(A)dS_{d+1}$ for arbitrary odd $d\geq 11$ can be presented as the following Taylor series expansion in the parameter $\rho$:
\be \label{manus-16092924-05add}
L_\CYMsm^\Hdr =   (-)^{\frac{d-1}{2}} \sum_{n=0,1,\ldots,\frac{d-3}{2} } G_n \rho^{\frac{d-3}{2}-n} I_{2n+4}^\Hdr\,, \qquad I_4^\Hdr: = I_4\,,
\ee
where coefficients $G_n$ coincide with the ones in \rf{manus-08042024-100-a3}, while the quantities $I_{2n+4}^\Hdr$ depend on the primary field and covariant derivative and do not depend on $d$. For $n=0,1,2,3$, the $I_{2n+4}^\Hdr$ are given in \rf{manus-16092924-04add}; {\bf b}) In the flat space limit, the $I_{2n+4}^\Hdr$ takes the same form as the higher-derivative Lagrangian of the conformal YM field in $R^{2n+3,1}$.

For $(A)dS_{d+1}$ with $d=5,7,9$, the statements of our conjecture are proved in this paper.

\newsection{\large Conformal Yang-Mills field in $(A)dS_{10}$ space. Decoupled formulation } \label{ten-fac}

{\bf Field content}. In the framework of gauge invariant decoupled
formulation, the conformal YM field in $(A)dS_{10}$ is described by four vector fields and three scalar fields:
\beq
\label{manus-03052024-01} && \varphi_1^A \qquad \varphi_3^A\qquad \varphi_5^A\qquad \varphi_7^A
\nonumber\\
&&
\nonumber\\[-35pt]
&& \hspace{0.7cm} \varphi_2 \qquad \varphi_4\qquad \varphi_6
\eeq
The vector fields are $y$-transversal \rf{manus-08042024-25}. Fields \rf{manus-03052024-01} are decomposed in $t^a$ as in \rf{manus-08042024-86add}.
Realization of $(A)dS_{10}$ algebra \rf{manus-08042024-10} on the fields takes the same form as in \rf{manus-08042024-16}.  Below, we show that the vector field $\varphi_1^A$ describes massless field, while the vector fields $\varphi_3^A$, $\varphi_5^A$, $\varphi_7^A$ accompanied by the scalar fields $\varphi_2$, $\varphi_4$, $\varphi_6$ describe massive vectors field with mass squares $m_{\varphi_3}^2$, $m_{\varphi_5}^2$, $m_{\varphi_7}^2$,
\be \label{manus-03052024-02}
m_{\varphi_1}^2 = 0\,, \qquad m_{\varphi_3}^2 = 6\rho\,, \qquad m_{\varphi_5}^2 = 10\rho\,, \qquad m_{\varphi_7}^2 = 12\rho\,,
\ee
where $\rho$ is given in \rf{manus-08042024-01}.  Namely, the vector field $\varphi_{2n+1}^A$ and the scalar field $\varphi_{2n}$ provide a gauge invariant description of massive field with the mass square $m_{\varphi_{2n+1}}^2$, $n=1,2,3$. Below, we show that the scalar fields $\varphi_2$, $\varphi_4$, and $\varphi_6$ play a role of the Stueckelberg fields.

\noinbf{Gauge invariant Lagrangian}. In the decoupled formulation, the ordinary-derivative Lagrangian of the conformal YM field in $(A)dS_{10}$ we find takes the form
\beq
\label{manus-03052024-03} &&   k_\rho^{-1} \LL_\CYMsm =  - \frac{1}{4} \FF_2 \FF_2   + \frac{1}{4} \FF_4 \FF_4 - \frac{1}{4} \FF_6 \FF_6 + \frac{1}{4} \FF_8 \FF_8
\nonumber\\
&& \hspace{1.6cm}  + \,\, 3\rho \FF_3 \FF_3 - 5\rho \FF_5 \FF_5 + 6\rho  \FF_7 \FF_7 \,, \qquad k_\rho := -720 \rho^3\,,
\eeq
where we use the shortcuts for the scalar products  defined in \rf{manus-08042024-88add}, while explicit expressions for field strengths are given below. It is instructive to represent Lagrangian \rf{manus-03052024-03} as
{\small
\beq
\label{manus-03052024-04} && \hspace{-0.7cm} k_\rho^{-1} \LL_\CYMsm = - \II_4 + \II_8 - \II_{12} + \II_{16}\,,
\nonumber\\
&& \II_4 := \frac{1}{4} \FF_2 \FF_2\,, \qquad \II_{4n+4} : = \frac{1}{4} \FF_{2n+2} \FF_{2n+2} + \frac{m_{\varphi_{2n+1}}^2}{2} \FF_{2n+1}\FF_{2n+1}\,,\qquad n=1,2,3\,,\qquad
\eeq
}
\!\!where the mass squares are given in \rf{manus-03052024-02}.  The field strengths entering Lagrangian \rf{manus-03052024-03} are expressed in terms of the fields \rf{manus-03052024-01} in the following way:
{\small
\beq
\label{manus-03052024-05} && \FF_2^{AB} := (\ppp^A - \rho y^A)\varphi_1^B - (\ppp^B - \rho y^B)\varphi_1^A +  [\varphi_1^A,\varphi_1^B] +  \FF_{2,\,\addsm}^{AB}\,,
\nonumber\\
&& \FF_{2n+2}^{AB}: = \DD^A \varphi_{2n+1}^B -  \DD^B \varphi_{2n+1}^A +  \FF_{2n+2,\,\addsm}^{AB}\,,
\nonumber\\
&& \FF_{2n+1}^A := \varphi_{2n+1}^A + \DD^A\varphi_{2n}  + \FF_{2n+1,\,\addsm}^A \,, \hspace{4cm} n=1,2,3\,, \qquad
\eeq
}
\!where the covariant derivative $\DD^A$ is given in \rf{manus-14092024-01add} when $d=9$,
while the expressions for $\FF_{2n+2,\addsm}^{AB}$ and $\FF_{2n+1,\addsm}^A$ appearing in \rf{manus-03052024-05} are given by
{\small
\beq
\label{manus-03052024-06} &&   \FF_{2,\,\addsm}^{AB} = -   [\varphi_3^A,\varphi_3^B] +   [\varphi_5^A,\varphi_5^B] -   [\varphi_7^A,\varphi_7^B]\,,
\nonumber\\
&&    \FF_{4,\,\addsm}^{AB}  = - 2 [\varphi_3^A,\varphi_5^B]  - 2[\varphi_5^A,\varphi_3^B] +  5 [\varphi_5^A,\varphi_7^B] + 5[\varphi_7^A,\varphi_5^B]  - 4c_5 [\varphi_7^A,\varphi_7^B] \,,
\nonumber\\
&&    \FF_{6,\,\addsm}^{AB}  =  2 [\varphi_3^A,\varphi_3^B]  - 14[\varphi_5^A,\varphi_5^B] - 5 [\varphi_3^A,\varphi_7^B] - 5[\varphi_7^A,\varphi_3^B]
\nonumber\\
&& \hspace{1.1cm} +\,\, 10 c_5\big( [\varphi_5^A,\varphi_7^B]+ [\varphi_7^A,\varphi_5^B]\big)  - 28 [\varphi_7^A,\varphi_7^B] \,,
\nonumber\\
&&    \FF_{8,\,\addsm}^{AB}  = 5[\varphi_3^A,\varphi_5^B] + 5 [\varphi_5^A,\varphi_3^B] - 10 c_5 [\varphi_5^A,\varphi_5^B] - 4 c_5  \big( [\varphi_3^A,\varphi_7^B]+ [\varphi_7^A,\varphi_3^B]\big)\qquad
\nonumber\\
&& \hspace{1.1cm}  +\,\, 28 \big( [\varphi_5^A,\varphi_7^B]+ [\varphi_7^A,\varphi_5^B]\big)  - 14c_5 [\varphi_7^A,\varphi_7^B] \,,
\nonumber\\
&&   \FF_{3,\,\addsm}^A = - \frac{5}{3} [\varphi_3^A,\varphi_4] - \frac{1}{3} [\varphi_5^A,\varphi_2] + \frac{10}{3}  [\varphi_5^A,\varphi_6] + \frac{5}{3} [\varphi_7^A,\varphi_4]- 2c_5 [\varphi_7^A,\varphi_6]\,, \qquad
\nonumber\\
&&   \FF_{5,\,\addsm}^A =  [\varphi_3^A,\varphi_2] - 4 [\varphi_3^A,\varphi_6] - 7   [\varphi_5^A,\varphi_4] -  [\varphi_7^A,\varphi_2]
\nonumber\\
&& \hspace{1.2cm} + \,\, 6c_5 [\varphi_5^A,\varphi_6] + 4c_5   [\varphi_7^A,\varphi_4] - 14 [\varphi_7^A,\varphi_6]
\nonumber\\
&&   \FF_{7,\,\addsm}^A =  \frac{10}{3} [\varphi_3^A,\varphi_4] + \frac{5}{3} [\varphi_5^A,\varphi_2] - 3c_5   [\varphi_3^A,\varphi_6] - 5c_5 [\varphi_5^A,\varphi_4]- c_5 [\varphi_7^A,\varphi_2]
\nonumber\\
&& \hspace{1.2cm} + \,\, \frac{49}{3} [\varphi_5^A,\varphi_6] + \frac{35}{3}   [\varphi_7^A,\varphi_4] - 7c_5 [\varphi_7^A,\varphi_6]\,,
\nonumber\\
&& \hspace{1cm} c_q : = \sqrt{q}\,.
\eeq
}
\!\!We note that the Lagrangian \rf{manus-03052024-03} is diagonal with respect to all field strengths.

\noinbf{Gauge transformations}.  In gauge invariant approach, each vector field \rf{manus-03052024-01} is accompanied by the corresponding gauge parameter. This implies that we use four gauge parameters,
\be \label{manus-03052024-07}
\eta_0\,, \qquad \eta_2\,, \qquad \eta_4\,, \qquad \eta_6\,,
\ee
which are decomposed in $t^a$ as in \rf{manus-08042024-87add}. Gauge transformations of fields \rf{manus-03052024-01} we find take the form
\beq
\label{manus-03052024-07add} && \delta_\eta  \varphi_{2n+1}^A = \DD^A\eta_{2n} + \delta_{\eta,\addsm}  \varphi_{2n+1}^A\,, \hspace{1.7cm}  n=0,1,2,3;
\nonumber\\
&& \delta_\eta  \varphi_{2n} = - \eta_{2n} + [\varphi_{2n},\eta_0] + \delta_{\eta,\addsm}  \varphi_{2n}\,,\qquad n=1,2,3\,,
\eeq
where the covariant derivative $\DD^A$ is given in \rf{manus-14092024-02add} when $d=9$
and the $\delta_{\eta,\addsm}$-contributions entering \rf{manus-03052024-07add} are given by
{\small
\beq
\label{manus-03052024-09} && \delta_{\eta,\addsm} \varphi_1^A =  -  [\varphi_3^A,\eta_2] +  [\varphi_5^A,\eta_4] -  [\varphi_7^A,\eta_6]\,,
\nonumber\\
&&  \delta_{\eta,\addsm} \varphi_3^A  = [\varphi_1^A,\eta_2]  - 2 [\varphi_3^A,\eta_4]  - 2 [\varphi_5^A,\eta_2] +    5[\varphi_5^A,\eta_6]+ 5 [\varphi_7^A,\eta_4]  - 4c_5 [\varphi_7^A,\eta_6] \,,\qquad
\nonumber\\
&&  \delta_{\eta,\addsm} \varphi_5^A  = [\varphi_1^A,\eta_4]  + 2 [\varphi_3^A,\eta_2] - 14[\varphi_5^A,\eta_4]
\nonumber\\
&& \hspace{1.8cm} - \,\,  5 \big( [\varphi_3^A,\eta_6] + [\varphi_7^A,\eta_2]\big) + 10c_5 \big( [\varphi_5^A,\eta_6]+ [\varphi_7^A,\eta_4]\big)  - 28 [\varphi_7^A,\eta_6] \,,
\nonumber\\
&&  \delta_{\eta,\addsm} \varphi_7^A  = [\varphi_1^A,\eta_6]  + 5 [\varphi_3^A,\eta_4] + 5[\varphi_5^A,\eta_2] - 10 c_5[\varphi_5^A,\eta_4]
\nonumber\\
&& \hspace{1.8cm}  - \,\,  4c_5 \big( [\varphi_3^A,\eta_6]+ [\varphi_7^A,\eta_2]\big) + 28 \big( [\varphi_5^A,\eta_6]+ [\varphi_7^A,\eta_4]\big)  - 14 c_5 [\varphi_7^A,\eta_6] \,,
\nonumber\\
&&   \delta_{\eta,\addsm} \varphi_2 =   - \frac{5}{3} [\varphi_4,\eta_2] - \frac{1}{3} [\varphi_2,\eta_4] + \frac{10}{3}  [\varphi_6,\eta_4] + \frac{5}{3} [\varphi_4,\eta_6]- 2c_5 [\varphi_6,\eta_6]\,, \qquad
\nonumber\\
&& \delta_{\eta,\addsm} \varphi_4 =    [\varphi_2,\eta_2] - 4 [\varphi_6,\eta_2] - 7   [\varphi_4,\eta_4] -  [\varphi_2,\eta_6]
\nonumber\\
&& \hspace{1.6cm} + \,\, 6c_5 [\varphi_6,\eta_4] + 4c_5   [\varphi_4,\eta_6] - 14 [\varphi_6,\eta_6]\,,
\nonumber\\
&&   \delta_{\eta,\addsm} \varphi_6 =  \frac{10}{3} [\varphi_4,\eta_2] + \frac{5}{3} [\varphi_2,\eta_4] - 3c_5   [\varphi_6,\eta_2] - 5c_5 [\varphi_4,\eta_4] - c_5 [\varphi_2,\eta_6]
\nonumber\\
&& \hspace{1.6cm} + \,\, \frac{49}{3} [\varphi_6,\eta_4] + \frac{35}{3}   [\varphi_4,\eta_6] - 7c_5 [\varphi_6,\eta_6]\,.
\eeq
}
We recall that the coefficient $c_q$ is defined in \rf{manus-03052024-06}. Gauge transformations of the corresponding field strengths \rf{manus-03052024-05} are given by
\beq
\label{manus-03052024-10} && \delta_\eta \FF_{2n+2}^{AB} = [\FF_{2n+2}^{AB},\eta_0] + \delta_{\eta,\addsm} \FF_{2n+2}^{AB}\,, \hspace{1cm} n=0,1,2,3\,;
\nonumber\\
&& \delta_\eta \FF_{2n+1}^A = [\FF_{2n+1}^A,\eta_0] + \delta_{\eta,\addsm} \FF_{2n+1}^A\,, \hspace{1cm}  n=1,2,3\,,
\eeq
where the $\delta_{\eta,\addsm}$-contributions entering \rf{manus-03052024-10} are given by
{\small
\beq
\label{manus-03052024-11} &&  \hspace{-1cm} \delta_{\eta,\addsm} \FF_2^{AB} =  -   [\FF_4^{AB},\eta_2] +   [\FF_6^{AB},\eta_4] -  [\FF_8^{AB},\eta_6] \,,
\nonumber\\
&&  \hspace{-1cm}  \delta_{\eta,\addsm} \FF_4^{AB}   = [\FF_2^{AB},\eta_2]   - 2 [\FF_4^{AB},\eta_4]  - 2 [\FF_6^{AB},\eta_2]  +   5[\FF_6^{AB},\eta_6] + 5 [\FF_8^{AB},\eta_4]  - 4c_5 [\FF_8^{AB},\eta_6] \,,
\nonumber\\
&& \hspace{-1cm}   \delta_{\eta,\addsm} \FF_6^{AB}  = [\FF_2^{AB},\eta_4]  + 2 [\FF_4^{AB},\eta_2]  - 14 [\FF_6^{AB},\eta_4]
\nonumber\\
&& \hspace{0.7cm} - \,\,  5 [\FF_4^{AB},\eta_6] - 5 [\FF_8^{AB},\eta_2] + 10 c_5 \big( [\FF_6^{AB},\eta_6]+ [\FF_8^{AB},\eta_4]\big)  - 28 [\FF_8^{AB},\eta_6] \,,
\nonumber\\
&& \hspace{-1cm}   \delta_{\eta,\addsm} \FF_8^{AB}  = [\FF_2^{AB},\eta_6]   +  5 [\FF_4^{AB},\eta_4] + 5 [\FF_6^{AB},\eta_2] - 10 c_5[\FF_6^{AB},\eta_4]
\nonumber\\
&& \hspace{0.5cm} - \,\,  4c_5 \big( [\FF_4^{AB},\eta_6]+ [\FF_8^{AB},\eta_2]\big) + 28 \big( [\FF_6^{AB},\eta_6]+ [\FF_8^{AB},\eta_4]\big)  - 14c_5 [\FF_8^{AB},\eta_6] \,,
\nonumber\\
&& \hspace{-1cm}  \delta_{\eta,\addsm} \FF_3^A =  - \frac{5}{3} [\FF_5^A,\eta_2] - \frac{1}{3}[\FF_3^A,\eta_4] + \frac{10}{3}  [\FF_7^A,\eta_4] + \frac{5}{3} [\FF_5^A,\eta_6]- 2c_5 [\FF_7^A,\eta_6]\,, \qquad
\nonumber\\
&&   \hspace{-1cm} \delta_{\eta,\addsm} \FF_5^A =    [\FF_3^A,\eta_2] - 4 [\FF_7^A,\eta_2] - 7   [\FF_5^A,\eta_4] -  [\FF_3^A,\eta_6]
\nonumber\\
&& \hspace{0.5cm} + \,\, 6c_5 [\FF_7^A,\eta_4] + 4c_5   [\FF_5^A,\eta_6] - 14 [\FF_7^A,\eta_6]\,,
\nonumber\\
&& \hspace{-1cm} \delta_{\eta,\addsm} \FF_7^A =   \frac{10}{3} [\FF_5^A,\eta_2] + \frac{5}{3} [\FF_3^A,\eta_4] - 3c_5   [\FF_7^A,\eta_2] - 5c_5 [\FF_5^A,\eta_4] - c_5 [\FF_3^A,\eta_6]
\nonumber\\
&& \hspace{0.5cm} + \,\, \frac{49}{3} [\FF_7,\eta_4] + \frac{35}{3}   [\FF_5,\eta_6] - 7c_5 [\FF_7,\eta_6]\,.
\eeq
}

From \rf{manus-03052024-07add}, we conclude that the vector fields are realized as gauge fields, while the scalar fields are realized as Stueckelberg fields.

\noinbf{$K^A$-transformations of fields and field strengths}. We find the following $K^A$-transformations of fields \rf{manus-03052024-01}:
{\small
\beq
\label{manus-03052024-12} && \delta_{K^A} \varphi_1^B = c_{\frac{1}{5}} K_7^A \varphi_3^B - 6c_{\frac{1}{5}}\theta^{AB}\varphi_2\,,
\nonumber\\
&& \delta_{K^A} \varphi_3^B = 2c_{\frac{1}{5}} K_6^A \varphi_5^B  - c_{\frac{1}{5}} K_1^A \varphi_1^B - 2c_5\theta^{AB}\varphi_4\,,
\nonumber\\
&& \delta_{K^A} \varphi_5^B = c_5 K_5^A \varphi_7^B - 2c_{\frac{1}{5}} K_2^A \varphi_3^B  - 4c_5 \theta^{AB}\varphi_6 + 2 c_{\frac{1}{5}} \theta^{AB}\varphi_2\,,
\nonumber\\
&& \delta_{K^A} \varphi_7^B =  4 K_4^A \varphi_7^B - c_5 K_3^A \varphi_5^B - 12\theta^{AB}\varphi_6 + 2c_5 \theta^{AB}\varphi_4\,,
\nonumber\\
&& \delta_{K^A} \varphi_2 = \frac{1}{3} c_5 K_6^A \varphi_4 + \frac{1}{3\rho} c_{\frac{1}{5}} \varphi_5^A - \frac{1}{\rho} c_{\frac{1}{5}} \varphi_1^A\,,
\nonumber\\
&& \delta_{K^A} \varphi_4 = 4 c_{\frac{1}{5}}  K_5^A \varphi_6 -  c_{\frac{1}{5}}  K_2^A \varphi_2
+ \frac{1}{\rho} c_{\frac{1}{5}} \varphi_7^A - \frac{1}{\rho} c_{\frac{1}{5}} \varphi_3^A\,,
\nonumber\\
&& \delta_{K^A} \varphi_6 = 3 K_4^A \varphi_6 -  \frac{2}{3}c_5 K_3^A \varphi_4 + \frac{1}{\rho} \varphi_7^A - \frac{1}{3\rho} c_5 \varphi_5^A\,,
\eeq
}
\!while the $K^A$-transformations of the corresponding field strengths \rf{manus-03052024-05} are given by
{\small
\beq
\label{manus-03052024-15} && \delta_{K^A} \FF_2^{BC} = c_{\frac{1}{5}} K_8^A \FF_4^{BC} + 6c_{\frac{1}{5}} \big( \theta^{AB} \FF_3^C  - \theta^{AC} \FF_3^B \big) \,,
\nonumber\\
&& \delta_{K^A} \FF_4^{BC} = 2c_{\frac{1}{5}} K_7^A \FF_6^{BC}  - c_{\frac{1}{5}} K_2^A \FF_2^{BC} + 2 c_5 \big( \theta^{AB} \FF_5^C -   \theta^{AC} \FF_5^B\big)\,,
\nonumber\\
&& \delta_{K^A} \FF_6^{BC} = c_5 K_6^A \FF_8^{BC} - 2 c_{\frac{1}{5}} K_3^A \FF_4^{BC}  + 4 c_5\big(\theta^{AB} \FF_7^C - \theta^{AC} \FF_7^B\big) - 2c_{\frac{1}{5}}\big(\theta^{AB} \FF_3^C - \theta^{AC} \FF_3^B\big)\,,\qquad
\nonumber\\
&& \delta_{K^A} \FF_8^{BC} =  4 K_5^A \FF_8^{BC} - c_5 K_4^A \FF_6^{BC} + 12\big(\theta^{AB} \FF_7^C - \theta^{AC} \FF_7^B\big) - 2c_5\big( \theta^{AB} \FF_5^C -\theta^{AB} \FF_5^C\big)\,,\qquad
\nonumber\\
&& \delta_{K^A} \FF_3^B  = \frac{1}{3} c_5 K_7^A \FF_5^B  - \frac{1}{3\rho} c_{\frac{1}{5}} \FF_6^{AB} + \frac{1}{\rho} c_{\frac{1}{5}} \FF_2^{AB}  \,,
\nonumber\\
&& \delta_{K^A} \FF_5^B = 4 c_{\frac{1}{5}} K_6^A \FF_7^B - c_{\frac{1}{5}} K_3^A \FF_3^B
- \frac{1}{\rho} c_{\frac{1}{5}} \FF_8^{AB} + \frac{1}{\rho} c_{\frac{1}{5}} \FF_4^{AB}\,,
\nonumber\\
&& \delta_{K^A} \FF_7^B = 3 K_5^A \FF_7^B -  \frac{2}{3} K_4^A \FF_5^B - \frac{1}{\rho} \FF_8^{AB} +  \frac{1}{3\rho}c_5 \FF_6^{AB}\,,
\eeq
}
\!\!where $c_q$ and $\theta^{AB}$ are defined in \rf{manus-01052024-06} and \rf{12062024-01} respectively. Action of operator $K_\Delta^A$ on the fields and fields strengths  \rf{manus-03052024-12}, \rf{manus-03052024-12} is defined in the same way as in \rf{manus-08042024-75}, \rf{manus-08042024-76}.

The following remarks are in order.

\noinbf{i}) To quadratic order in the fields, Lagrangian \rf{manus-03052024-04} takes the form
\be \label{manus-03052024-15a1}
k_\rho^{-1} \LL_\CYMsm^\smtwo =  - \II_4^\smtwo  +  \II_8^\smtwo -  \II_{12}^\smtwo + \II_{16}^\smtwo\,,
\ee
where $\II_{4n+4}^\smtwo$, $n=0,1,2,3$ are obtained by setting $d=9$ in \rf{manus-08042024-106-b3}. From \rf{manus-03052024-15a1}, we see that the free Lagrangian describes the four set of decoupled fields. The first set is given by the single field $\varphi_1^A$ which describes massless field, while the remaining three sets are given by the fields $\varphi_{2n+1}^A$, $\varphi_{2n}$, $n=1,2,3$ which describe the respective massive fields with the mass square $m_{\varphi_{2n+1}}^2$.

\noinbf{ii}) For $AdS_{10}$, we note $k_\rho > 0 $ and hence the massless field and the massive field with the mass square $m_{\varphi_5}^2$ are realized with the correct sign of kinetic terms, while the massive fields with the mass squares $m_{\varphi_3}^2$ and $m_{\varphi_7}^2$ are realized with the wrong sign of the kinetic terms. For $dS_{10}$, we note $k_\rho<0$ and hence the massless field and the massive field with the mass square $m_{\varphi_5}^2$ are realized with the wrong sign of kinetic terms, while the massive fields with the mass squares $m_{\varphi_3}^2$ and $m_{\varphi_7}^2$ are realized with the correct sign of the kinetic terms. For $AdS_{10}$, all massive fields have negative mass squares, while, for $dS_{10}$, all massive fields have positive mass squares.

\noinbf{Extended gauge algebra}.  For the case under consideration, the extended gauge algebra consists of four generators denoted as
\be \label{manus-03052024-20}
\TT_0^\asf\,, \quad \TT_2^\asf\,, \quad \TT_4^\asf\,, \quad \TT_6^\asf\,.
\ee
Considering commutators of gauge transformations \rf{manus-03052024-07add} or  \rf{manus-03052024-10}, we find the commutators for generators \rf{manus-03052024-20},
{\small
\beq
\label{manus-03052024-26} &&   [\TT_0^\asf,\TT_0^\bsf] =  f^{\asf\bsf\csf}\TT_0^\csf \,,
\nonumber\\
&&   [\TT_0^\asf,\TT_2^\bsf] =  f^{\asf\bsf\csf}\TT_2^\csf \,, \quad [\TT_0^\asf,\TT_4^\bsf] =  f^{\asf\bsf\csf}\TT_4^\csf \,, \quad   [\TT_0^\asf,\TT_6^\bsf] =  f^{\asf\bsf\csf}\TT_6^\csf \,,
\nonumber\\
&&  [\TT_2^\asf,\TT_2^\bsf] = -   f^{\asf\bsf\csf}\TT_0^\csf +  2 f^{\asf\bsf\csf} \TT_4^\csf\,,
\nonumber\\
&& [\TT_2^\asf,\TT_4^\bsf] = - 2 f^{\asf\bsf\csf}\TT_2^\csf +  5 f^{\asf\bsf\csf} \TT_6^\csf\,,
\nonumber\\
&&  [\TT_2^\asf,\TT_6^\bsf] = - 5 f^{\asf\bsf\csf}\TT_4^\csf - 4c_5 f^{\asf\bsf\csf} \TT_6^\csf\,,
\nonumber\\
&&  [\TT_4^\asf,\TT_4^\bsf] =   f^{\asf\bsf\csf}\TT_0^\csf - 14 f^{\asf\bsf\csf} \TT_4^\csf- 10c_5 f^{\asf\bsf\csf} \TT_6^\csf\,,
\nonumber\\
&& [\TT_4^\asf,\TT_6^\bsf] = 5 f^{\asf\bsf\csf}\TT_2^\csf + 10c_5  f^{\asf\bsf\csf} \TT_4^\csf + 28 f^{\asf\bsf\csf} \TT_6^\csf\,,
\nonumber\\
&& [\TT_6^\asf,\TT_6^\bsf] = - f^{\asf\bsf\csf}\TT_0^\csf - 4c_5 f^{\asf\bsf\csf} \TT_2^\csf -  28 f^{\asf\bsf\csf} \TT_4^\csf -
14 c_5 f^{\asf\bsf\csf} \TT_6^\csf\,,
\eeq
}
\!\!where $c_q$ is given in \rf{manus-03052024-06}. Using commutators \rf{manus-03052024-26}, we can apply the general setup of the extended gauge algebra described in Sec.\ref{not-conv}. Namely, setting $d=9$ in \rf{manus-08042024-91ab}, we introduce
{\small
\be \label{manus-03052024-27}
\varphi^A : = \sum_{n=0,1,2,3} \varphi_{2n+1}^{A\asf} \TT_{2n}^\asf\,, \qquad \FF^{AB} : = \sum_{n=0,1,2,3} \FF_{2n+2}^{AB\asf} \TT_{2n}^\asf\,, \qquad \eta = \sum_{n=0,1,2,3} \eta_{2n}^\asf \TT_{2n}^\asf\,.
\ee
}
\!\!Now, using $\varphi^A$, $\FF^{AB}$, and $\eta$, we note that the expressions for $\FF_{2n+2}^{AB}$, $n=0,1,2,3$, given in \rf{manus-03052024-05} and the gauge transformations for the fields $\varphi_{2n+1}^A$ and field strengths $\FF_{2n+2}^{AB}$, $n=0,1,2,3$ given in \rf{manus-03052024-07add} and \rf{manus-03052024-10} can be represented in a more compact way as in \rf{manus-08042024-120}.

Following the general pattern in Sec.\ref{not-conv}, we then represent the set of generators \rf{manus-03052024-20} as
{\small
\beq
\label{manus-03052024-28} && \hspace{-1cm} \TT_0^\asf\,, \quad \Scl_2^\asf\,, \quad \Scl_4^\asf\,,  \quad \Scl_6^\asf\,;
\nonumber\\
&& \Scl_2^\asf := -\frac{8}{9} c_5 \TT_0^\asf -  \frac{52}{9} \TT_2^\asf - 3 c_5 \TT_4^\asf - \frac{37}{9} \TT_6^\asf\,,
\nonumber\\
&& \Scl_4^\asf := \frac{52}{9} \TT_0^\asf +  \frac{46}{9} c_5 \TT_2^\asf + 9 \TT_4^\asf + \frac{13}{9} c_5 \TT_6^\asf\,,
\nonumber\\
&& \Scl_6^\asf := - \frac{89}{9} c_{\frac{1}{5}} \TT_0^\asf -  \frac{59}{9} \TT_2^\asf -   3c_{\frac{1}{5}} \TT_4^\asf + \frac{25}{9} \TT_6^\asf\,,
\eeq
}
\!\!where, in \rf{manus-03052024-28}, we identify the generators $\Scl_2^\asf$, $\Scl_4^\asf$ , $\Scl_6^\asf$ which have not been fixed in \rf{manus-08042024-94ab}. In the basis \rf{manus-03052024-28}, the commutators \rf{manus-03052024-26} are represented as
{\small
\beq
\label{manus-03052024-29}  &&  [\TT_0^a,\TT_0^b] = f^{\asf\bsf\csf} \TT_0^\csf  \,, \quad [\Scl_2^a,\TT_0^b] = f^{\asf\bsf\csf} \Scl_2^\csf\,, \quad    [\Scl_4^a,\TT_0^b] = f^{\asf\bsf\csf} \Scl_4^\csf\,,\quad    [\Scl_6^a,\TT_0^b] = f^{\asf\bsf\csf} \Scl_6^\csf\,,\qquad
\nonumber\\
&&   [\Scl_2^\asf,\Scl_2^\bsf] = \frac{1}{9}c_5 f^{\asf\bsf\csf} \Scl_2^\csf  -\frac{5}{3} f^{\asf\bsf\csf} \Scl_4^\csf  - \frac{8}{9} c_5 f^{\asf\bsf\csf} \Scl_6^\csf\,,
\nonumber\\
&& [\Scl_2^\asf,\Scl_4^\bsf] = \frac{25}{9}f^{\asf\bsf\csf} \Scl_2^\csf  + \frac{11}{3} c_5 f^{\asf\bsf\csf} \Scl_4^\csf  + \frac{70}{9}  f^{\asf\bsf\csf} \Scl_6^\csf\,,
\nonumber\\
&&  [\Scl_2^\asf,\Scl_6^\bsf] = -\frac{16}{9} c_5 f^{\asf\bsf\csf} \Scl_2^\csf  - \frac{28}{3} f^{\asf\bsf\csf} \Scl_4^\csf  - \frac{34}{9} c_5 f^{\asf\bsf\csf} \Scl_6^\csf\,,
\nonumber\\
&& [\Scl_4^\asf,\Scl_4^\bsf] = - \frac{55}{9} c_5 f^{\asf\bsf\csf} \Scl_2^\csf  - \frac{85}{3} f^{\asf\bsf\csf} \Scl_4^\csf  - \frac{100}{9} c_5  f^{\asf\bsf\csf} \Scl_6^\csf\,,
\nonumber\\
&&  [\Scl_4^\asf,\Scl_6^\bsf] = \frac{140}{9} f^{\asf\bsf\csf} \Scl_2^\csf  + \frac{40}{3} c_5 f^{\asf\bsf\csf} \Scl_4^\csf  + \frac{230}{9}  f^{\asf\bsf\csf} \Scl_6^\csf\,,
\nonumber\\
&&  [\Scl_6^\asf,\Scl_6^\bsf] = - \frac{68}{9} c_5 f^{\asf\bsf\csf} \Scl_2^\csf  - \frac{92}{3} f^{\asf\bsf\csf} \Scl_4^\csf  - \frac{104}{9} c_5  f^{\asf\bsf\csf} \Scl_6^\csf\,.
\eeq
}
Using commutators \rf{manus-03052024-29}, we verify that the generators $S_2^\asf$, $S_4^\asf$, $S_6^\asf$ constitute a radical of the extended gauge algebra. In other words, we have the Levy-Maltsev decomposition,
\be
\TT_0^\asf\,, \ \Scl_2^\asf\,, \ \Scl_4^\asf\,, \ \Scl_6^\asf\,\, =\,\, \Scl_2^\asf\,, \ \Scl_4^\asf \,, \ \Scl_6^\asf \,\, \subsetext\,\, \TT_0^\asf\,.
\ee

In order to represent gauge transformations of the Stueckelberg fields $\varphi_{2n}$ and the corresponding field strengths $\FF_{2n+1}^A$, $n=1,2,3$, in the framework of extended gauge algebra we introduce
\beq
\label{manus-03052024-30} &&  \varphi := \varphi_2^\asf \Scl_2^\asf + \varphi_4^\asf \Scl_4^\asf + \varphi_6^\asf \Scl_6^\asf\,, \hspace{1.8cm} \FF^A : = \FF_3^{A\asf}  \Scl_2^\asf + \FF_5^{A\asf} \Scl_4^\asf+ \FF_7^{A\asf} \Scl_6^\asf\,,
\nonumber\\
&& \varphi_\subsm^A: = \varphi_3^{A\asf} \Scl_2^\asf + \varphi_5^{A\asf} \Scl_4^\asf + \varphi_7^{A\asf} \Scl_6^\asf\,, \qquad \eta_\subsm =  \eta_2^\asf \Scl_2^\asf +  \eta_4^\asf \Scl_4^\asf +  \eta_6^\asf \Scl_6^\asf\,.\qquad
\eeq

Using the definitions given in \rf{manus-03052024-27}, \rf{manus-03052024-30}, we find that the expressions for $\FF_3^A$, $\FF_5^A$, $\FF_7^A$ in \rf{manus-03052024-05} and the gauge transformations of $\varphi_{2n}$ and $\FF_{2n+1}^A$, $n=1,2,3$, given in \rf{manus-03052024-07add}, \rf{manus-03052024-10} can shortly be represented as in \rf{manus-08042024-126}. To this end the following commutators turn out to be helpful:
{\small
\beq
\label{manus-03052024-35} &&   [\Scl_2^\asf,\TT_0^\bsf] = f^{\asf\bsf\csf} \Scl_2^\csf\,,\qquad   [\Scl_4^\asf,\TT_0^\bsf] = f^{\asf\bsf\csf} \Scl_4^\csf\,,\qquad  [\Scl_6^\asf,\TT_0^\bsf] = f^{\asf\bsf\csf} \Scl_6^\csf\,,
\nonumber\\
&&   [\Scl_2^\asf,\TT_2^\bsf] = f^{\asf\bsf\csf} \Scl_4^\csf\,,
\nonumber\\
&&  [\Scl_2^\asf,\TT_4^\bsf] = -\frac{1}{3} f^{\asf\bsf\csf} \Scl_2^\csf + \frac{5}{3} f^{\asf\bsf\csf} \Scl_6^\csf\,,
\nonumber\\
&&   [\Scl_2^\asf,\TT_6^\bsf] = -   f^{\asf\bsf\csf} \Scl_4^\csf - c_5 f^{\asf\bsf\csf} \Scl_6^\csf\,,
\nonumber\\
&&   [\Scl_4^\asf,\TT_2^\bsf] = - \frac{5}{3} f^{\asf\bsf\csf} \Scl_2^\csf + \frac{10}{3} f^{\asf\bsf\csf} \Scl_6^\csf\,,
\nonumber\\
&&    [\Scl_4^\asf,\TT_4^\bsf] = - 7 f^{\asf\bsf\csf} \Scl_4^\csf - 5 c_5
f^{\asf\bsf\csf} \Scl_6^\csf\,,
\nonumber\\
&& [\Scl_4^\asf,\TT_6^\bsf] =  \frac{5}{3} f^{\asf\bsf\csf} \Scl_2^\csf  + 4c_5 f^{\asf\bsf\csf} \Scl_4^\csf + \frac{35}{3} f^{\asf\bsf\csf} \Scl_6^\csf\,,
\nonumber\\
&&  [\Scl_6^\asf,\TT_2^\bsf] =  - 4 f^{\asf\bsf\csf} \Scl_4^\csf  - 3 c_5 f^{\asf\bsf\csf} \Scl_6^\csf\,,
\nonumber\\
&&  [\Scl_6^\asf,\TT_4^\bsf] =  \frac{10}{3} f^{\asf\bsf\csf} \Scl_2^\csf  + 6 c_5  f^{\asf\bsf\csf} \Scl_4^\csf  + \frac{49}{3}  f^{\asf\bsf\csf} \Scl_6^\csf\,,
\nonumber\\
&&  [\Scl_6^\asf,\TT_6^\bsf] = - 2c_5 f^{\asf\bsf\csf} \Scl_2^\csf  - 14 f^{\asf\bsf\csf} \Scl_4^\csf  - 7c_5  f^{\asf\bsf\csf} \Scl_6^\csf\,.
\eeq
}
\ Finally, we note that Lagrangian \rf{manus-03052024-03} can be represented as in  \rf{manus-08042024-117}, \rf{manus-08042024-100add}, where all non-zero values of the invariant bilinear forms are given by
\beq
\label{manus-03052024-45} &&  \GG_{00} = k_\rho\,,\quad  \quad \GG_{22} = - k_\rho \,,\quad \GG_{44} = k_\rho \,,\quad \GG_{66} = - k_\rho \,,
\nonumber\\
&& \GG_{22}^\subsm = - 6\rho k_\rho\,, \quad \GG_{44}^\subsm = 10\rho k_\rho\,, \quad
\GG_{22}^\subsm = - 12\rho k_\rho\,,
\eeq
where $k_\rho$ is given in \rf{manus-03052024-03}. For $d=9$, the conjectured expressions \rf{manus-08042024-106} agree with \rf{manus-03052024-45}.

\newsection{\large Conclusions } \label{concl}

Using the embedding space method, we developed the ordinary-derivative formulation of conformal YM field in $(A)dS_{d+1}$ when $d=5,7,9$. We studied two ordinary-derivative formulations which we refer to as generic and decoupled formulations. The key feature of the generic formulation is that this formulation is obtainable from its flat space cousin in relatively straightforward way. Namely, the Lagrangian, gauge transformation, and conformal algebra transformations of conformal YM field in $(A)dS$ are regular in the flat space limit. Therefore, a knowledge of the ordinary-derivative formulation in the flat space allows us in a relatively straightforward way to restore its $(A)dS$ cousin. Other key feature of the generic formulation shared with the decoupled  formulation is that the two-derivative contributions to kinetic operators for  scalar and vector fields are realized as Klein-Gordon and Maxwell kinetic operators. Undesirable feature of the generic formulation is that some  kinetic terms are non-diagonal with respect to fields. This undesirable feature of the generic formulation is overcome in the decoupled formulation. In the decoupled formulation, two-derivative contributions to kinetic terms take the form of the standard Klein-Gordon and Maxwell {\it diagonal kinetic terms}, and importantly, the mass terms are also diagonal. Our results might have the following interesting applications and generalization.

{\bf i}) In the decoupled formulation of conformal YM field in $(A)dS$, the kinetic terms are governed by the standard diagonal kinetic terms of the massless and massive $(A)dS$ fields, where some kinetic terms enter with the wrong sign. This implies that a quantization of the conformal field in the framework of the decoupled formulation can be carried out by using more or less standard methods for massless and massive fields in $(A)dS$.

{\bf ii}) Scattering amplitudes for conformal fields in flat space were studied in Refs.\cite{Joung:2015eny,Beccaria:2016syk,Adamo:2018srx}.%
\footnote{ Scattering amplitudes for conformal fields can also be investigated by using twistors (see Ref.\cite{Adamo:2016ple}). Recent interesting applications of twistors to YM-like theories and higher-spin gravity may be found in Refs.\cite{Steinacker:2023zrb}-\cite{Basile:2022mif}.
}
We believe that our decoupled formulation creates new interesting possibilities to realize $S$-matrix program for conformal fields. In the decoupled formulation of conformal $(A)dS$ field,  the kinetic terms turn out to be standard kinetic terms of massless and massive $(A)dS$ fields. Therefore we can try to apply various methods for a study of scattering amplitudes of conformal fields, which were discussed for $(A)dS$ fields in the literature (see, e.g., Refs.\cite{Penedones:2010ue}-\cite{Melville:2023kgd} and references therein). We note also hidden conformal symmetries of scattering amplitudes for Einstein graviton field in Ref.\cite{Loebbert:2018xce} and the investigation of hidden conformal symmetries of higher-spin gravity in Refs.\cite{Vasiliev:2007yc}.

{\bf iii}) Conformal interactions between matter fields and conformal fields have actively been studied by many researchers (see,e.g., Refs.\cite{Manvelyan:2006bk,Bekaert:2010ky,Kuzenko:2022hdv} and references therein).
In Ref.\cite{Metsaev:2016rpa}, by using a framework of light-cone approach, we constructed all cubic interaction vertices for conformal vector and scalar fields in flat space of arbitrary even dimension. We expect that our approach in this paper creates the possibility for the relatively simple generalization of the light-cone gauge vertices to their Lorentz covariant and gauge invariant cousins to all orders in fields. It seems to be of some interest to apply our ordinary-derivative approach to the study of Lorentz covariant and gauge invariant conformal interactions between conformal YM vector field and matter fields.

{\bf iv}) In this paper, we studied conformal field in $(A)dS$ space of six, eight, and ten dimensions. In the recent time, conformal fields in three dimensions have attracted some interest (see, e.g., Refs.\cite{Grigoriev:2019xmp}-\cite{Ponomarev:2021xdq}).
Whether or not and in which way an application of our formulation to conformal fields in $(A)dS_3$ might be helpful remains to be understood.

{\bf v}) One-loop UV behaviour of  conformal (supersymmetric) YM theory in flat 6d was investigated in Refs.\cite{Ivanov:2005qf,Buchbinder:2020tnc,Casarin:2019aqw} (see also Ref.\cite{Casarin:2023ifl}). We think that representation of the $6d$ conformal YM theory in terms of one massless and massive vector fields in $(A)dS_6$ provides new interesting setup for the study of quantum properties of $6d$ conformal YM theory.

{\bf vi}) In the framework of the $AdS/CFT$ duality, an action of $AdS$ field
evaluated on the solution of the Dirichlet problem is referred to as effective action.
Action of conformal fields is realized then as UV divergence of the effective action.%
\footnote{
For example, for free graviton and higher-spin bulk fields, it was shown in Refs.\cite{Liu:1998ty} and Ref.\cite{Metsaev:2009ym} that UV divergence of effective action coincides with the free action of Weyl graviton and higher-spin conformal field. For higher-spin massive field bulk, computation of the effective action may be found in Ref.\cite{Metsaev:2011uy}, while, in Ref.\cite{Metsaev:2016oic}, it was shown that, for critical values of masses, the UV divergence of such effective action coincides with action of some general conformal field. Interrelation between ambient space $AdS$ fields and conformal fields are discussed in Ref.\cite{Bekaert:2012vt}. Interesting discussion of $AdS/CFT$ in the framework of bi-local field approach may be found in Refs.\cite{deMelloKoch:2023ylr}.
}
For other popular method for finding the action of conformal fields, which is based on the use of induced action approach, see, e.g., Refs.\cite{Bekaert:2010ky,Kuzenko:2022qeq} and references therein. In the literature, the Dirichlet problem is adopted for the computation of the action of conformal fields in the flat space. It will be interesting to adopt the Dirichlet problem for the computation of the action of conformal fields in $AdS$ space.

{\bf vii}) Shift symmetries were intensively studied in the literature (see, e.g., Refs.\cite{Bonifacio:2018zex,Bonifacio:2019hrj}). It would be interesting to learn about whether or not and in which way the shift symmetries are realized for the conformal YM field in $(A)dS_{d+1}$ studied in this paper.%
\footnote{For $(A)dS_4$, we note the interesting coincidence of the mass square for shift symmetric $k=0$ abelian massive vector field in Ref.\cite{Bonifacio:2019hrj} and mass square of the abelian massive vector field entering unitary subsector of $4d$ conformal gravity in Sec.6 in Ref.\cite{Metsaev:2007fq}.}

{\bf viii}) In Refs.\cite{Takata:2012mpa,Metsaev:2015yyv}, various ordinary-derivative BRST-BV formulations of free arbitrary spin conformal fields in flat space were developed. In Ref.\cite{Metsaev:2015yyv}, we shown how the BRST-BV Lagrangian of the conformal fields is related to the metric-like  Lagrangian of the conformal fields obtained in Ref.\cite{Metsaev:2007rw}. Generalization of the approach in Ref.\cite{Metsaev:2015yyv} to interacting conformal fields in $(A)dS$ along BRST-BV method in Refs.\cite{Dempster:2012vw}-\cite{Reshetnyak:2023oyj} could be of some interest. Recent application of the BRST-BV metric-like method to the study of interacting $AdS$ fields may be found in Refs.\cite{Reshetnyak:2023oyj}, while, for free $AdS$ fields, the embedding space BRST-BV method  was studied in Ref.\cite{Bekaert:2023uve}.

\setcounter{section}{0}\setcounter{subsection}{0}
\appendix{\large Relations for embedding and intrinsic spaces}

\noinbf{Useful relations of embedding space method}.  Tangential derivative $\ppp^A$ with respect to coordinates $y^A$ satisfies the following commutators:
\beq
\label{12062024-01} && [\ppp^A,y^B] = \theta^{AB}\,,\hspace{3.1cm} \theta^{AB} : =
\eta^{AB} - \rho y^A y^B\,,
\nonumber\\
&& [\ppp^A,\ppp^B] = \rho L^{AB} \,, \hspace{2.8cm} L^{AB} : = y^A\ppp^B - y^B\ppp^A \,.
\nonumber\\
&& [\ppp^2,y^A] =  2 \ppp^A  - (d+1) \rho  y^A\,, \qquad p^2:= p^A p^A\,,
\nonumber\\
&& y^A \ppp^A = 0 \,,\qquad \ppp^A y^A = d+1\,.
\eeq

Helpful integral relations involving the derivative $\ppp^A$ and arbitrary rank tensor fields denoted as $f_1$, $f_2$, and $f$ are given by
\beq
\label{12062024-10} && \int d\sigma\,\, f_1 \ppp^A f_2  =    \int d\sigma\,\,  \big( -f_2 \ppp^A f_1  + (d+1)\rho y^A f_1 f_2 \big) \,,
\nonumber\\
&& \int d\sigma\,\, \ppp^A f  =   (d+1)\rho \int d\sigma\,\, y^A f \,, \qquad d\sigma := d^{d+2}y\,\, \delta(y^Ay^A-\frac{1}{\rho})\,.
\eeq

\noinbf{Interrelations between embedding and intrinsic formulations}. Let $x^\mun$, $\mun=0,1,\ldots, d$, be the intrinsic coordinates in $(A)dS_{d+1}$ and let $y^A(x)$ be embedding map, where the $y^A(x)$ obey restriction \rf{manus-08042024-01}. Introducing the notation
\be \label{manus-12062024-15}
y_\mun^A := \partial_\mun y^A\,, \qquad  y^{A\mun} :=  g^{\mun\nun}y_\nun^A\,,\qquad \partial_\mun:=\partial/\partial x^\mun\,,
\ee
we obtain the following relations for the intrinsic geometry metric tensor $g_{\mun\nun}$ and the symbol $\theta^{AB}$:
\be \label{manus-12062024-20}
g_{\mun\nun} = y_\mun^A y_\nun^A\,, \qquad  \theta^{AB} = y^{A\mun} y_\mun^B\,.
\ee
We note then following helpful relations:
\beq
\label{manus-12062024-25 }&& \ppp^A x^\mun = y^{A\mun}\,, \hspace{1cm} y_\mun^A \ppp^A x^\nun = \delta_m^n\,, \hspace{1cm}  D_\mun y_\nun^A = - \rho g_{\mun\nun} y^A\,,
\nonumber\\
&& \ppp^A y_\mun^B = \Gamma_{\mun\nun}^l y^{A\nun} y_l^B - \rho y_\mun^A y^B\,,\qquad \ppp^A y_\mun^A = \Gamma_{\mun\lun}^\lun\,, \qquad \ppp^A y^{A\mun} = - \Gamma_{\nun\lun}^\mun g^{\nun\lun}\,,\qquad
\eeq
where $\Gamma_{\mun\nun}^\lun$ stands for the Cristoffel symbol with respect to the metric  $g_{\mun\nun}$, while $D_\mun$ is the covariant derivative, $D_\mun = \partial_\mun + \Gamma_{\mun \star }^\star$. The relationships between embedding space vector $\phi^A$ satisfying the constraint $y^A \phi^A=0$ and intrinsic coordinates vector $\phi_\mun$ are given by
\be \label{manus-12062024-30}
\phi^A = y_\mun^A \phi^\mun\,, \qquad \phi^\mun = y^{A\mun}\phi^A\,, \qquad \phi^\mun = g^{\mun\nun}\phi_\nun\,.
\ee
We note the following helpful relations for the vectors $\phi^A$, $\phi^\mun$, and a scalar $\phi$:
\beq
\label{manus-12062024-35}&& \ppp^A \phi^B = y^{A\mun} y^{B\nun} D_\mun\phi_\nun - \rho y^B \phi^A\,,
\qquad  y_\mun^B \ppp^A \phi^B = y^{A\nun}  D_\nun\phi_\mun \,,
\nonumber\\
&& y_\mun^A y_\nun^B \ppp^A \phi^B = D_\mun\phi_\nun\,,
\nonumber\\
&& y_\mun^A \ppp^2 \phi^A  = D^\nun D_\nun\phi_\mun - \rho \phi_\mun\,, \qquad \ppp^2 \phi  = D^\nun D_\nun \phi\,.
\eeq
For the scalar products, we get the relations
\be
\ppp^A \phi^B\ppp^A \phi^B =  D^\mun \phi^\nun D_\mun \phi_\nun  +  \rho \phi^\mun \phi_\mun\,,
\qquad \ppp^A \phi^B \ppp^B \phi^A =  D^\mun \phi^\nun D_\nun \phi_\mun \,.
\ee
For the intrinsic space cousins of the field strengths $F^{AB}$ \rf{manus-08042024-92} and $F^A$ \rf{manus-08042024-96} defined as $F_{\mun\nun} = y_\mun^A y_\nun^B F^{AB}$ and $F_\mun = y_\mun^A F^A$ respectively, we get the relations
\beq
&& F_{\mun\nun} = \partial_\mun \phi_\nun - \partial_\nun\phi_\mun + [\phi_\mun,\phi_\nun] \,, \qquad F_\mun = \phi_{\subsm\, \mun} + \partial_\mun \phi + [\phi_\mun,\phi]\,, \qquad
\nonumber\\
&& F^{AB} F^{AB} = F^{\mun\nun} F_{\mun\nun}\,, \hspace{2.7cm} F^A F^A  = F^\mun F_\mun\,.
\eeq

\appendix{\large Extended gauge algebra and field redefinitions}

\noinbf{Extended gauge algebra of conformal YM field in $R^{d,1}$}. We start with the update of our presentation of extended gauge algebra for conformal YM field in $R^{d,1}$ found in Ref.\cite{Metsaev:2023qif}.  Let us use the following notation for generators of the extended gauge algebra:
\be \label{01062024-01}
\Tch_{2m}^\asf\,, \qquad m = 0,1,\ldots,N\,, \qquad N:= \half(d-3)\,.
\ee
Then, in terms of the generators $\Tch_{2m}^\asf$, the commutator of two gauge transformations given in (6.14) in Ref.\cite{Metsaev:2023qif}, amounts to the following commutator:
{\small
\be \label{01062024-01-a1}
[\Tch_{2m}^\asf,\Tch_{2n}^\bsf] = e_{m,n} f^{\asf\bsf\csf} \Tch_{2m+2n}^\csf\,, \qquad e_{m,n} := \frac{(m+n)!}{m!n!} \frac{(N-m)! (N-n)!}{N! (N-m-n)!} \,,
\ee
}
\!where, in commutator \rf{01062024-01-a1}, we use the convention $\Tch_{2m+2n}^\asf=0$ for $m+n> N$. Introducing new basis of the generators denoted as $\Tbf_{2m}^\asf$,
{\small
\be \label{01062024-02}
\Tbf_{2m}^\asf : = e_m \Tch_{2m}^\asf\,, \qquad e_m : = \frac{ N! \, m! }{(N-m)! } \,, \qquad e_{m,n} = \frac{e_{m+n}}{e_me_n} \,,
\ee
}
\!we represent the commutators \rf{01062024-01-a1}  in a more simple form
\be \label{01062024-05}
[\Tbf_{2m}^\asf,\Tbf_{2n}^\bsf] = f^{\asf\bsf\csf} \Tbf_{2m+2n}^\csf\,,\qquad \Tbf_{2m+2n}^\asf := 0 \ \ \hbox{ for} \  \ m+n> N.
\ee

\noinbf{ Change of basis of the extended gauge algebra generators and field redefinitions in generic formulation}.%
\footnote{In various contexts, the discussions of field redefinitions for conformal fields may be found in Refs.\cite{Metsaev:2007fq,Beccaria:2016syk,Basile:2022nou}. Interesting general method for the analysis of field redefinitions is discussed in Ref.\cite{Spirin:2024zgy}.
}
First, we show that by a suitable change of basis for the generators of the extended gauge algebra $T_{2n}^\asf$ the commutators \rf{manus-02052024-26} can be cast into the form of commutators for extended gauge algebra in $R^{d,1}$ given in \rf{01062024-01-a1}. To this end, in place of the generators $T_{2n}^\asf$, we introduce new generators $\Tch_{2n}^\asf$ given by
\be \label{01062024-10}
\Tch_0^\asf=T_0^\asf\,,\qquad \Tch_2^\asf=T_2^\asf - \frac{14}{3}\rho T_4^\asf\,, \qquad
\Tch_4^\asf=T_4^\asf\,, \qquad
\Tch_6^\asf=T_6^\asf\,.
\ee
Using commutators \rf{manus-02052024-26}, we then verify that commutators for generators $\Tch_{2n}^\asf$ coincide with the commutators in $R^{d,1}$ given in \rf{01062024-01-a1}, when $d=9$. In other words, the commutators for $\Tch_{2n}^\asf$ takes the same form as in \rf{manus-02052024-26} when $\rho=0$.

Second, we show that the change of the basis for the generators $T_{2n}^\asf$ \rf{01062024-10} amount to field redefinitions. Namely, we note that the expression for $\phi^A$ defined in \rf{manus-02052024-25} can be represented as
{\small
\beq
\label{01062024-15} && \phi^A = \phich^A\,, \qquad \phich^A := \sum_{n=0,1,2,3} \phich_{2n+1}^{A\asf} \Tch_{2n}^\asf\,,
\nonumber\\
&& \phich_1^A := \phi_1^A\,,\qquad \phich_3^A := \phi_3^A\,, \qquad
\phich_5^A := \phi_5^A + \frac{14}{3}\rho \phi_3^A\,, \qquad
\phich_7^A := \phi_7^A\,.
\eeq
}
\!The 1st relation in \rf{01062024-15} demonstrates explicitly that the connection $\phi^A$ can be represented in terms of the new connection $\phich^A$ which is expanded in terms of the generators $\Tch_{2n}^\asf$ which obey the commutators of the extended gauge algebra in $R^{d,1}$ \rf{01062024-01-a1}. Note however the use of the new auxiliary fields $\phich_{2n+1}^{A\asf}$ breaks the representation of the Lagrangian in terms of the quantities $I_{2n+4}$ \rf{manus-02052024-03}. In this paper, we prefer to use the quantities $I_{2n+4}$ to build Lagrangian \rf{manus-02052024-03}. For this reason,  we prefer to use the generators and commutators given in \rf{manus-02052024-26}.

\appendix{\large Map of fields of generic formulation to fields of decoupled formulation}

In the framework of the generic formulation, we collect the vector fields, their field strengths, and gauge parameters of the extended gauge algebra into $N+1\, \times 1$ matrices,
{\small
\be \label{02062024-01}
\phi_{_{N+1\,\times 1}}^A = \left(
\begin{array}{c}
\phi_1^A
\\[5pt]
\phi_3^A
\\[5pt]
\vdots
\\[5pt]
\phi_{_{2N+1}}^A
\end{array}\right),
\qquad
F_{_{N+1\,\times 1}}^{AB} = \left(
\begin{array}{c}
F_2^{AB}
\\[5pt]
F_4^{AB}
\\[5pt]
\vdots
\\[5pt]
F_{_{2N+2}}^{AB}
\end{array}\right),
\qquad
\xi_{_{N+1\,\times 1}} = \left(
\begin{array}{c}
\xi_0
\\[5pt]
\xi_2
\\[5pt]
\vdots
\\[5pt]
\xi_{_{2N}}
\end{array}\right),\qquad
\ee
}
\!\!while the Stueckelberg fields, their field strengths, gauge parameters of the radical, and vector fields of the radical are collected into $N\times 1$ matrices,
{\small
\be \label{02062024-05}
\phi_{_{N\,\times 1}} \,\,\, = \,\, \, \left(\!\!
\begin{array}{c}
\phi_2
\\[5pt]
\phi_4
\\[5pt]
\vdots
\\[5pt]
\phi_{_{2N}}
\end{array}\!\!\right),
\quad \ \ \ \
F_{_{N\,\times 1}}^A \,\,\, =\,\,\, \left(\!\!
\begin{array}{c}
F_3^A
\\[5pt]
F_5^A
\\[5pt]
\vdots
\\[5pt]
F_{_{2N+1}}^A
\end{array}\!\!\right),
\quad
\xi_{_{\subsm\, N\times 1}} \,\, =\,\, \left(\!\!
\begin{array}{c}
\xi_2
\\[5pt]
\xi_4
\\[5pt]
\vdots
\\[5pt]
\xi_{_{2N}}
\end{array}\!\!\right),
\quad
\phi_{_{\subsm\, N\times 1}}^A \,\, =\,\, \left(\!\!
\begin{array}{c}
\phi_3^A
\\[5pt]
\phi_5^A
\\[5pt]
\vdots
\\[5pt]
\phi_{_{2N+1}}^A
\end{array}\!\!\right),\qquad
\ee
}
\!\!where $N=\frac{d-3}{2}$. Also we introduce the $N+1\times 1$ and $N\times 1$ matrices for generators of the extended gauge algebra and the generators of the radical,

{\small
\be \label{02062024-10}
T_{_{N+1\,\times 1}}^\asf = \left(
\begin{array}{c}
T_0^\asf
\\[5pt]
T_2^\asf
\\[5pt]
\vdots
\\[5pt]
T_{_{2N}}^\asf
\end{array}\right),
\qquad
S_{_{N\times 1}}^\asf \,\, =\,\, \left(
\begin{array}{c}
S_2^\asf
\\[5pt]
S_4^\asf
\\[5pt]
\vdots
\\[5pt]
S_{_{2N}}^\asf
\end{array}\right).
\ee
}

For the decoupled formulation, the corresponding $N+1\times 1$ and $N \times 1$ matrices are defined analogically,
{\small
\beq
\label{02062024-15} && \hspace{-0.5cm} \varphi_{_{N+1\,\times 1}}^A\! =\! \left(\!\!
\begin{array}{c}
\varphi_1^A
\\[5pt]
\varphi_3^A
\\[5pt]
\vdots
\\[5pt]
\varphi_{_{2N+1}}^A
\end{array}\!\!\right),
\quad
\FF_{_{N+1\,\times 1}}^{AB} = \left(\!\!
\begin{array}{c}
\FF_2^{AB}
\\[5pt]
\FF_4^{AB}
\\[5pt]
\vdots
\\[5pt]
\FF_{_{2N+2}}^{AB}
\end{array}\!\!\right),
\quad
\eta_{_{N+1\,\times 1}} = \left(\!\!
\begin{array}{c}
\eta_0
\\[5pt]
\eta_2
\\[5pt]
\vdots
\\[5pt]
\eta_{_{2N}}
\end{array}\!\!\right),
\nonumber\\
&& \hspace{-0.5cm} \varphi_{_{N\,\times 1}} \,\,\, = \,\, \, \left(\!\!
\begin{array}{c}
\varphi_2
\\[5pt]
\varphi_4
\\[5pt]
\vdots
\\[5pt]
\varphi_{_{2N}}
\end{array}\!\!\right)\,,
\quad
\FF_{_{N\,\times 1}}^A \,\,\, =\,\,\, \left(\!\!
\begin{array}{c}
\FF_3^A
\\[5pt]
\FF_5^A
\\[5pt]
\vdots
\\[5pt]
\FF_{_{2N+1}}^A
\end{array}\!\!\right)\,,
\quad
\eta_{_{\subsm\, N\times 1}} \,\, =\,\, \left(\!\!
\begin{array}{c}
\eta_2
\\[5pt]
\eta_4
\\[5pt]
\vdots
\\[5pt]
\eta_{_{2N}}
\end{array}\!\!\right)\,,
\quad
\varphi_{_{\subsm\, N\times 1}}^A \,\, =\,\, \left(\!\!
\begin{array}{c}
\varphi_3^A
\\[5pt]
\varphi_4^A
\\[5pt]
\vdots
\\[5pt]
\varphi_{_{2N}}^A
\end{array}\!\!\right)\,,
\nonumber\\
&& \hspace{-0.5cm} \TT_{_{N+1\,\times 1}}^\asf = \left(
\begin{array}{c}
\TT_0^\asf
\\[5pt]
\TT_2^\asf
\\[5pt]
\vdots
\\[5pt]
\TT_{_{2N}}^\asf
\end{array}\right)\,,
\quad
\Scl_{_{N\times 1}}^\asf \,\, =\,\, \left(
\begin{array}{c}
\Scl_2^\asf
\\[5pt]
\Scl_4^\asf
\\[5pt]
\vdots
\\
\Scl_{_{2N}}^\asf
\end{array}\right)\,.
\eeq
}

Vector fields, their field strengths, and gauge parameters of the generic formulation and the ones of the decoupled formulation are connected by the relations
{\small
\beq
\label{02062024-20} && \phi_{_{N+1\,\times 1}}^A = L_{_{N+1\, \times N+1\, }}^\Tsm \varphi_{_{N+1\,\times 1}}^A\,, \quad F_{_{N+1\,\times 1}}^{AB} = L_{_{N+1\, \times N+1\, }}^\Tsm \FF_{_{N+1\,\times 1}}^{AB}\,,
\nonumber\\
&& \hspace{2cm} \xi_{_{N+1\,\times 1}} = L_{_{N+1\, \times N+1\, }}^\Tsm \eta_{_{N+1\,\times 1}}\,,
\eeq
}
\!\!while Stueckelberg fields, their field strengths, and gauge parameters of the radical, and vector fields of the radical entering the generic and decoupled formulations are connected by the relations
{\small
\beq
\label{02062024-25} &&\phi_{_{N\,\times 1}} = M_{_{N\,\times N}}^\Tsm K_{_{N\,\times N}} \varphi_{_{N\,\times 1}}\,,   \hspace{1.5cm}  F_{_{N\,\times 1}}^A = M_{_{N\,\times N}}^\Tsm K_{_{N\,\times N}} \FF_{_{N\,\times 1}}^A\,,
\nonumber\\
&& \xi_{_{\subsm\, N\,\times 1}} = M_{_{N\,\times N}}^\Tsm K_{_{N\,\times N}} \eta_{_{\subsm\, N\,\times 1}}\,, \qquad \phi_{_{\subsm\, N\,\times 1}}^A = M_{_{N\,\times N}}^\Tsm K_{_{N\,\times N}} \varphi_{_{\subsm\,N\,\times 1}}^A\,,
\eeq
}
\!\!where $L_{_{N+1\,\times N+1}}$ stands for $N+1\times N+1$ matrix given by
{\small
\beq
\label{02062024-30}  && L_{_{N+1\,\times N+1}} = \left(
\begin{array}{cc}
1   &\,\, 0_{_{1\times N}}
\\[7pt]
0_{_{N\times 1}} &\,\, K_{_{N\times N}}
\end{array}\right)
\left(
\begin{array}{cc}
1   &\,\, 0_{_{1\times N}}
\\[7pt]
1_{_{N\,\times 1}} &\,\, M_{_{N\,\times N}}
\end{array}\right)\,,
\qquad
1_{_{N\times 1}}  \, :=\, \left(
\begin{array}{c}
1
\\[5pt]
\vdots
\\[5pt]
1
\end{array}\right)\,.\qquad
\eeq
}
In \rf{02062024-25}, \rf{02062024-30}, the $M_{_{N\times N}}$ is a $N\times N$ matrix, while the $K_{_{N\times N}}$ is $N\times N$ diagonal matrix. The $L_{_{N+1\times N+1}}^\Tsm$ and $M_{_{N\times N}}^\Tsm$ stand for transposed matrices.

The relationships between the generators of the extended gauge algebra and the generators of the radical entering the generic and decoupled formulations are given by
\be \label{02062024-35} T_{_{N+1\, \times 1}}^\asf = L_{_{N+1\,\times N+1}}^{-1} \TT_{_{N+1\,\times 1}}^\asf\,, \hspace{1.3cm} S_{_{N\,\times 1}}^\asf = M_{_{N\,\times N}}^{-1} K_{_{N\,\times N}}^{-1} \Scl_{_{N\,\times 1}}^\asf\,.
\ee
Transformation rules given in \rf{02062024-20}-\rf{02062024-30} imply the following invariance relations:
\beq
&& \phi^A = \varphi^A\,, \qquad F^{AB}=\FF^{AB}\,, \qquad \xi =\eta\,,
\nonumber\\
&& \phi =\varphi\,, \hspace{1.3cm} F^A=\FF^A\,, \hspace{1.3cm} \xi_\subsm =\eta_\subsm\,,\qquad  \phi_\subsm^A = \varphi_\subsm^A\,,
\eeq
where we use the notation given in \rf{manus-08042024-91}, \rf{manus-08042024-95} and \rf{manus-08042024-91ab}, \rf{manus-08042024-95ab}.

For $(A)dS_{d+1}$, with $d=5,7,9$, we verified relations given in \rf{02062024-20}-\rf{02062024-30}, while, for $d\geq 11$, these relations should be considered as our conjecture. For $d=5,7,9$, we now present the explicit expressions for matrices $L_{_{N+1\,\times N+1}}$, $M_{_{N\,\times N}}$, and $K_{_{N\,\times N}}$.

\noinbf{$(A)dS_6$, $d=5$}. For this case, we find
\beq
&& L_{_{2\times 2}} = \left(
\begin{array}{cc}
1   &\,\, 0
\\[7pt]
-1 &\,\, -2\rho
\end{array}\right),
\qquad
M_{1\times 1} = 2\rho\,, \qquad K_{1\times 1} = -1\,.
\eeq

\noinbf{$(A)dS_8$, $d=7$}. Using the notation $c_q:=\sqrt{q}$, our solution for this case is given by
{\small
\be
\label{07052024-45} L_{_{3\times 3}} = \left(
\begin{array}{ccc}
1   &  0   & 0
\\[7pt]
-c_3   &  -4\rho c_3   & - 4^2\rho^2 c_3
\\[7pt]
c_2   & 6\rho c_2   &  6^2 \rho^2 c_2
\end{array}\right),
\quad
M_{_{2\times 2}} := \left(
\begin{array}{cc}
4\rho & 4^2\rho^2
\\[5pt]
6\rho & 6^2\rho^2
\end{array}\right),
\quad  K_{_{2\times 2}} = \left(
\begin{array}{cc}
-c_3   & 0
\\[7pt]
0 &  c_2
\end{array}\right).
\ee
}

\noinbf{$(A)dS_{10}$, $d=9$}. Using the notation $c_q:=\sqrt{q}$, our solution for this case is given by
{\small
\beq
&& L_{_{4\times 4}} := \left(
\begin{array}{cccc}
1& 0 & 0 & 0
\\
-c_5 & -6\rho c_5& - 6^2\rho^2 c_5 & - 6^3\rho^3 c_5
\\[5pt]
3 & 3\cdot 10\rho & 3\cdot 10^2\rho^2 & 3\cdot 10^3\rho^3
\\[5pt]
-c_5  & -12\rho c_5 & - 12^2\rho^2 c_5 & - 12^3\rho^3 c_5
\end{array}\right)\,,
\nonumber\\[5pt]
&& M_{_{3\times 3}} := \left(
\begin{array}{ccc}
6\rho & 6^2\rho^2 & 6^3\rho^3
\\[3pt]
10\rho & 10^2\rho^2 & 10^3\rho^3
\\[3pt]
12\rho & 12^2\rho^2 & 12^3\rho^3
\end{array}\right)\,,
\qquad
K_{_{3\times 3}} = \left(
\begin{array}{ccc}
-c_5   & 0   & 0
\\[3pt]
0 & 3   & 0
\\[3pt]
0 & 0   & - c_5
\end{array}\right).
\eeq
}

\appendix{\large Flat space limit of $K^A$- transformations}

First, we show how the generators of conformal algebra $K^A$, $J^{AB}$ \rf{manus-08042024-60} are related to their flat space cousins. Second, we show how $K^A$-transformations for primary $(A)dS$ field $\phi_1^A$ \rf{manus-09042024-50} are related to conformal boost and dilatation transformations of primary vector field in the flat space.

\noinbf{i}) Using the decomposition
\beq
\label{manus-21062024-01} && J^{AB} =  J^{0'a},\, J^{ab}\,, \qquad K^A   = K^a,\, K^{0'}\,, \qquad \hbox{ for } \ AdS;
\nonumber\\
&& J^{AB} =  J^{1'a},\, J^{ab}\,, \qquad K^A = K^a,\, K^{1'},\,\qquad \hbox{ for } \ dS;\qquad \qquad  a,b = 0,1,\ldots , d\,,\qquad
\eeq
we introduce translation generators in $(A)dS$ denoted as $P^a$ and generator $K$ which is $(A)dS$ cousin of the standard dilatation generator in flat space,
{\small
\beq
\label{manus-21062024-05} && P^a := R^{-1} J^{0'a}\,, \qquad D :=  R^{-1} K^{0'}\,,\qquad \hbox{ for } \ AdS;
\nonumber\\
&& P^a := R^{-1} J^{1'a}\,,\qquad  D : = R^{-1}  K^{1'}\,,\qquad \hbox{ for } \  dS\,.
\eeq
}
\!Using notation \rf{manus-21062024-05}, the generators of the conformal algebra given in \rf{manus-08042024-40} are represented as
\be \label{manus-21062024-10}
P^a\,, \qquad K^a\,, \qquad D\,, \qquad J^{ab} \,, \hspace{1cm} \hbox{ for } \ (A)dS_{d+1}\,,
\ee
where indices the $a,b$ take values as in \rf{manus-21062024-01}.
Note however that we prefer to use the basis of generators which is the slight modification of  basis \rf{manus-21062024-10},
\be \label{manus-21062024-15}
\Psf^a := P^a\,, \hspace{1cm} \Jsf^{ab} := J^{ab}\,, \qquad \Ksf^a := K^a + \rho^{-1} P^a\,,\qquad \Dsf := D\,.
\ee
Second, we proceed with the use of the  generators of the conformal algebra of $(A)dS_{d+1}$ defined in \rf{manus-21062024-15},
\be \label{manus-21062024-20}
\Psf^a\,, \qquad \Ksf^a\,, \qquad \Dsf\,, \qquad \Jsf^{ab} \,, \hspace{1cm} \hbox{ for } \ (A)dS_{d+1}\,.
\ee
In terms of the generators \rf{manus-21062024-20}, commutators \rf{manus-08042024-60} take the form
\beq
\label{manus-21062024-25} &&  [\Psf^a,\Jsf^{bc}]=\eta^{ab} \Psf^c -\eta^{ac} \Psf^b\,, \hspace{1cm} [\Psf^a,\Psf^b] = - \rho \Jsf^{ab}\,, \hspace{0.7cm} [\Jsf^{ab}, \Jsf^{ce}]=\eta^{bc} \Jsf^{ae}+3\hbox{ terms} \,,\qquad
\nonumber\\
&& [\Ksf^a, \Jsf^{bc}] = \eta^{ab} \Ksf^c - \eta^{ac} \Ksf^b\,, \hspace{1cm} [\Psf^a, \Ksf^b] = \eta^{ab} \Dsf - \Jsf^{ab}\,,
\nonumber\\
&& [\Psf^a,\Dsf] = \Psf^a - \rho\Ksf^a \,, \hspace{2cm}  [\Ksf^a,\Dsf] = - \Ksf^a\,,
\eeq
where $[\Ksf^a,\Ksf^b]=0$. In the flat space limit $\rho \rightarrow 0$, we get
\be
\label{manus-21062024-30} \Psf^a|_{_{\rho\rightarrow 0}} \longrightarrow  \Pch^a\,,\hspace{0.7cm} \Jsf^{ab}|_{_{\rho\rightarrow 0}} \longrightarrow  \Jch^{ab}\,, \hspace{0.7cm} \Dsf|_{_{\rho\rightarrow 0}} \longrightarrow  \Dch\,,\hspace{0.7cm} \Ksf^a|_{_{\rho\rightarrow 0}} \longrightarrow   \Kch^a\,,
\ee
where the checked generators are the standard conformal algebra generators in the flat space whose non-trivial commutators are given by
{\small
\beq
\label{manus-21062024-35} && \hspace{-1cm} [\Pch^a,\Jch^{bc}]=\eta^{ab} \Pch^c -\eta^{ac} \Pch^b\,,\hspace{0.5cm} [\Dch,\Pch^a]=-\Pch^a\,, \hspace{1cm} [\Jch^{ab}, \Jch^{ce}]=\eta^{bc} \Jch^{ae}+3\hbox{ terms} \,,
\nonumber\\
&&  \hspace{-1cm}  [\Kch^a, \Jch^{bc}] = \eta^{ab} \Kch^c - \eta^{ac} \Kch^b\,, \hspace{0.5cm}  [\Dch,\Kch^a] = \Kch^a\,, \hspace{1cm} [\Pch^a,\Kch^b] = \eta^{ab} \Dch - \Jch^{ab}\,.
\eeq
}
\!Relations \rf{manus-21062024-05}-\rf{manus-21062024-25} show explicitly how the generators $K^A$, $J^{AB}$ space defined for $(A)dS$ are related to their flat space cousins.

\noinbf{ii}) To consider the flat space limit of the $K^A$-transformations of the primary $(A)dS$ field $\phi_1^A$ \rf{manus-09042024-50} we find it convenient to use the intrinsic Lorentz coordinates $x^\mun$ defined by the relations
{\small
\beq
\label{manus-21062024-40} && y^{0'} = RY \quad \hbox{for } \ \ AdS\,; \hspace{2cm}  y^{1'} = RY  \quad \hbox{for } \ \ dS\,;
\nonumber\\
&& y^a = \frac{x^a}{1 + \frac{1}{4} \rho x^2}\,, \qquad Y := \frac{1- \frac{1}{4}\rho x^2}{1 + \frac{1}{4} \rho x^2}\,, \qquad x^2:= \eta_{\mun\nun} x^\mun x^\nun\,,\qquad x^a:=\delta_\mun^a x^\mun\,, \qquad
\eeq
}
\!where $\eta_{\mun\nun}$ is a flat metric, while $\delta_\mun^a$ is the delta Kronecker symbol.  We recall that, in the frame of Lorentz coordinates, a square of the line element of the $(A)dS$ space is given by
\be
ds^2 = \big(1+\frac{1}{4} \rho x^2\big)^{-2} \eta_{\mun\nun} dx^\mun dx^\nun\,.
\ee
In the frame of the intrinsic coordinates, a cousin of the embedding space vector $\phi_1^A$ is denoted as $\phi_1^\mun$ and defined by the relation $\phi_1^\mun : =y^{A\mun} \phi_1^A$. Using \rf{manus-08042024-16}, \rf{manus-09042024-50}, and \rf{manus-12062024-35}, we find that $\Ksf^a$- and $\Dsf$- transformations of $\phi_1^\mun$ take the form
\beq
\label{manus-21062024-45}  && \hspace{-1.3cm} \delta_{\Ksf^a} \phi_1^\mun = \rho^{-1}  y_\nun^a \big( (Y-1) D^\nun \phi_1^\mun + Y^\mun \phi_1^\nun \big) + y^a \phi_1^\mun  -   \frac{1}{\rho} Y_\nun (y^a D^\nun\phi_1^\mun + y^{a\mun} \phi_1^\nun\big)\,,\qquad
\nonumber\\
&& \hspace{-1.3cm}  \delta_{\Dsf} \phi_1^\mun =  \big(- \rho^{-1} Y^\nun  D_\nun  + Y\big) \phi_1^\mun\,,
\nonumber\\
&& \delta_{P^a} \phi_1^\mun = y_\nun^a \big(Y D^\nun \phi_1^\mun + Y^\mun \phi_1^\nun\big) - Y_\nun (y^a D^\nun\phi_1^\mun + y^{a\mun} \phi_1^\nun\big)\,,
\nonumber\\
&& \delta_{K^a} \phi_1^\mun = - \rho^{-1} y_\nun^a D^\nun \phi_1^\mun + y^a \phi_1^\mun\,,\qquad Y_\mun:=\partial_\mun Y\,,    \qquad Y^\mun:= g^{\mun\nun} Y_\nun\,,
\eeq
where $\delta_{\Ksf^a} \phi_1^\mun$ is obtained by using the definition of $\Ksf^a$ in \rf{manus-21062024-15} and the relations for $\delta_{P^a} \phi_1^\mun$ and $\delta_{K^a} \phi_1^\mun$ in \rf{manus-21062024-45}. In the flat space limit, $\rho \sim 0$, using the asymptotic  relations
\be \label{manus-21062024-50}
Y-1 \sim  - \half \rho x^2 \,, \qquad Y^\mun \sim - \rho x^\mun \,, \qquad y^a \sim \delta_\mun^a x^\mun \,, \qquad \qquad y_\mun^a \sim \delta_\mun^a\,,
\ee
we find the following $\Ksf^a$- and $\Dsf$- transformations of the primary field $\phi_1^\mun$:
\beq
\label{manus-21062024-55} && \hspace{-0.5cm} \delta_{\Ksf^a} \phi_1^\mun \sim  \big( -\half x^2 \partial^a + x^a \Dch \big) \phi_1^\mun + \eta^{a\mun} x_\nun \phi_1^\nun - x^\mun \phi_1^a\,,\qquad \delta_{\Dsf^a} \phi_1^\mun \sim \Dch \phi_1^\mun\,, \qquad
\nonumber\\
&& \hspace{1cm} \Dch = x^\nun\partial_\nun + 1\,, \qquad \phi_1^a:= \delta_\mun^a \phi_1^\mun\,,\qquad \partial^a := \eta^{a\mun}\partial_\mun\,,\qquad
\eeq
which tell us that, in flat space limit, the $\Ksf^a$- and $\Dsf$- transformations of primary field in $(A)dS$ coincide with the conformal boost and dilatation transformations of primary field in flat space.

Note that $\Ksf^a$-transformations \rf{manus-21062024-15} are regular in the flat space limit. Relations \rf{manus-21062024-15} as well as the explicit expressions for the operator $K_\Delta^A$ \rf{manus-08042024-75} then imply that
\be \label{manus-21062024-56}
\rho K_\Delta^a  \sim  - \partial^a\,, \quad \hbox{for} \quad \rho \sim 0\,.
\ee
Using \rf{manus-21062024-56} and repeating our analysis above-given, it is easy to verify that the $K^A$-transformations for the auxiliary field and the Stueckelberg field given in \rf{manus-09042024-50} amount to their flat space cousins given in Sec.3 in Ref.\cite{Metsaev:2023qif}.


\end{document}